\documentclass[review]{elsarticle}

\usepackage{lineno,hyperref}
\usepackage{tikz}
\usepackage{geometry}
\usepackage{framed,multirow}
\usepackage{amsmath,amssymb}
\usepackage{latexsym}
\usepackage{url}
\usepackage{xcolor}
\usepackage{enumitem} 
\usepackage{graphicx}
\usepackage{tikz}
\usetikzlibrary{hobby}
\usetikzlibrary{calc}
\graphicspath{figures/}
\usepackage{subcaption}
\usepackage{multirow}
\usepackage{tabularx}
\usepackage{longtable}
\usepackage{rotating}
\usepackage{float}
\usepackage{todonotes}

\modulolinenumbers[5]

\journal{Elsevier}

\begin{document}

\begin{frontmatter}

\title{Alpine Permafrost Modeling: On the influence of topography driven lateral fluxes.}

\author[1]{Jonas Beddrich\corref{cor1}}
\ead{beddrich@ma.tum.de}
\cortext[cor1]{Corresponding author: Tel.: +49-151-40014508;}
\author[2]{Shubhangi Gupta}
\author[1]{Barbara Wohlmuth}
\author[3]{Gabriele Chiogna}

\address[1]{Chair of Numerical Mathematics, Technical University of Munich, Boltzmannstr. 3, 85748 Garching bei M\"unchen, Germany}
\address[2]{GEOMAR Helmholtz Center for Ocean Research Kiel, Wischhofstra\ss{}e 1-3, 24148 Kiel, Germany }
\address[3]{Chair of Hydrology and River Basin Management, Technical University of Munich, Arcisstra\ss{}e 21, 80333 M\"unchen}

\begin{abstract}

Alpine permafrost environments are highly vulnerable and sensitive to changes in regional and global climate trends.
Thawing and degradation of permafrost has numerous adverse environmental, economic, and societal impacts.
Mathematical modeling and numerical simulations provide powerful tools for predicting the degree of degradation and evolution of subsurface permafrost as a result of global warming. 
A particularly significant characteristic of alpine environments is the high variability in their topography and geomorphology which drives large lateral thermal and fluid fluxes. 
Additionally, harsh winds, extreme weather conditions, and various degrees of saturation have to be considered. 
The combination of large lateral fluxes and unsaturated ground makes alpine systems markedly different from Arctic permafrost environments and general geotechnical ground freezing applications, and therefore, alpine permafrost demands its own specialized modeling approaches. 
In this research work, we present a multi-physics permafrost model tailored to alpine regions.
In particular, we resolve the ice-water phase transitions, unsaturated conditions, and capillary actions, and account for the impact of the evolving pore volume on fluid-matrix interactions.
Moreover, the approach is multi-dimensional, and therefore, inherently resolves fluxes along topographic gradients.
Through numerical cases studies based on the elevation profiles of the two prominent peaks of the Zugspitze (DE) and the Matterhorn (CH), we show the strong influence of topography driven thermal and fluid fluxes on active layer dynamics and the distribution of permafrost. 

\end{abstract}


\end{frontmatter}

\section{Introduction}
\label{sec:introduction}

The cryosphere of the European mountains is expected to experience drastic changes within the next decades, such as the disappearance of glaciers, the rise of snow lines, the expansion of vegetation, and the degradation of permafrost at lower altitudes \citep{beniston2018european}. 
While the former three can be observed visually, the current state of permafrost has to be monitored in form of borehole measurements \citep{etzelmuller2013recent} or under application of eletrical resistivity tomography (e.g. \cite{you2013application}). 
Permafrost is defined as ground where temperatures have remained continuously below 0\,$^\circ$C for a period of at least two consecutive years (e.g. \cite{riseborough2008recent}). 
The EU-funded Permafrost and Climate in Europe project (PACE) has been monitoring the mountain permafrost in Europe over the last twenty years.
The results show clear warming trends at all investigated sites across Europe down to a depths of 50 meters and an increase of up to 200\% in the thickness of the active layer \citep{etzelmuller2020twenty}, i.e.,  the area of seasonal freezing and thawing. The degradation of permafrost has a major influence on slope stability (e.g. \cite{krautblatter2013permafrost}, \cite{mccoll2012paraglacial}), construction and engineering (e.g. \cite{bommer2010practical, arenson2009geotechnical}), hydrological systems and water management (e.g. \cite{liu2007effect}, \cite{quinton2011permafrost}), and greenhouse gas emissions (e.g. \cite{schuur2015climate}, \cite{anthony2012geologic}). Modeling and simulation of mountain permafrost are essential to assess the impacts of global climate change and to predict, manage, and mitigate the associated risks.

Alpine regions are characterized by their high spatial variability of elevation, subsurface composition, geomorphology, and micro-climatology \citep{gruber2009mountain} with differences of up to 6$^\circ$C in mean annual ground surface temperature (MAGST) within 300\,m distance \citep{gubler2011scale}. 
For this reason, permafrost modeling in alpine environments requires high resolutions and detailed weather information since regional climate models are typically not fine enough to capture temperature variations in mountainous terrains accurately \citep{etzelmuller2013recent}. 
Additionally, the topographic variations lead to lateral thermal and fluid fluxes which require multi-dimensional modelling \citep{etzelmuller2013recent}. This stands in contrast to the common simplification of one-dimensional subsurface heat transport \citep{gruber2004interpretation}. 

Finally, dry climate as well as the topography of alpine regions can lead to unsaturated subsurface conditions \citep{gruber2017inferring}. Rock glaciers may contain up to 25\% air \citep{arenson2005mathematical} and talus slopes stay unsaturated throughout the year due to their high permeability \citep{rogger2017impact}. 
The air circulation within talus slopes can even lead to a cooling effect that promotes the occurrence of permafrost \citep{wicky2017numerical}. 
Thus, alpine permafrost models also need to account for the presence of air and freezing of unsaturated soils, and therefore, demand more specialized and detailed multiphysics approaches compared to the general permafrost and geotechnical soil freezing models.

In recent years, numerous approaches to describe the evolution of permafrost have been developed. Modern empirical-statistical models combine permafrost evidence and climate data to estimate the occurrence of permafrost. Statistical approaches such as by \cite{boeckli2012statistical} use inventories about active and relict rock glacier sites in combination with mean annual rock surface temperature (MARST) measurements to assess the state of permafrost. The model by \cite{sattler2016estimating} additionally considers solar radiation in snow free months to obtain a probability for the presence of permafrost. Recently, \cite{deluigi2017data} applied machine learning algorithms to a combination of thermal and geoelectrical field data in combination with rock glacier inventories to estimate the occurrence of permafrost achieving an accuracy of 88\%. 

In contrast, physics-based models formulate thermal and hydraulic processes in the form of mathematical expressions of conservation laws and conduct simulations by application of numerical methods. Concessions are made either by limiting the number of considered processes or in form of simplifying assumptions, such as neglecting lateral fluxes. The latter justifies the usage of one-dimensional numerical schemes, and thus, reduces the complexity and computational cost (e.g. \cite{pruessner2021framework}). Multiple regional and global-scale models (e.g. \cite{luetschg2008sensitivity, hipp2012modelling, jafarov2012numerical, ekici2014simulating, staub2015ground, westermann2016simulating, pruessner2021framework}) rely on this approach. The so-called 2.5D models Alpine3D (\cite{lehning2006alpine3d}) and GEOtop2.0 (\cite{fiddes2015large, endrizzi2014geotop}) combine three-dimensional hydrological surface models with one-dimensional modeling of subsurface processes. Their possible applications range from permafrost development to avalanche warnings and flood forecasting. 
A surface energy balance coupled with three-dimensional subsurface heat transfer has been applied to highly idealized mountain settings by \cite{noetzli2007three} and to a realistic mountain profile by \cite{noetzli2009transient} showing the existence of permafrost even without indication of the ground surface temperature. 
Applications of purely three-dimensional thermo-hydraulic concepts to alpine permafrost are rare and most established models are not intended for the use in alpine regions. 
In the case of SUTRA-ICE \citep{mckenzie2007groundwater}, problems occur when modeling the freezing of unsaturated soil, and its simulation of active layer dynamics is not accurate \citep{bui2020review}. Marsflo \citep{painter2011three} considers a simplified ice-water phase transition and PFLOTRAN-ICE \citep{karra2014three} assumes full saturation of the modeled soil, and thus, considers only water as fluid phase. 
Also, modern soil freezing models (e.g. \cite{sweidan2021experimental, nishimura2019simple}), which are primarily focused on geotechnical applications such as frost heave and the formation of ice lenses, assume fully saturated soil and ignore the strong capillary effects in the fringe zones, which in the presence of complex morphologies can lead to inaccurate convective fluxes, and thus, to incorrect active-layer dynamics. 

For a detailed overview of the concepts of permafrost modeling, we refer to the work of \cite{riseborough2008recent} and to the articles by \cite{bui2020review} and \cite{gao2021permafrost} for reviews of hydrological models that have been applied to the permafrost-dominated arctic regions and the Qinghai-Tibet plateau.

In order to address the summarized gaps in the modeling of alpine permafrost, we have developed a detailed thermo-hydraulic model tailored to the special geophysical and geomorphological conditions of alpine regions.
In particular, we extend the common continuum-based permafrost model concepts to include the unsaturated flow and the associated capillary actions and the soil freezing model such that it is compatible with three-phase conditions in the fringe layers. Moreover, we resolve the pore-water to pore-ice phase transitions in a consistent manner so that our model has a much higher flexibility in the choice of soil freezing curves. Most importantly, our model is multi-dimensional, and therefore, naturally resolves the fluxes along topographic gradients.

The rest of the paper is structured as follows:
The mathematical model and numerical scheme are presented in detail in Sections \ref{sec:model} and \ref{sec:numerical_solution}. 
The thermo-hydraulic behavior of the model is verified comprehensively in Section \ref{sec:validation} using the benchmark test cases proposed in the INTERFROST project \citep{ruhaak2015benchmarking}. 
Finally, in Section \ref{sec:field_scale}, a series of numerical case studies are performed based on the elevation profiles of the two distinctly different and prominent peaks of the Zugspitze (DE) and the Matterhorn (CH), whereby we focus on the influences of mountainous terrain on the distribution of permafrost and the dynamics of the active layer. 
\section{Mathematical Model}
\label{sec:model}
The main phenomenological process of interest for our permafrost model is the influence of long range (e.g. global warming) and short range (e.g. seasonal or daily) temperature changes on freezing and thawing of soil, with the objective of analyzing the sensitivity of alpine permafrost to cyclic or continuous temperature changes. 
At a process level, the model focuses on resolving the complex pore-water to pore-ice phase transition and the associated transport of heat and pore-fluids. 

In mountain regions, the upper sediment layers are typically variably saturated and the exchange of pore-filling air and the atmosphere is locally restricted by varying surface covers such as glaciers and snow. To account for the presence of air and its effects on pore-water and heat transport, our model explicitly considers a mobile air phase as well as irreducible air in the pore spaces.  

Our model is based on the macroscopic continuum theory of multi-phase flow in porous media.
It considers two mobile and immiscible fluid phases, namely air (subscript `a') and water (subscript `w'), and two immobile and undeformable solid phases, namely ice (subscript `i') and granular sediment (subscript `s'). 
The solid phases constitute the skeleton of the porous medium, while the fluid phases occupy the void spaces. 
The mathematical description of the fundamental conservation laws is based on the following homogenized variables defined locally over an arbitrary representative elementary volume (REV, see Figure \ref{fig:REV_schematic}) contained within a domain of interest $\Omega \subset \mathbb{R}^d$ with $d=\{2,3\}$: porosity $\phi\left(\mathbf{x},t\right)$, fluid phase saturations $S_a\left(\mathbf{x},t\right)$ and $S_w\left(\mathbf{x},t\right)$, and solid phase concentrations $c_i\left(\mathbf{x},t\right)$ and $c_s\left(\mathbf{x},t\right)$, with position $\mathbf{x}\in \Omega$ and time $t\subset \mathbb{R}$.
The following summation conditions hold: 

\begin{align}
    \label{eqn:sum_saturations}
    S_a + S_w = 1, \quad c_i + c_s = 1.     
\end{align}

\usetikzlibrary{decorations.pathreplacing}

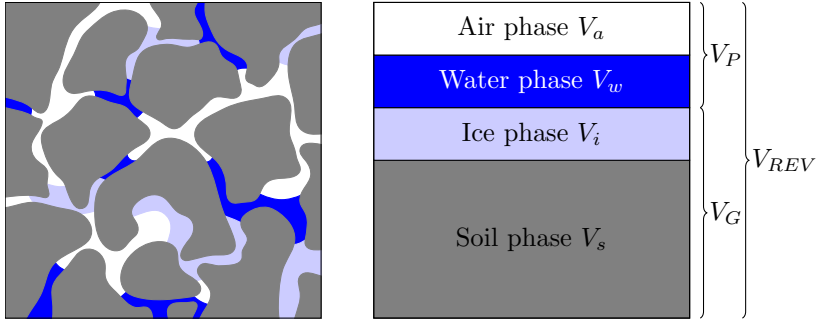
\begin{figure}
\centering

\begin{tikzpicture}[scale=0.7]

\begin{scope}
\clip (0,0) rectangle (6,6); 

\draw[use Hobby shortcut, fill=blue, color=blue, closed] (0.7,1.7) .. (1,1).. (1.7,0.7) .. (0,0);  
\draw[use Hobby shortcut, fill=blue, color=blue, closed] (1.5,0) .. (2,0.25) .. (2.5,0.6) .. (3,0.5) .. (3.5,0);
\draw[use Hobby shortcut, fill=blue, color=blue, closed] (2.5,0) .. (3,1) .. (3.5,0.5) .. (4,0.25) .. (4.5,0.4) .. (5,0);  
\draw[use Hobby shortcut, fill=blue, color=blue, closed] (3.5,3) .. (3.9,3).. (3.8,3.5) .. (4.5,3.5) .. (5,3) .. (4.5,2.5) .. (4,2) .. (3.5,2.5);  
\draw[use Hobby shortcut, fill=blue, color=blue, closed] (4,2.5) .. (4.5,2.5) .. (5.25,2.25) .. (6,2.5) .. (6.5,2.5) .. (5.25,1);
\draw[use Hobby shortcut, fill=blue, color=blue, closed] (1.5,4.5) .. (1.7,4.3) .. (1.7,4) .. (2.3,3.7) .. (2.8,4.2) .. (2,5.5) .. (1.75,5); 
\draw[use Hobby shortcut, fill=blue, color=blue, closed] (4,5) .. (4,4.5) .. (4.25,4.25) .. (4.25,4.5) .. (4.5,4.75) .. (4.75,4.75) .. (5,5) .. (4.5,5.5); 
\draw[use Hobby shortcut, fill=blue, color=blue, closed] (1.5,3) .. (2,3.5) .. (2.25,3.5) .. (2.4,3) .. (2.75,3) .. (2.75,2.75) .. (2.5,2.5) .. (2,2.5); 
\draw[use Hobby shortcut, fill=blue, color=blue, closed] (0,3.5) .. (0,4) .. (0,4.5) .. (0.5,4); 
\draw[fill=blue, color=blue] (3.5,5) rectangle (4.5,6); 

\draw[use Hobby shortcut, fill=blue!20, color=blue!20, closed] (2,2) .. (3,2.5) .. (4,2) .. (5,1) .. (4,0.8) .. (3,0.8) .. (3,2) .. (2,2);
\draw[use Hobby shortcut, fill=blue!20, color=blue!20, closed] (5,5.1) .. (5.5,4.8) .. (6,5) .. (6,6) .. (5,6) .. (5,5.5); 
\draw[use Hobby shortcut, fill=blue!20, color=blue!20, closed] (0.5,2.5) .. (1, 2) .. (2,2.5) .. (1,2.5);  
\draw[use Hobby shortcut, fill=blue!20, color=blue!20, closed] (2,5.5) .. (3,5.5) .. (3.5,5.5) .. (3,4.5); 
\draw[use Hobby shortcut, fill=blue!20, color=blue!20, closed] (5,-1) .. (6,-1) .. (6,1.7) .. (5.5,1.4) .. (5,1.5) .. (4.5,0); 

\draw[use Hobby shortcut, fill=gray, color=gray, closed] (1.8,6) .. (1.5, 5.7) .. (1.3,5.5) .. (1.1,5) .. (1, 4.7) .. (0.9, 4.5) .. (0.8,4) .. (0.7, 3.9) .. (0.5,4) .. (0, 4.2) .. (-0.5,4.4); 
\draw[use Hobby shortcut, fill=gray, color=gray, closed] (-0.5, 4.1) .. (0, 4) .. (0.5,3.85) .. (0.6, 3.8) .. (0.7,3.75) .. (0.7, 3.7).. (0.6,3.5) .. (0.5, 3.3) .. (0.6,3) .. (0.8, 2.8) .. (0.8, 2.5) .. (0.8, 2.3) .. (0.7, 2.3) .. (0, 2.3) .. (-0.5, 2.3); 
\draw[use Hobby shortcut, fill=gray, color=gray, closed]  (-0.5, 2.1) .. (0, 2.1) ..  (0.5, 2.2) ..  (1, 2.1) .. (1.4,2) .. (1.6,1.5) .. (1.5,1.4) .. (1,1.1) .. (0.9, 1) .. (0.7,0.5) .. (0.4,0) .. (0.1, -0.5); 

\draw[use Hobby shortcut, fill=gray, color=gray, closed] (5.5, -0.5) .. (5.5,0) .. (5.6,0.5) .. (5.6,0.8) .. (6,0.8) .. (7, 0.8); 
\draw[use Hobby shortcut, fill=gray, color=gray, closed] (7, 1.3) .. (6,1.3) .. (5.7,1.4) .. (5.7,1.5) .. (5.7,2) .. (5.6, 2.2) .. (5.7, 2.5) .. (5.8, 2.7) .. (5.7, 3) .. (5.6, 3.2) .. (6, 3.3) .. (7, 3.3); 
\draw[use Hobby shortcut, fill=gray, color=gray, closed] (7, 3.6) .. (6, 3.5) .. (5.8, 3.5) .. (5.6, 3.6) .. (5.5,4) .. (5.4, 4.5) .. (5.4,5) .. (5.5, 5.2) .. (6, 5.4) .. (6.5, 5.6); 

\draw[use Hobby shortcut, fill=gray, color=gray, closed] (0.9,-1) .. (0.9,0) .. (0.9,0.3) .. (0.9,0.6) .. (1.2, 1) .. (1.5, 1.3) .. (1.9,1) .. (1.9, 0.8) .. (2, 0.6) .. (2.1,0.5) .. (2.3,0.3) .. (2.5,0.2) .. (2.7, 0.2) .. (2.8,0); 
\draw[use Hobby shortcut, fill=gray, color=gray, closed] (3.1,0) .. (3.1, 0.2) .. (3.5, 0.3) .. (3.6, 0.3) .. (3.55, 0) .. (3.5, -0.3); 
\draw[use Hobby shortcut, fill=gray, color=gray, closed] (4,-1) .. (3.7, 0) .. (3.9, 0.4) .. (3.9, 0.5) .. (3.7,1) .. (4,1) .. (4.2,1) .. (4.5,1.15) .. (4.5,1.25) .. (4.6,1.5) .. (4.5, 1.7) .. (4.5, 1.9) .. (4.6, 2) .. (5,2.1) .. (5.2, 2) .. (5.5,1.5) .. (5.3, 1) .. (5.2, 0.5) .. (5.2, 0) .. (5.2, -1);

\draw[use Hobby shortcut, fill=gray, color=gray, closed] (2.75,6) .. (2.95, 5.5) .. (2.9,5.3) .. (3,5.3) .. (3.5,5.3) .. (3.6, 5.3) .. (3.7, 5.3) .. (3.75,5.4) .. (3.6, 5.9) .. (3.65,6); 
\draw[use Hobby shortcut, fill=gray, color=gray, closed] (3.8,6).. (3.8,5.8)  .. (3.9, 5.5) .. (4, 5.3) .. (4.2, 5.1) .. (4.4,5) .. (4.6, 4.8) .. (4.6, 4.5) .. (4.7, 4.45) .. (4.8,4.4) .. (5, 4.5) .. (5.2, 4.85) .. (5.2, 5) .. (5.5,5.3) .. (5.7,5.5) .. (6, 5.6); 
\draw[use Hobby shortcut, fill=gray, color=gray, closed] (2,5.8) .. (2.5,5.8) .. (2.7,5.5) .. (2.6, 5.25) .. (2.5, 5.1) .. (2.35,5) .. (2.3,4.5) .. (2.29, 4.375) .. (2.26, 4.25) .. (2.13,4.25) .. (2,4.3) .. (1.5, 4.3) .. (1.2, 4.3) .. (1.1, 4.5) .. (1.3,5) .. (1.5, 5.4) .. (1.6,5.5); 

\draw[use Hobby shortcut, fill=gray, color=gray, closed] (3,5.2) .. (3.15,5.2) .. (3.5,5.15) .. (4,5.2) .. (4.1, 5) .. (4.2,4.9) .. (4.35, 4.8) .. (4.45, 4.5) .. (4.46, 4.25) .. (4, 4.2) .. (3.75,4) .. (3.5, 3.8) .. (3.2, 3.6) .. (3,3.7) .. (2.5,4) .. (2.4,4.1) .. (2.4,4.5) .. (2.5,4.9) .. (2.55,5);  
\draw[use Hobby shortcut, fill=gray, color=gray, closed] (1,3.75) .. (1.3,4) .. (1.5,4.2) .. (2,4.1) .. (2.3,4) .. (2.4,3.9) .. (2.5, 3.8) .. (2.75,3.5) .. (2.5, 3.25) .. (2.2,3) .. (2,2.8) .. (1.7,2.6) .. (1.6,2.4) .. (1.5,2.1) .. (1.15,2.5) .. (1.1,2.6) .. (1,2.7) .. (0.8,3) .. (0.7,3.3).. (0.7,3.5); 
\draw[use Hobby shortcut, fill=gray, color=gray, closed] (3.6, 3.5) .. (4,3.9) .. (4.2,4) .. (4.5,4.1) .. (5,4.2) .. (5.3,4.2) .. (5.4,3.5) .. (5.5, 3.2) .. (5.55,3) .. (5.6,2.8) .. (5.5,2.7) .. (5,2.6) .. (4.9, 2.5) .. (4.75,2.25) .. (4.5,2.2) .. (4.3,2.2) .. (4.2,2.5) .. (4,3); 
 
\draw[use Hobby shortcut, fill=gray, color=gray, closed] (1.7,2) .. (1.8,2.5) .. (2,2.7) .. (2.5,3) .. (3,3.4) .. (3.5, 3.3) .. (3.8, 3) .. (4,2.5) .. (4.15,2) .. (4.35,1.5) .. (4.35,1.3) .. (4,1.2) .. (3.5,1.1) .. (3.4, 1.1) .. (3.5,1.5) .. (3.5,1.9) .. (3.3,2) .. (3,2.3) .. (2.5,2.2) .. (2.4,2) .. (2.5,1.8) .. (2.5,1.6) .. (2.4,1.5) .. (2.2, 1.3) .. (2,1.3) .. (1.8,1.5) .. (1.7, 2); 
\draw[use Hobby shortcut, fill=gray, color=gray, closed] (3, 0.35) .. (2.5, 0.4) .. (2.3, 0.5) .. (2.25, 1) .. (2.5, 1.3) .. (3, 1.45) .. (3.25, 1.3) .. (3.3, 1) .. (3.5, 0.8) .. (3.7, 0.5) .. (3.5, 0.4) .. (3, 0.35); 

\draw[] (0,0) rectangle (6,6); 
\end{scope}

\begin{scope}
\draw[fill=gray] (7,0) rectangle (13,3) node[pos=0.5] {Soil phase $V_s$};
\draw[fill=blue!20] (7,3) rectangle (13,4) node[pos=0.5] {Ice phase $V_i$}; 
\draw[fill=blue] (7,4) rectangle (13,5) node[text=white, pos=0.5] {Water phase $V_w$};
\draw[fill=white] (7,5) rectangle (13,6) node[pos=0.5] {Air phase $V_a$};
\draw[] (7,0) rectangle (13,6); 
\end{scope}

\draw[-,decorate,decoration=brace] (13.2,6) -- node[right] {$V_P$} (13.2,4);
\draw[-,decorate,decoration=brace] (13.2,4) -- node[right] {$V_G$} (13.2,0);
\draw[-,decorate,decoration=brace] (14,6) -- node[right] {$V_{REV}$} (14,0);
 
\end{tikzpicture}
\caption{Illustration of a representative elementary volume (REV).
    The homogenized variables which characterize the prototypical porous medium in this model are: local porosity $\phi\left(\mathbf{x},t\right):= V_p / V_{REV}$, 
    local fluid phase saturations $S_\alpha\left(\mathbf{x},t\right):=V_\alpha/V_p$ for $\alpha=\{a,w\}$, and 
    local solid phase concentrations $c_\beta\left(\mathbf{x},t\right):=V_\beta/V_G$ for $\beta=\{i,s\}$, 
    where,  $V_{REV}\subset\Omega$ is the volume of the REV, $V_p\subset V_{REV}$ is the volume of void space where fluid flow is possible, and $V_G\subset V_{REV}$ is the volume of the solids, such that, $V_p+V_G=V_{REV}$. Furthermore, $V_a,V_w \subset V_{p}$ are the volumes of the fluid phases with $V_w+V_a=V_p$, and $V_i,V_s \subset V_{G}$ are the volumes of the solid phases with $V_i+V_s=V_G$. Except of $V_{REV}$, all volume terms are time dependent and change over the course of the simulation. 
    }
\label{fig:REV_schematic}

\end{figure}

\subsection{Governing equations:}
Under the assumptions of local thermal equilibrium within an REV, local thermodynamic equilibrium at the ice-water phase interface, and low Reynolds numbers for the flow of pore-fluids, the transport process within the porous medium can be expressed in form of the following conservation laws. We consider phase-wise mass conservation: 

\begin{align}
    \label{eqn:mb_air}
    &\partial_t \phi \rho_a S_a + \nabla \cdot \rho_a \mathbf{v}_a = 0,
    \\ \label{eqn:mb_water}
    &\partial_t \phi \rho_w S_w + \nabla \cdot \rho_w \mathbf{v}_w = -Q_f,
    \\ \label{eqn:mb_ice}
    &\partial_t \left(1-\phi\right) \rho_i c_i = Q_f,
    \\ \label{eqn:mb_soil}
    &\partial_t \left(1-\phi\right) \rho_s c_s = 0,
\end{align}

where $\rho_\alpha \left(\mathbf{x},t\right)$ denotes the phase densities, $\mathbf{v}_\alpha\left(\mathbf{x},t\right)$ the phase velocities, for $\alpha \in \left\{ a,w,i,s \right\}$, and $Q_f\left(\mathbf{x},t\right)$ is the rate of mass exchange from pore-water to pore-ice during freezing. The momentum of the solid phases is always conserved due to their immobility, i.e., $\mathbf{v}_i = \mathbf{v}_s = 0$, and in the case of the fluid phases, it can be reduced to Darcy's law, (see, e.g. \cite{helmig1997multiphase}). 

\begin{align}
    \label{eqn:darcy_air}
    &\mathbf{v}_a = -\mathbf{K}\frac{k_{ra}}{\mu_a} \left(\nabla P_a + \rho_a \mathbf{g} \right), 
    \\ \label{eqn:darcy_water}
    &\mathbf{v}_w = -\mathbf{K}\frac{k_{rw}}{\mu_w} \left(\nabla P_w + \rho_w \mathbf{g} \right), 
\end{align}

where $\mathbf{K}\left(\mathbf{x},t\right)$ denotes the absolute hydraulic permeability of the medium, $\mu_\alpha \left(\mathbf{x},t\right)$ the phase viscosities, $k_{r\alpha} \left(\mathbf{x},t\right)$ the relative phase permeabilities, $P_{\alpha}\left(\mathbf{x},t\right)$ the phase pressure, for $\alpha \in \left\{a,w\right\}$, and $\mathbf{g}$ the gravity vector. Under the assumption of a local thermal equilibrium, the energy balance can be described for the bulk medium: 

\begin{align}
    \label{eqn:energy}
    &\sum\limits_{\alpha\in\{a,w\}} \left( \partial_t \phi \rho_\alpha S_\alpha C^v_\alpha T + \nabla \cdot \rho_\alpha C^p_\alpha T \mathbf{v}_\alpha \right)
    + \sum\limits_{\sigma\in\{i,s\}} \partial_t \left(1-\phi\right) \rho_\sigma c_\sigma C^v_\sigma T - \nabla \cdot k^T \nabla T 
    = L_f Q_f, 
\end{align}

where $C^v_{\alpha}\left(\mathbf{x},t\right)$ denotes the phase-wise isochoric specific heat capacities, $C^p_{\alpha}\left(\mathbf{x},t\right)$ the isobaric specific heat capacities of the fluid phases, $k^T_{\alpha}\left(\mathbf{x},t\right)$ the phase-wise thermal conductivity, for $\alpha \in \left\{ a,w,i,s\right\}$, and $k^T\left(\mathbf{x},t\right)$ the effective thermal conductivity of the REV, such that, $k^T = \phi \left(S_a k^T_a + S_w k^T_w\right) + (1-\phi)\left( c_i k^T_i + c_s k^T_s\right)$.
Additionally, $L_f$ is the latent heat of fusion which is assumed to be constant.

We condense these conservation laws by substituting the Darcy velocities \eqref{eqn:darcy_air}-\eqref{eqn:darcy_water} into the Eqns. \eqref{eqn:mb_air}, \eqref{eqn:mb_water}, and \eqref{eqn:energy}, and by summation of Eqns. \eqref{eqn:mb_water} and \eqref{eqn:mb_ice}, we get 

\begin{align}
    \label{eqn:mb_ice_water}
    \partial_t \phi \rho_w S_w + \nabla \cdot \rho_w \mathbf{v}_w + \partial_t \left(1-\phi\right) \rho_i c_i = 0. 
\end{align}

Finally, we replace $Q_f$ in Eqn. \eqref{eqn:energy} exploiting Eqn. \eqref{eqn:mb_water}. 

\subsection{Constitutive relations:}

The constitutive relations of our model describe the interactions between the phases, such as the pressure difference between air and water, the ice-water phase transition, and the impact of evolving pore volumes on fluid flow. 

\subsubsection{Hydraulic properties:}

The capillary pressure $P_c \left( \mathbf{x},t \right):=P_a-P_w$ describes the pressure difference across the interface of the gaseous and the aqueous phases. Typically, capillary pressure is parameterized as a function of the wetting phase saturation. We use the Brooks--Corey parameterization \citep{brooks1966properties},

\begin{align}
    \label{eqn:cap_pressure}
    P_c = P_e S_{we}^{-\frac{1}{\lambda}}, 
\end{align}

where $P_e \left( \mathbf{x},t \right)$ denotes the air entry pressure, $\lambda$ the pore size distribution index, and $S_{we} = \frac{S_w - S_{wr}}{1 - S_{wr} - S_{ar}}$ the effective water saturation which can be replaced by capillary action. Here, $S_{ar} \left( \mathbf{x},t \right)$ and $S_{wr} \left( \mathbf{x},t \right)$ are the residual (or irreducible) phase saturations. The relative phase permeabilities are also parameterized according to the Brooks--Corey model \citep{helmig1997multiphase}, 

\begin{align}
    \label{eqn:rel_permeabiliy}
    k_{ra} \left( S_a \right) = {S_{ae}}^2\left( 1-{S_{we}}^{\frac{2}{\lambda}+1} \right), \quad k_{rw}\left( S_w \right) = {S_{we}}^{\frac{2}{\lambda}+3},  
\end{align} 

where the parameter $\lambda$ is the same as in Eqn. \eqref{eqn:cap_pressure}.

The absolute permeability is an intrinsic property of the porous medium which depends on the properties of the granular material as well as the characteristics of the pore network. 
The freezing and thawing processes alter the local pore volumes, which in turn influence the absolute permeability. We parameterize these changes according to Civan's Power Law \citep{civan2001scale},

\begin{align}
    \label{eqn:CivansPowerLaw}
    \mathbf{K} = \frac{\phi}{\phi_0} \left( \frac{\phi \left( 1 - \phi_0 \right)}{\phi_0 \left( 1 - \phi \right)}\right) ^ {2 \beta} \mathbf{K}_0, 
\end{align}

where $\phi_0$ denotes the reference porosity, $\mathbf{K}_0$ the corresponding reference permeability, and $\beta$ an empirically derived parameter. 

\subsubsection{Ice-water phase transition:}

The freezing behavior of bulk water and pore-water differs due to the particle surface energy of water \citep{tian2014freezing}. In a highly idealized setting, the ice-water phase transition in a porous medium can be derived based on the assumption of a thermo-hydraulic equilibrium \citep{loch1978thermodynamic}, while in practice empirically derived soil freezing curves are used to calculate the amount of unfrozen pore-water in dependence of the temperature and soil properties. \cite{hu2020review} provide an overview about recent advances in the parameterization of liquid water in frozen soil, including the soil freezing curve which is utilized in our model: 

\begin{align}
    \label{eqn:SFC}
    SFC\left(T\right) = 
    \begin{cases}
        1, & \text{if}\ T \geq T_f \\  
        0.05 + (0.95) * \exp \left( \frac{T-T_f}{W} \right), & \text{otherwise,}
    \end{cases}
\end{align}

where $W$ as a parameter controls the steepness of the curve, and $T_f$ denotes the freezing point of water. The relative amount of water that stays unfrozen is set to $0.05$ (e.g. \cite{mckenzie2007groundwater}). Under the assumption that the amount of unfrozen water is greater or equal than the soil freezing curve, whereby equality holds in presence of ice, i.e., $c_i > 0$, we obtain a set of Karush--Kuhn--Tucker complementary conditions \citep{kuhn2014nonlinear} to describe the ice-water phase transition: 

\begin{align}
     \label{eqn:NCP}
    c_i \geq 0, \quad \psi = \frac{S_w \phi}{1 - S_a \phi - \left(1-\phi \right)c_s} - SFC(T) \geq 0, \quad c_i \psi = 0.  
\end{align}

This generalized formulation of the ice-water phase transitions also accounts for the presence of air, and thus, is applicable to the freezing of unsaturated soil. 
\section{Numerical Solution}
\label{sec:numerical_solution}

The condensed mathematical model consists of five governing equations: PDEs \eqref{eqn:mb_air}, \eqref{eqn:energy}, and \eqref{eqn:mb_ice_water}, ODE \eqref{eqn:mb_soil}, and NCP \eqref{eqn:NCP}. 
Exploiting Eqns. \eqref{eqn:sum_saturations} and \eqref{eqn:cap_pressure}, we choose the following primary variables: $\mathcal{P} = \left[P_a,S_w,c_i,\phi,T\right]^T.$ 
The governing equations are discretized in space using a fully-upwinded, cell-centered finite volume method, and in time using a fully implicit finite difference method. The fluxes across the cell-interfaces are approximated using a linear two-point finite difference approximation defined on an orthogonal mesh with rectangular P0-elements. 

For the linearization of the system of governing PDEs, we have implemented a semi-smooth Newton solver \citep{lauser2011new}, which can handle the ice-water phase transitions and the appearance and disappearance of phases in a mathematically consistent manner. 
The time step sizes are adapted heuristically based on the performance of the Newton solver (e.g. \cite{helmig2006multifield}). 
It is increased by $10\%$ if less than 4 Newton iterations were taken for the last time integration step, and decreased by $10\%$ if more than eight iterations were necessary to reach the minimal reduction threshold. The approach has already been applied successfully to resolve phase-transitions in the context of submarine methane hydrates \citep{gupta2020all}.

The numerical scheme is implemented within the C++ based DUNE-PDELab framework (version 2.8) \citep{bastian2010generic}. The resulting linear system is solved by a built-in Algebraic Multi-Grid (AMG) parallel solver with a stabilized bi-conjugate gradient method as preconditioner.
\section{Validation studies}
\label{sec:validation}

In order to build confidence in the coupled thermo-hydraulic behavior of our mathematical model, we run two benchmark test cases: Melting of a Frozen Inclusion (TH2) and Talik Opening/Closure (TH3), both of which were developed as part of the INTERFROST project \citep{ruhaak2015benchmarking} and provide useful settings for validating ice-water phase changes and advective mass and heat fluxes. 
Using the performance measures proposed in \cite{grenier2018groundwater}, we make detailed comparisons of our simulation results with those of the other well-established numerical codes, which were the original participants of the Grenier et al. benchmark study. These include: 
ATS \citep{doecode_28622}, 
Cast3M (www-cast3m.cea.fr), 
COMSOL (www.comsol.com/comsol-multiphysics), 
DarcyTools \citep{svensson2014darcytools}, 
FEFLOW \citep{diersch2013feflow}, 
FlexPDE (www.pdesolutions.com), 
GEOAN \citep{holmen2011modelling}, 
Ginette \citep{riviere2014experimental}, 
MELT \citep{frederick2014taliks}, 
PermaFoam (www.openfoam.com), 
PFLOTRAN-ICE \citep{karra2014three}, 
SMOKER \citep{molson1992thermal}, 
and
SUTRA \citep{mckenzie2007groundwater}.
Note that the results of the above codes, which are used in this manuscript, are freely available on the website of the INTERFROST project (\url{https://wiki.lsce.ipsl.fr/interfrost}).
Also, our intention is not to compare our approach against any individual code, but rather, to show that under standard settings our results are consistent with those of the well-established models.

\subsection{Melting of a Frozen Inclusion}
\label{subsec:TH2}

\begin{figure}
    \centering
    \begin{subfigure}[b]{0.49\textwidth}
        \centering
        \includegraphics[width=\textwidth]{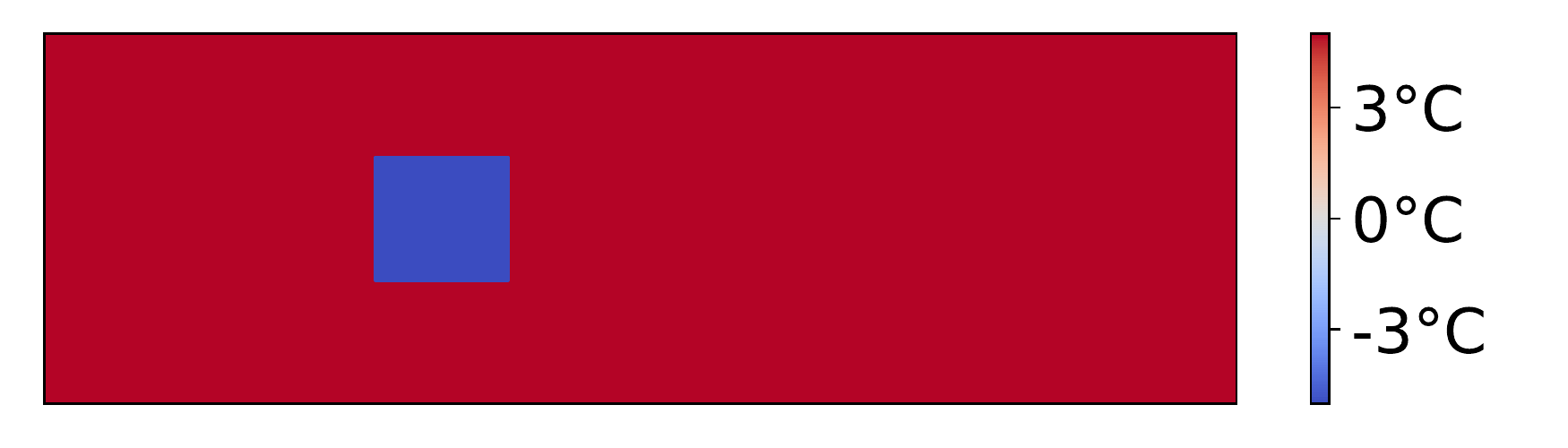}
        \caption{Initial conditions}
        \label{fig:TH2_TF_initial}
    \end{subfigure}
    \begin{subfigure}[b]{0.49\textwidth}
        \centering
        \includegraphics[width=\textwidth]{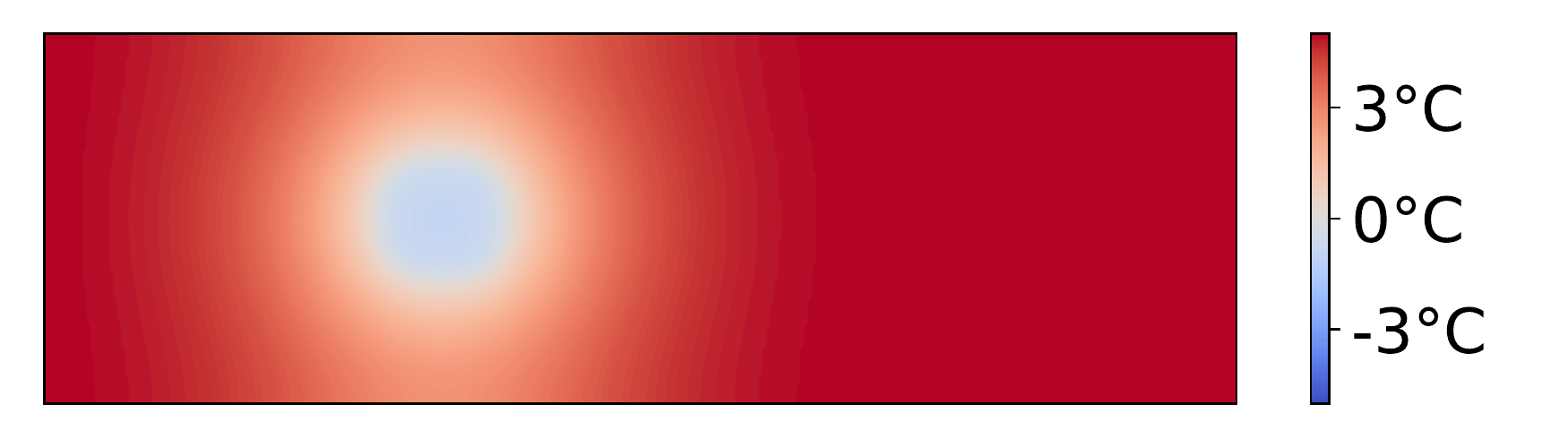}
        \caption{0\% case after 8\,h}
        \label{fig:TH2_TF_00_8h}
    \end{subfigure}
    \begin{subfigure}[b]{0.49\textwidth}
        \centering
        \includegraphics[width=\textwidth]{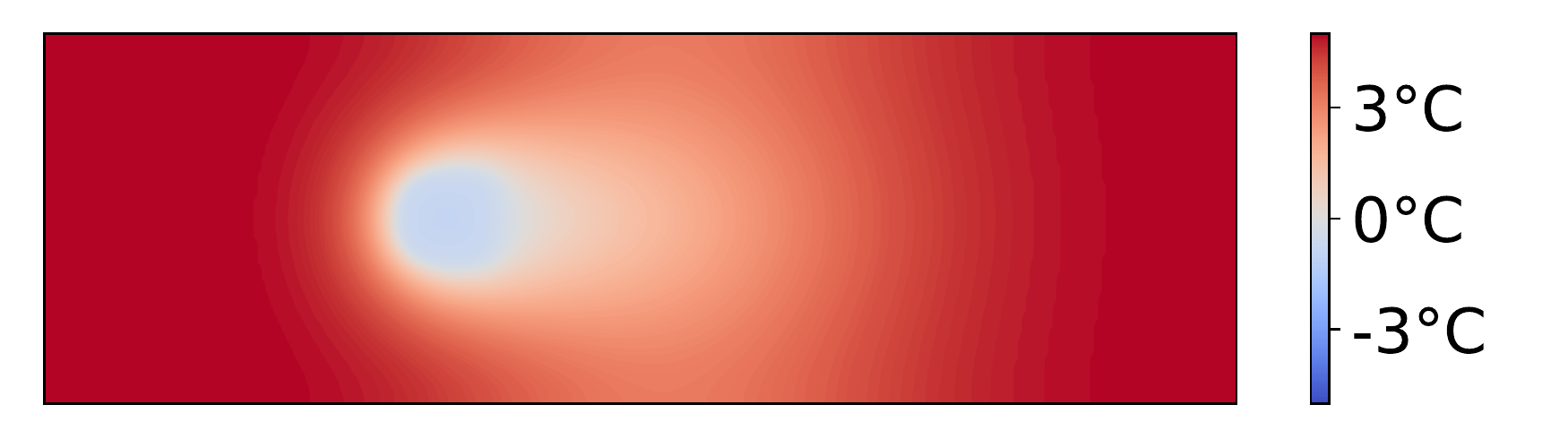}
        \caption{3\% case after 8\,h}
        \label{fig:TH2_TF_03_8h}
    \end{subfigure}
    \begin{subfigure}[b]{0.49\textwidth}
        \centering
        \includegraphics[width=\textwidth]{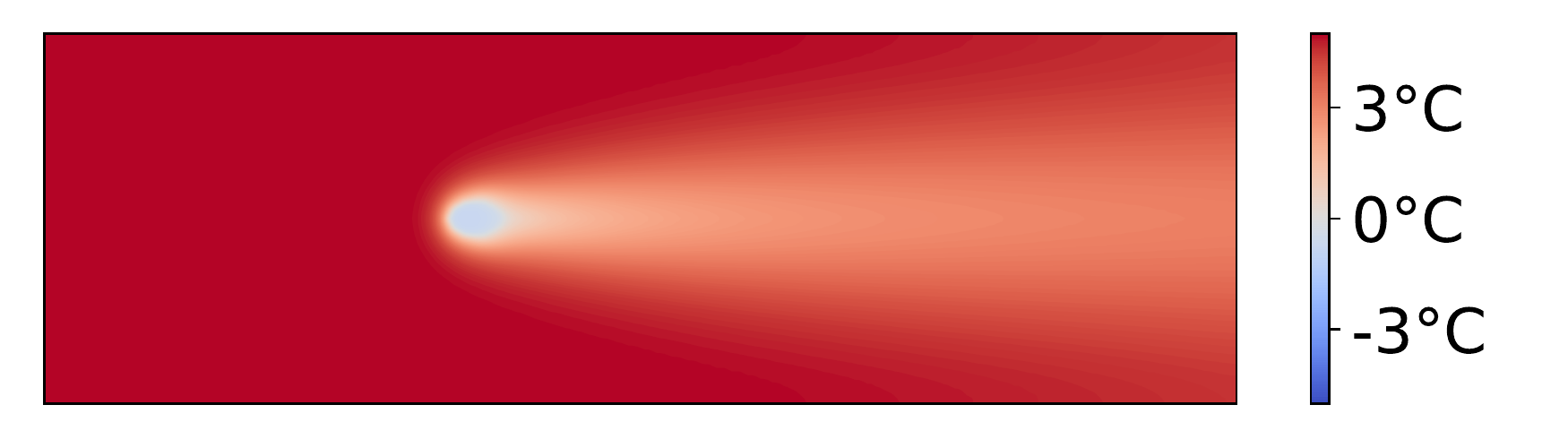}
        \caption{15\% case after 8\,h}
        \label{fig:TH2_TF_15_8h}
    \end{subfigure}
    \caption{Temperature fields of TH2: Melting of a Frozen Inclusion. Figure (a) displays the initial conditions and (b), (c), and (d) visualize snapshots of the scenarios with hydraulic gradients of 0\%, 3\%, and 15\% after 8\,hours.}
    \label{fig:TH2_TF}
\end{figure}

\begin{figure}
    \centering
    \begin{subfigure}[b]{\textwidth}
        \centering
        \includegraphics[width=0.32\textwidth]{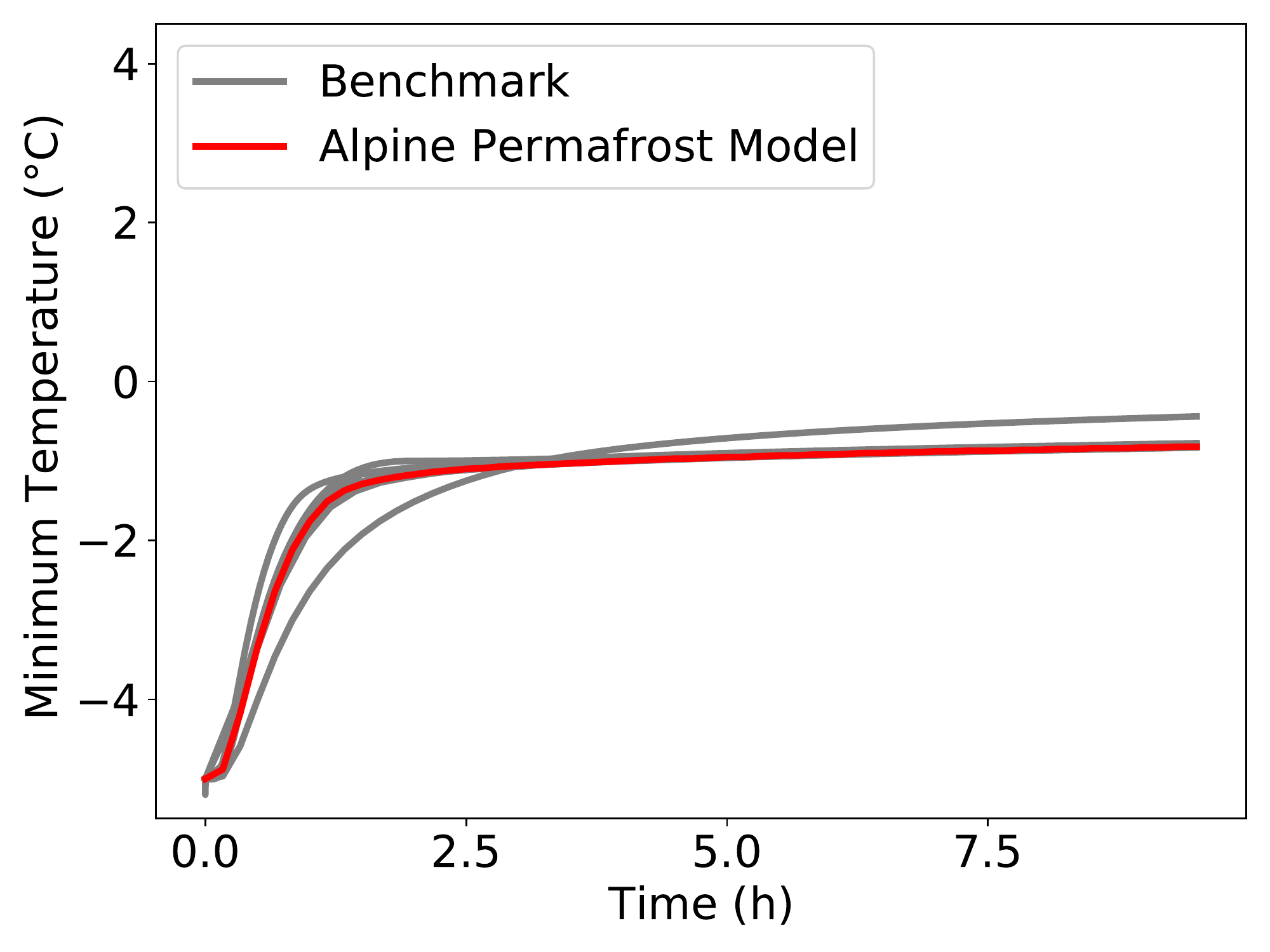}
        \hfill
        \includegraphics[width=0.32\textwidth]{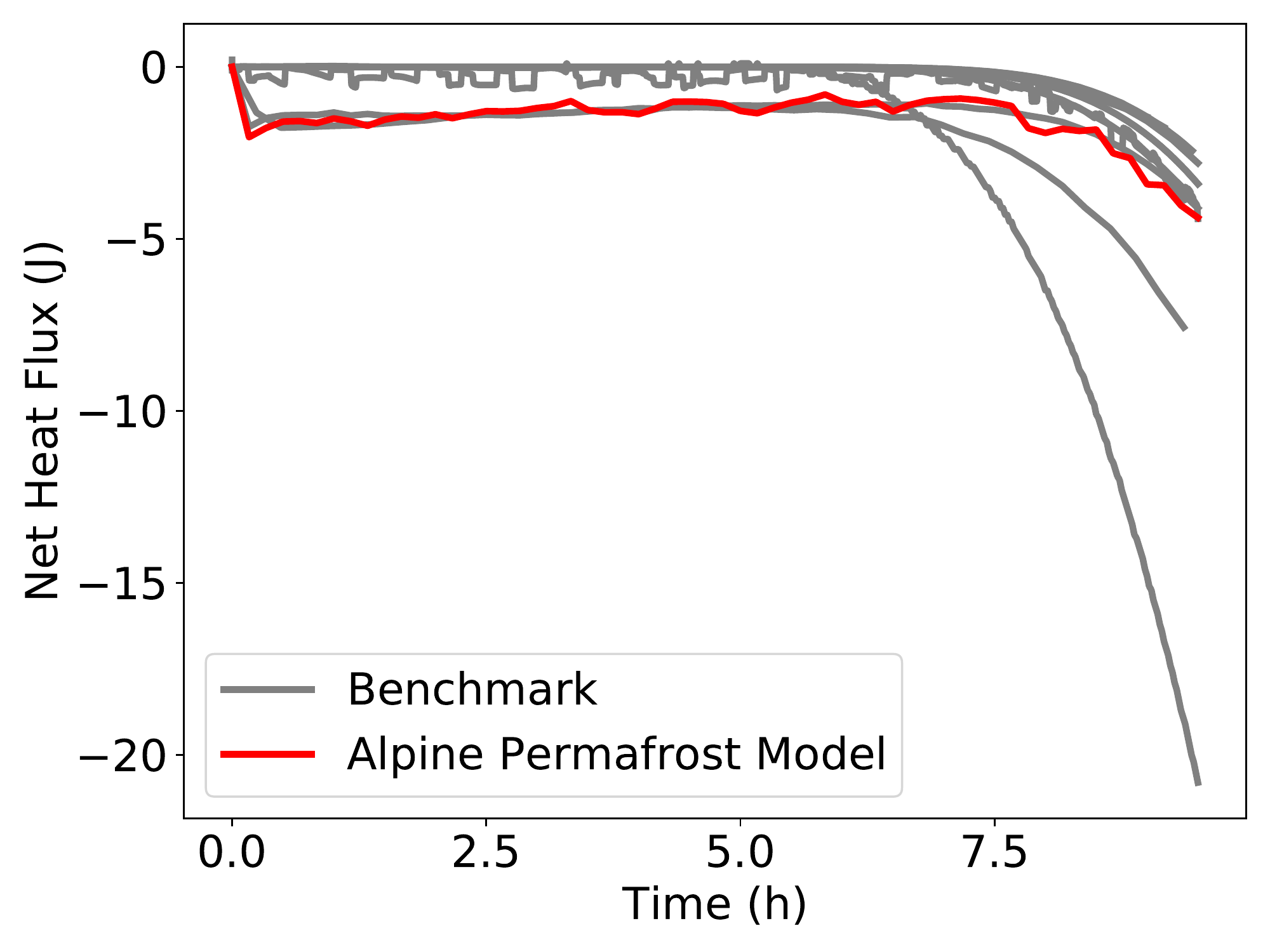}
        \hfill
        \includegraphics[width=0.32\textwidth]{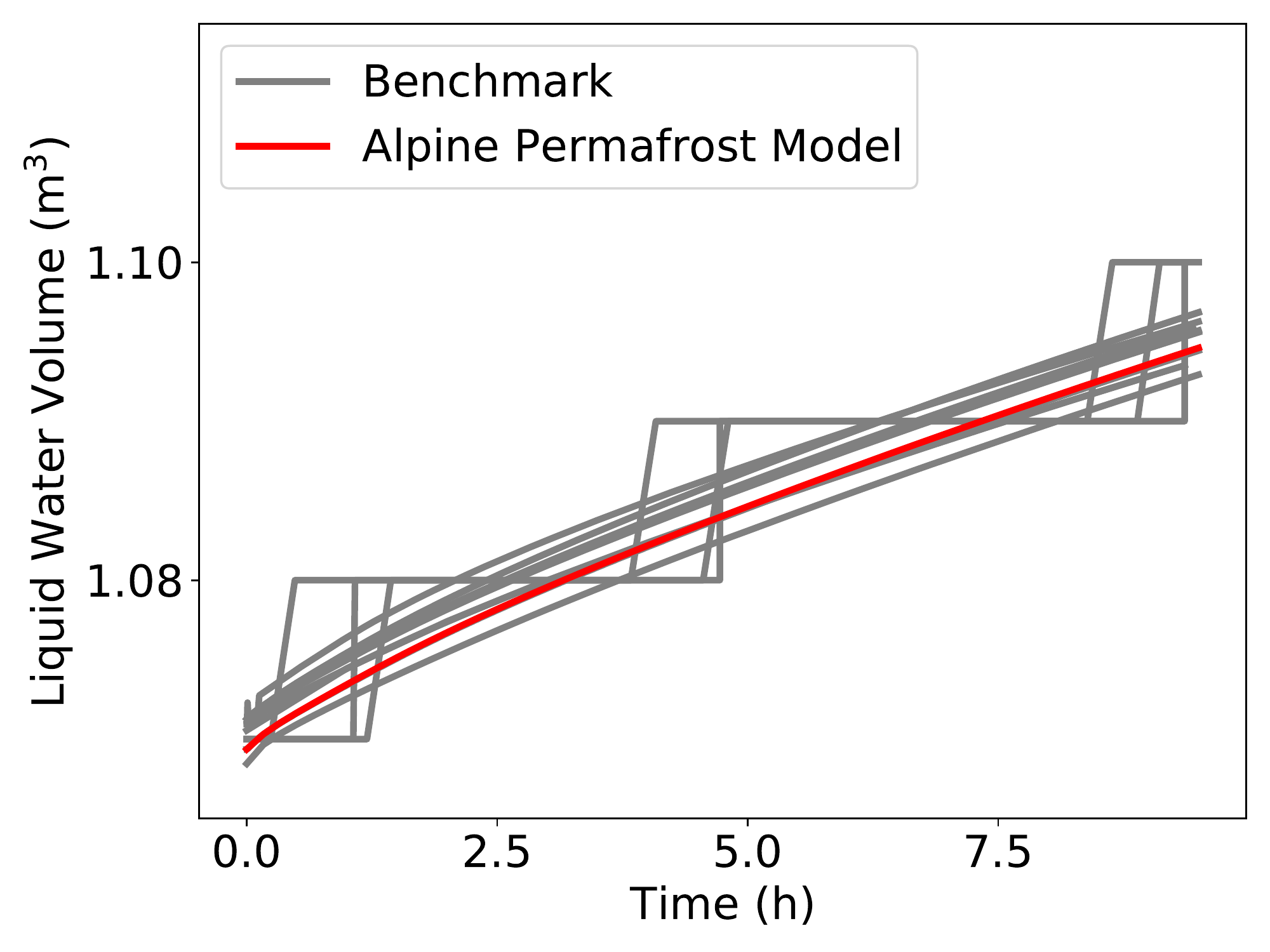}
        \caption{Hydraulic gradient: 0\%}
        \label{fig:TH2_PM_00}
    \end{subfigure}
    
    \begin{subfigure}[b]{\textwidth}
        \centering
        \includegraphics[width=0.32\textwidth]{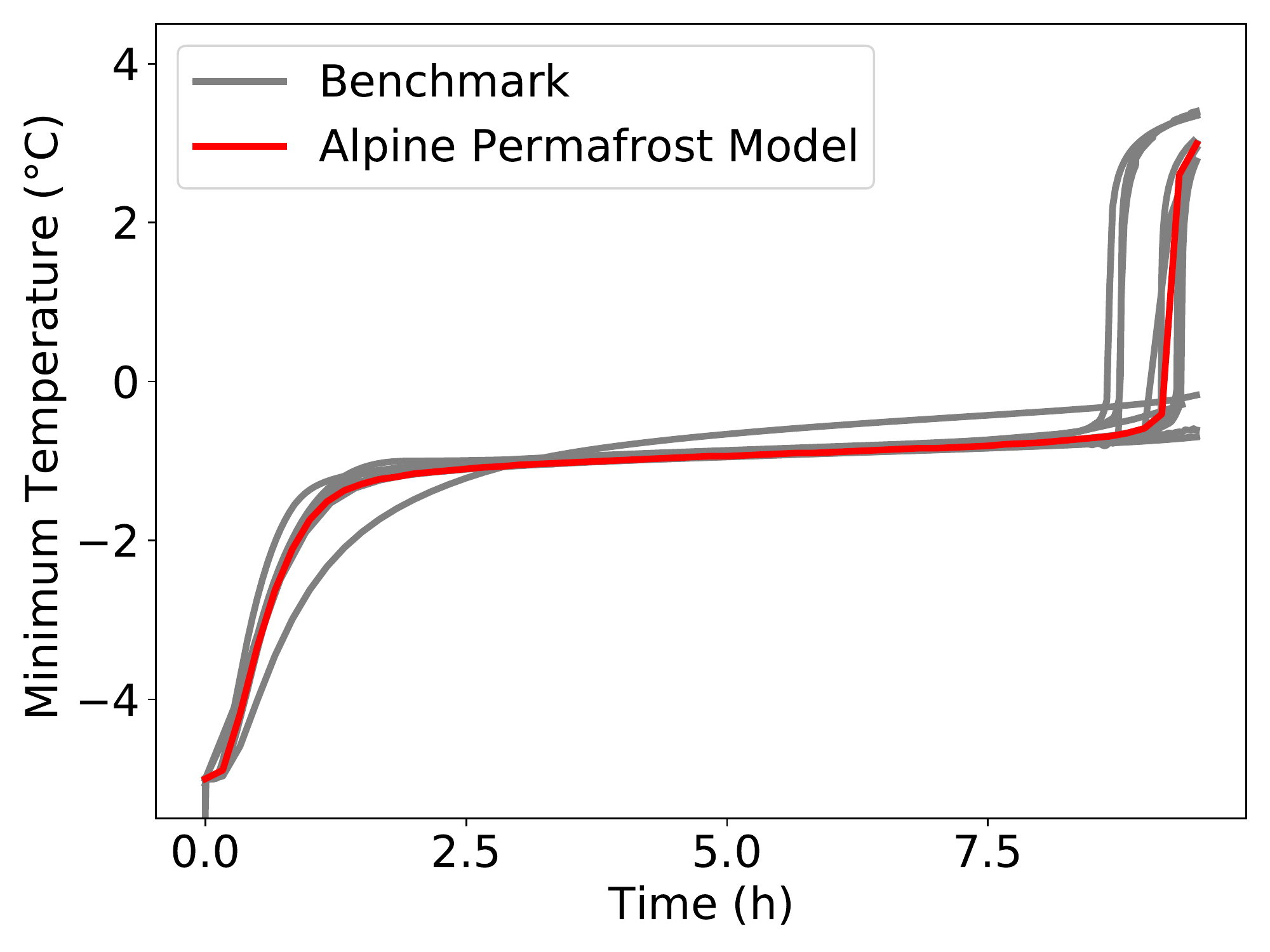}
        \hfill
        \includegraphics[width=0.32\textwidth]{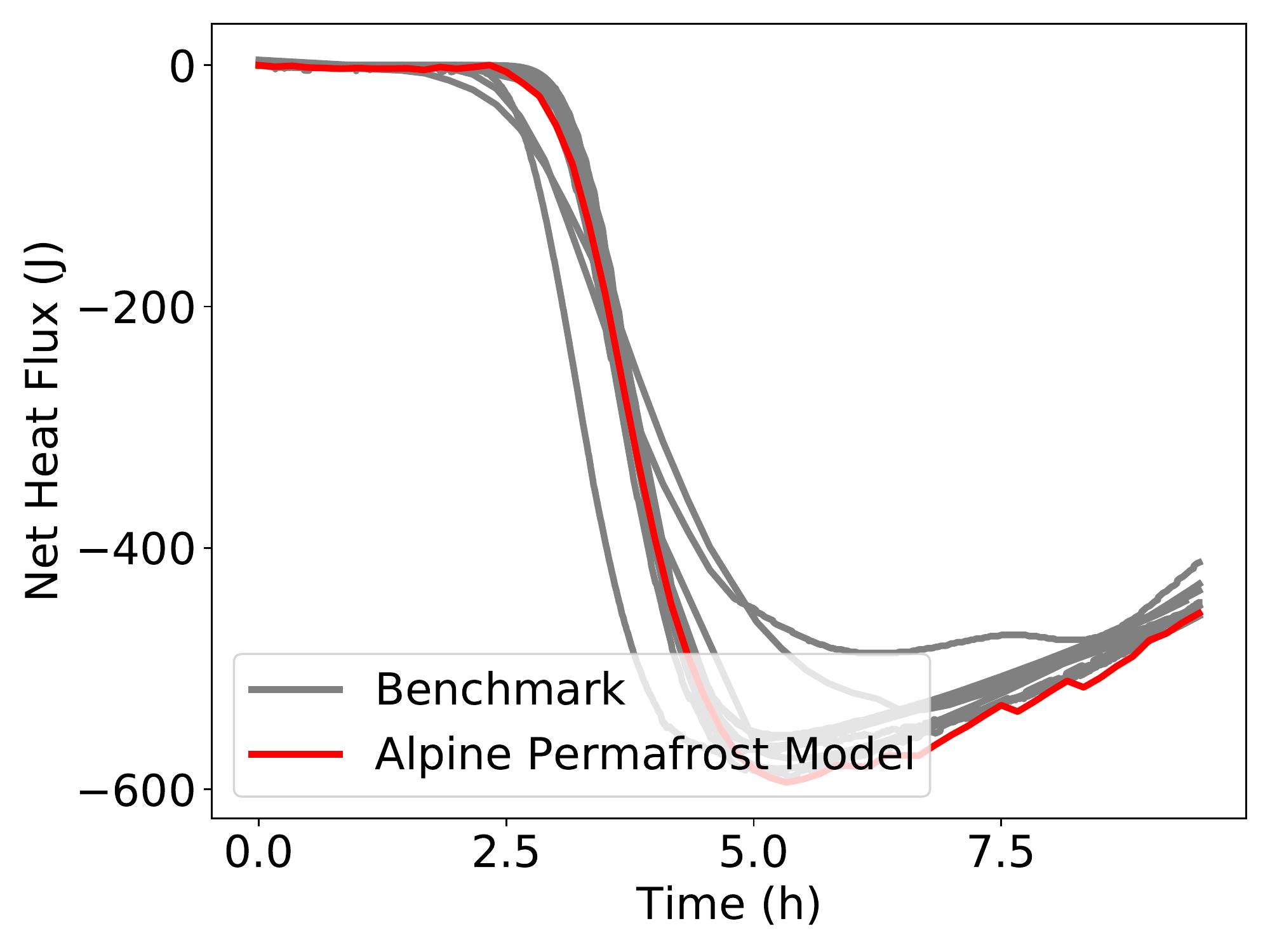}
        \hfill
        \includegraphics[width=0.32\textwidth]{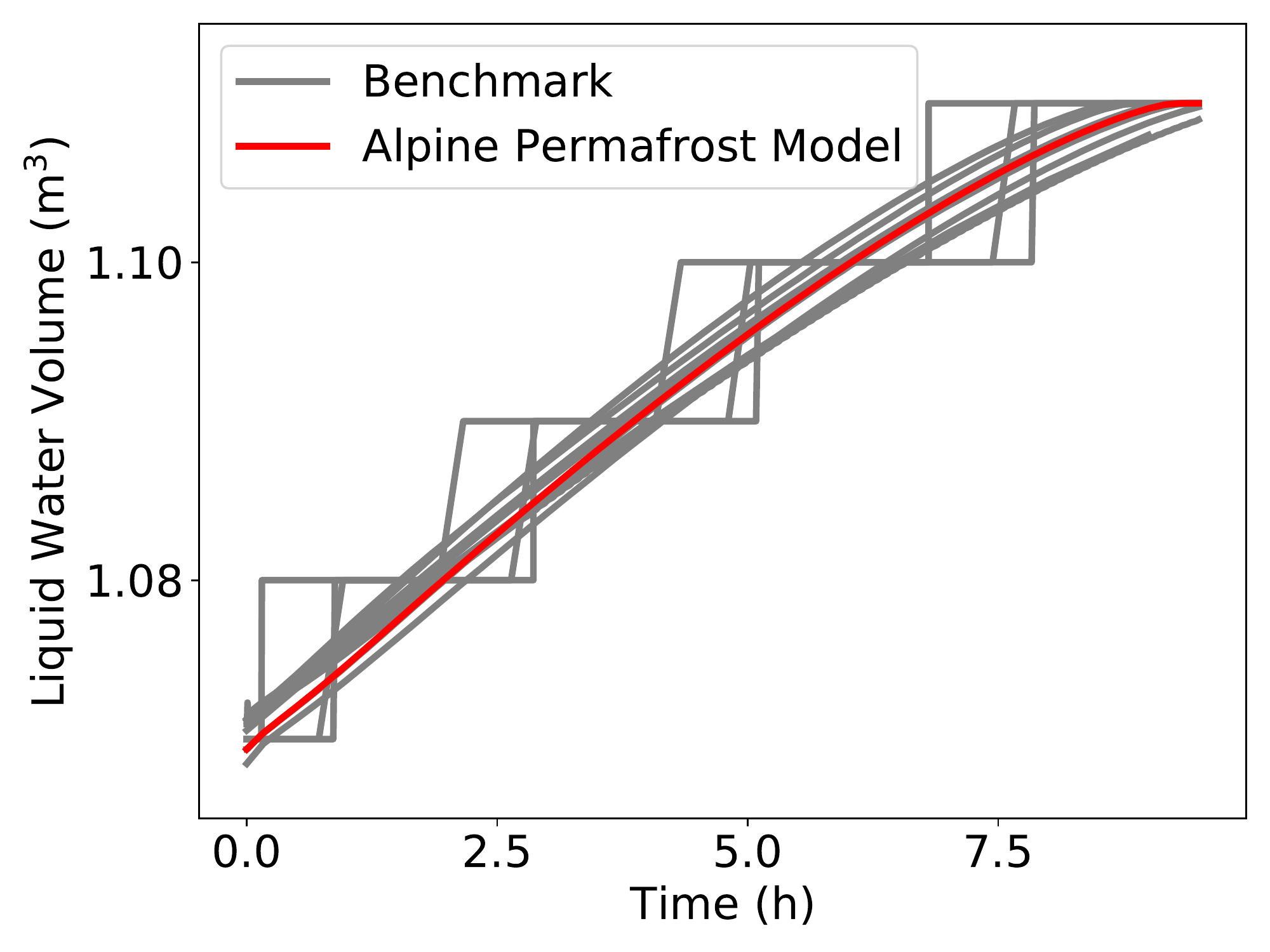}
        \caption{Hydraulic gradient: 15\%}
        \label{fig:TH2_PM_15}
    \end{subfigure}
    \caption{Performance Measures for TH2: Melting of a Frozen Inclusion. The graphics on the left show the development of the minimal temperature, the figures in the middle display the total heat flux exiting the system and the ones on the right visualize the evolution of the total water volume in the domain. The upper row (a) represents the no flow scenario, i.e., a hydraulic gradient of 3\% and (b) visualizes the 15\% case.}
    \label{fig:TH2_PM_comparison}
\end{figure}

Due to the seasonal thawing of the upper soil layers and the resulting runoff of melt water, the bi-directional coupling between thermal and hydraulic processes is of high importance for the correct description of active layer dynamics. 
The first benchmark specifically targets the thermo-hydraulic coupling through simulation of the thawing of an idealized frozen inclusion under the influence of different hydraulic gradients.

The 3\,m by 1\,m computational domain has an initial temperature of 5\,$^\circ$C, except for the frozen inclusion of size 33\,cm x 33\,cm, which is at -5\,$^\circ$C. The initial temperature field is visualized in Figure \ref{fig:TH2_TF_initial}. The initial water saturation, porosity, and ice concentration are determined by the prescribed temperature field and the applied soil freezing curve. No air is present in the domain and the soil is assumed to be saturated. 

There are no mass and heat fluxes across the bottom and the top boundaries of the domain. 
At the left and right boundaries, hydraulic heads of $H_0 + \Delta H$ and $H_0$ are imposed, where $\Delta H$ is chosen to create hydraulic gradients of 0\%, 3\%, 9\%, and 15\%, resulting in four scenarios with flows of different strengths. While there is no conductive-flux through the other boundaries, we impose a constant temperature of 5\,$^\circ$C at the left side. 
We discretize the domain into 30000 square elements and, as suggested in the benchmark study, the simulated time span is $9.6$\,hours. The initial, the minimal, and the maximal time steps are set to $0.1$\,seconds, $0.1$\,seconds, and $60$\,seconds. 

In all four scenarios, the temperature inside the frozen inclusion rises quickly up to -1\,$^\circ$C and remains at that level as the melting progresses from the outside of the frozen inclusion towards the inside.
The hydraulic gradient determines the downstream transport of water, which in turn controls the convective thermal fluxes and the rate of ice-water phase change. 
These effects are clearly demonstrated in Figures \ref{fig:TH2_TF_00_8h}-\ref{fig:TH2_TF_15_8h}. For the 0\% hydraulic gradient, the heat transport is purely conduction-driven, while as the hydraulic gradient increases, the effects of convective heat transfer on the rate of thaw become increasingly dominant. 

Figure \ref{fig:TH2_PM_comparison} shows the comparison of our results with the other participating codes against the following performance measures: the minimal temperature in the domain, the total heat flux exiting the system, and the total liquid water volume. 
Here, we exemplary show the results of the hydraulic gradients of 3\% and 15\%. 
For completeness, the outcomes of all simulations are included in \ref{apx:performance measures}.

In all scenarios, the minimal temperature starts at -5\,$ ^\circ$C, raises within the first hour up to -1\,$ ^\circ$C, and increases slowly thereafter due to latent heat effects. 
After 9 hours, the minimal temperature of the 15\% case raises steeply up to 3\,$^\circ$C. 
At this point, the frozen inclusion has melted entirely. 
The total heat flux exiting the system stays close to 0\,J until the downstream moving melt water of the frozen inclusion reaches the right boundary of the domain and declines afterwards as more and more cold water leaves the system. 
For the hydraulic gradients of 3\% and 15\%, this takes 8 and 3\,hours, respectively. 
After 5\,hours, the total heat flux of the 15\% scenario starts to increase. 
By then, the melting of the frozen inclusion has progressed so far, that the amount of melt water reaching the right boundary of the domain is already declining. 
For the total liquid water volume, the simulations start with a value of $1.069$\,m$^3$. With the thawing of the frozen inclusion the amount of liquid water increases. 
The higher the hydraulic gradient, the higher is the rate of thaw, and therefore, the faster is the increase in liquid water volume. At the time when the temperature sharply rises, the total liquid water volume reaches and remains at its maximal value of $1.11$\,m$^3$ indicating that the entire frozen inclusion has melted.

\subsection{Talik Opening/Closure}
\label{subsec:TH3}

The term talik describes perennial unfrozen ground in permafrost areas which, for example, occurs in between the bottoms of lakes and the underlying permafrost layer \citep{rowland2011role}. The correct distinction between permafrost and unfrozen ground as well as the prediction of the opening and closure of these gaps in otherwise permanently frozen soil is an essential feature of alpine permafrost models. In this benchmark case, we examine whether an idealized talik widens or narrows and subsequently closes depending on the magnitude of the flow passing through the opening. 

\begin{figure}
    \centering
    \begin{subfigure}[b]{0.3\textwidth}
        \centering
        \includegraphics[width=\textwidth]{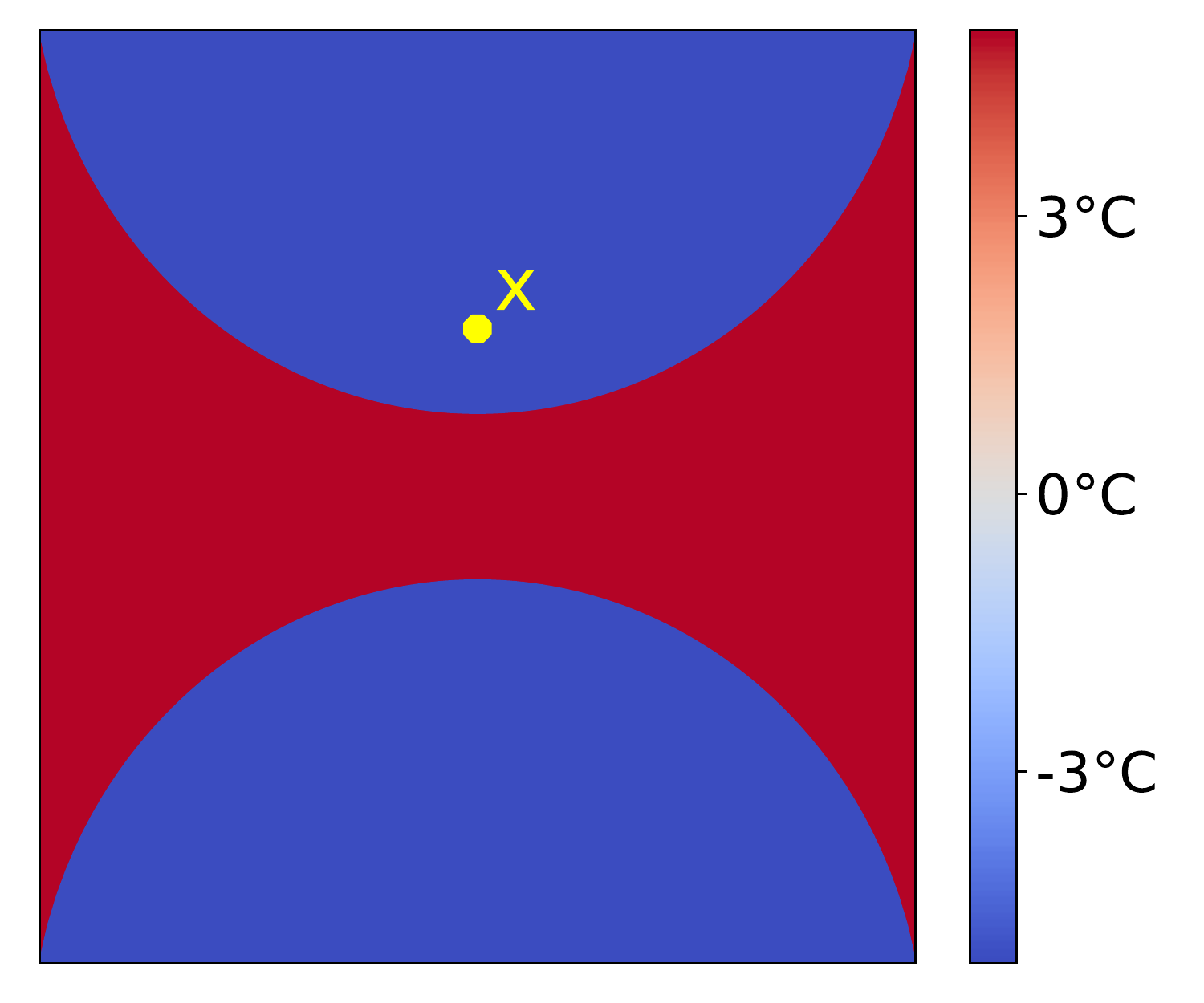}
        \caption{Initial conditions}
        \label{fig:TH3_TF_initial}
    \end{subfigure}
    \hfill
    \begin{subfigure}[b]{0.3\textwidth}
        \centering
        \includegraphics[width=\textwidth]{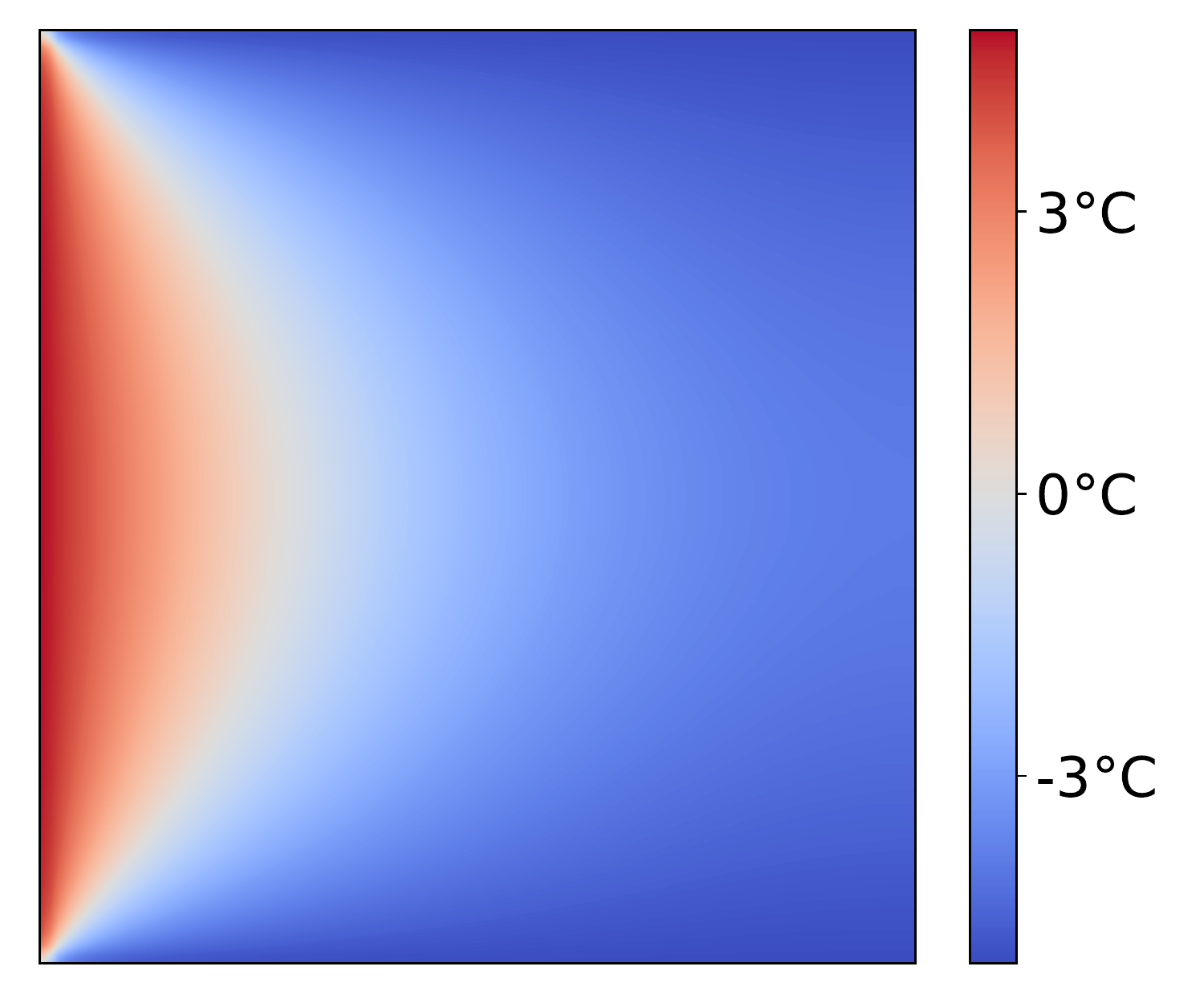}
        \caption{6\% case after 200\,h}
        \label{fig:TH3_TF_06_200h}
    \end{subfigure}
    \hfill
    \begin{subfigure}[b]{0.3\textwidth}
        \centering
        \includegraphics[width=\textwidth]{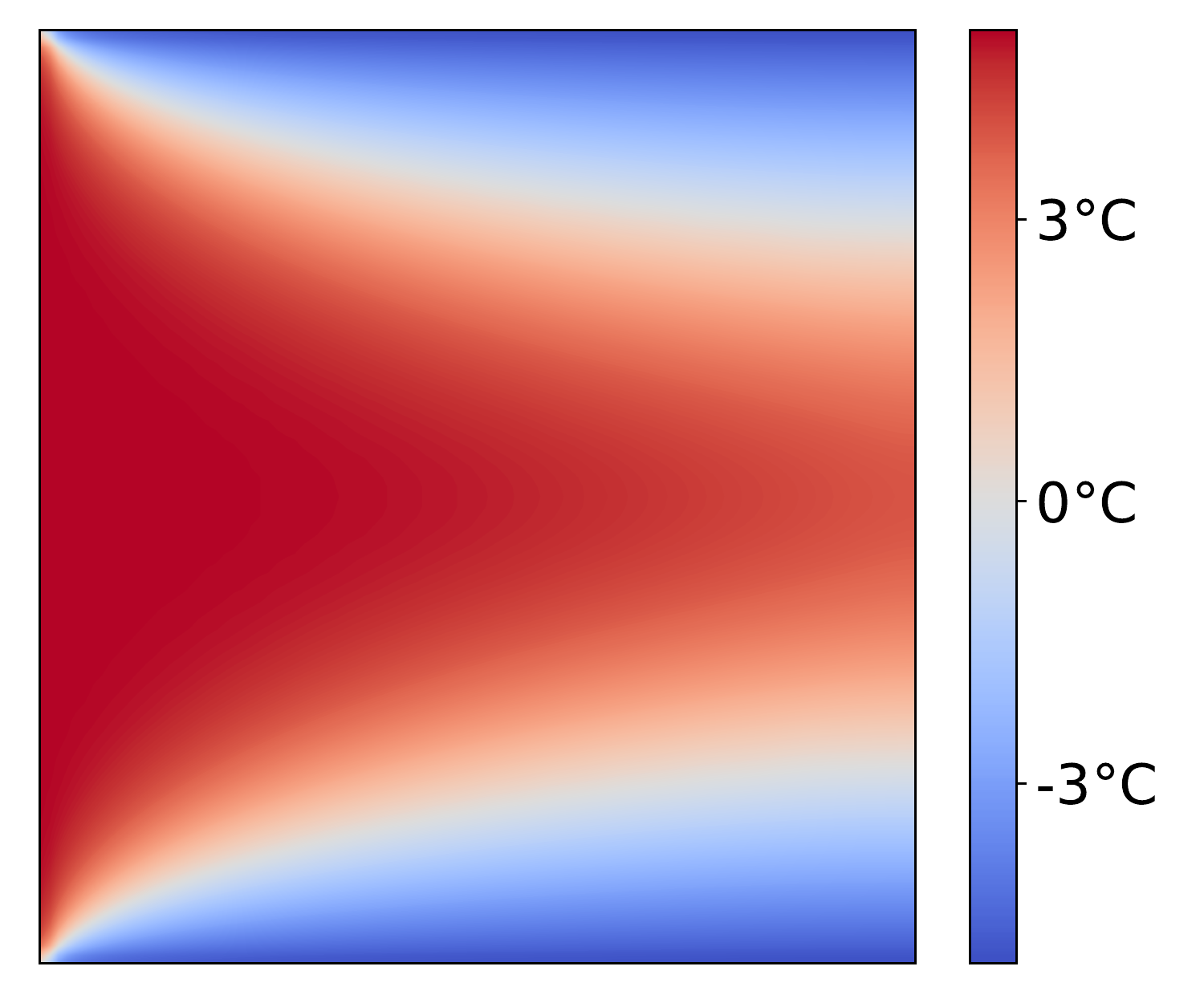}
        \caption{9\% case after 200\,h}
        \label{fig:TH3_TF_09_200h}
    \end{subfigure}
    \caption{Temperature fields of TH3: Talik Opening/Closure. Figure (a) displays the initial conditions and (b) and (c) visualize snapshots of the scenarios with hydraulic gradients of 6\% and 9\% at the end of the simulation. 
    }
    \label{fig:TH3_TF}
\end{figure}

In a square domain, measuring 1\,m by 1\,m, with an initial temperature of 5\,$^\circ$C, there are two symmetric -5\,$^\circ$C cold hemispheres at the top and the bottom. This setting is visualized in Figure \ref{fig:TH3_TF_initial}. Again, the starting values of the porosity, the ice concentration, and the water saturation are determined by the soil freezing curve. While there is no mass transfer through the top or the bottom of the domain, we impose hydraulic heads at the sides to create a flow from left to right. The values of the hydraulic gradients are 3\%, 6\%, 9\%, and 15\%. There is no conductive flux through the right side and we impose temperatures of 5\,$^\circ$C at the left and -5\,$^\circ$C at the top and the bottom boundaries. 
Initially, we use a time step of $0.1$\,seconds and we bound it by  $0.1$\,seconds from below and by $600$\,seconds from above. 
For the discretization, the domain was split into 10000 square elements and the simulated time span is 200\,hours. 

At the end of the simulation, all scenarios have reached a steady state, which are exemplary visualized for the hydraulic gradients of 6\% and 9\% in Figures \ref{fig:TH3_TF_06_200h} and \ref{fig:TH3_TF_09_200h}. For the two weakest fluxes, our model predicts a closure of the talik, while the remaining two scenarios result in an opening. In the case of a closure, the water in the middle of the domain cools down until it reaches freezing temperature, the two hemispheres connect, and the talik closes. From that point, the entire right side of the domain freezes. At the left side, some water remains fluent due to the positive temperature applied at the left boundary. For the higher hydraulic gradients, the cold water is flushed away and replaced with warm water from the left before it reaches freezing temperature. Over the course of the simulation, the hemispheres deform and the talik widens until a balance between the warm stream in the middle and the cold temperatures at the top and the bottom of the domain is reached. 

For the comparison with the other participants of the benchmark study, we use the following performance measures: the total heat, the temperature at the center of the domain, and the temperature 18\,cm above the center of the domain. Figure \ref{fig:TH3_PM_comparison} exemplary displays the evolution of the performance measures over the course of the simulation for the hydraulic gradients of 6\% and 9\%. Again, the results of all scenarios can be found in \ref{apx:performance measures}. 
In the 6\% case, the total heat increases from its initial value of $667$\,MJ up to $677$\,MJ within the first $15$\,hours before it drops down to $625$\,MJ and remains at that level with the simulation approaching steady state. The temperature at the center of the domain starts at $5^\circ$C and quickly drops to $1^\circ$C. Afterwards, with the closure of the talik it approaches a value of $-2.2^\circ$C. The temperature measured at point $x$ is initially at $-5^\circ$C, rises quickly up to $-1^\circ$C, and stagnates at this value before it eventually decreases and settles at a temperature of $-2.66^\circ$C towards the end of the simulation. 
For the hydraulic gradient of 9\%, again, the measured temperatures at the center and above the center quickly approach $1^\circ$C and $-1^\circ$C before both increase and level out at 4.79\,$^\circ$C and 3.72\,$^\circ$C. In this scenario, the total heat rises from its initial value continuously for the first 100\,hours before approaching $749$\,MJ. 

\begin{figure}
    \centering
    \begin{subfigure}[b]{\textwidth}
        \centering
        \includegraphics[width=0.32\textwidth]{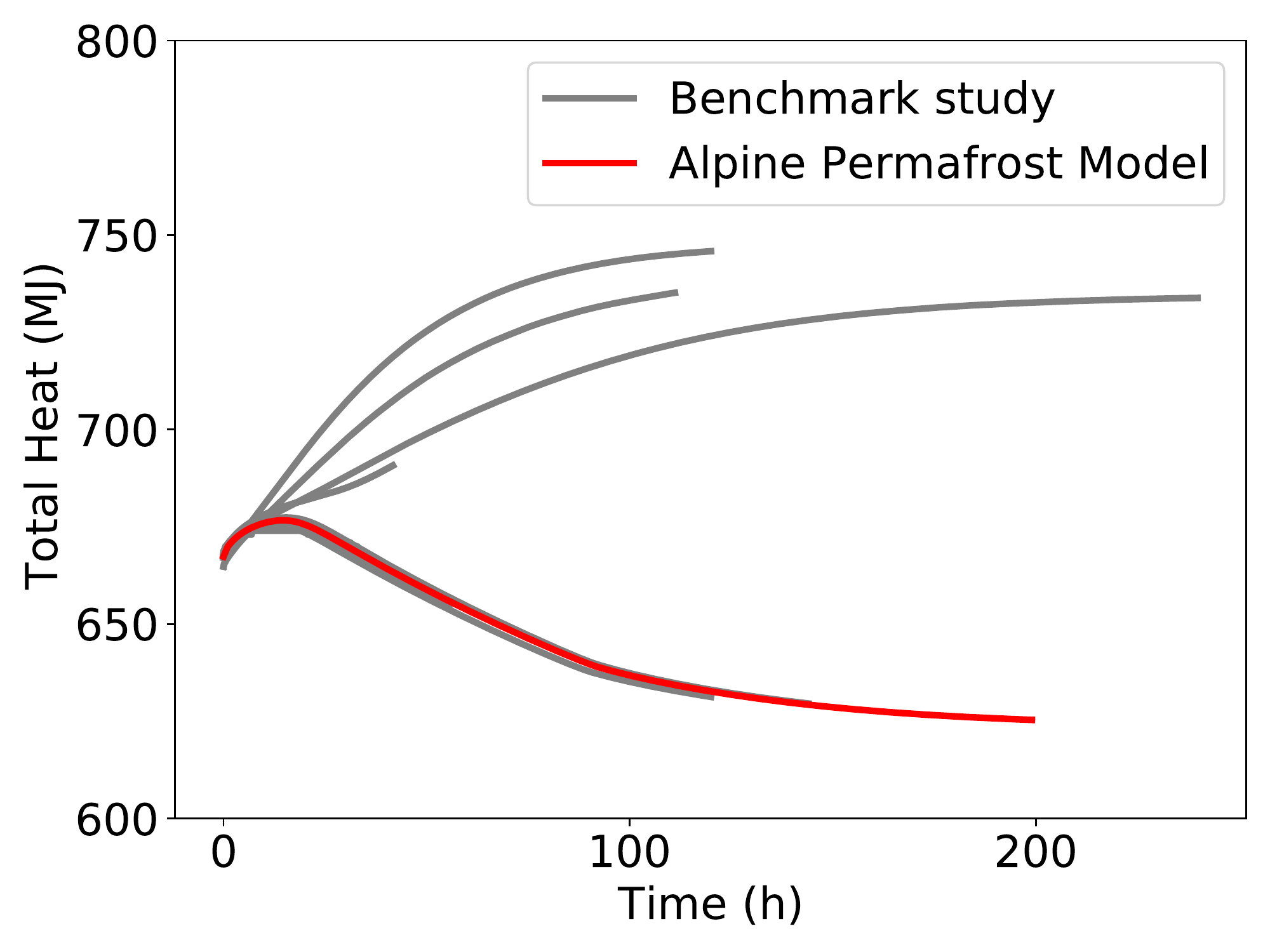}
        \hfill
        \includegraphics[width=0.32\textwidth]{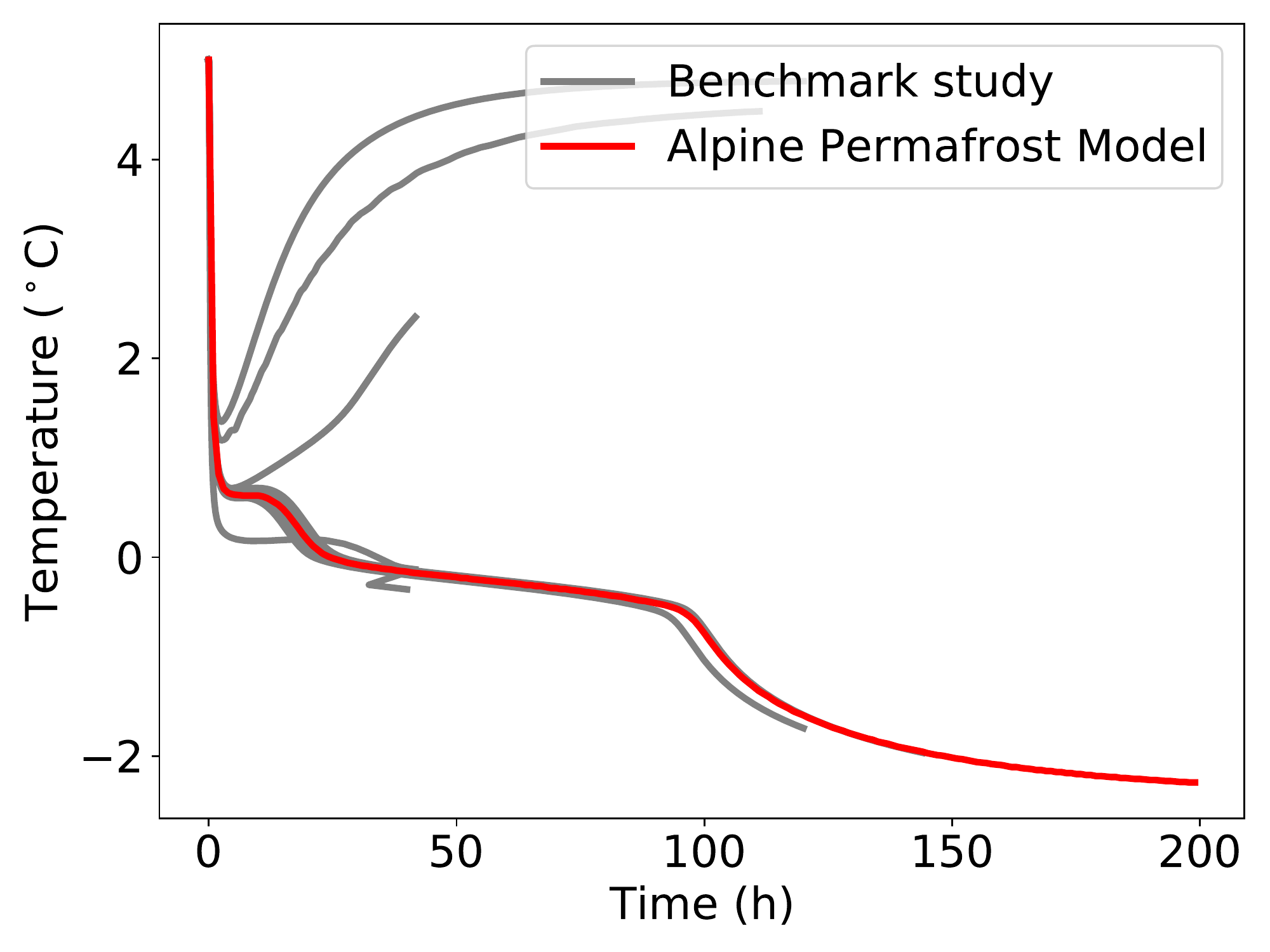}
        \hfill
        \includegraphics[width=0.32\textwidth]{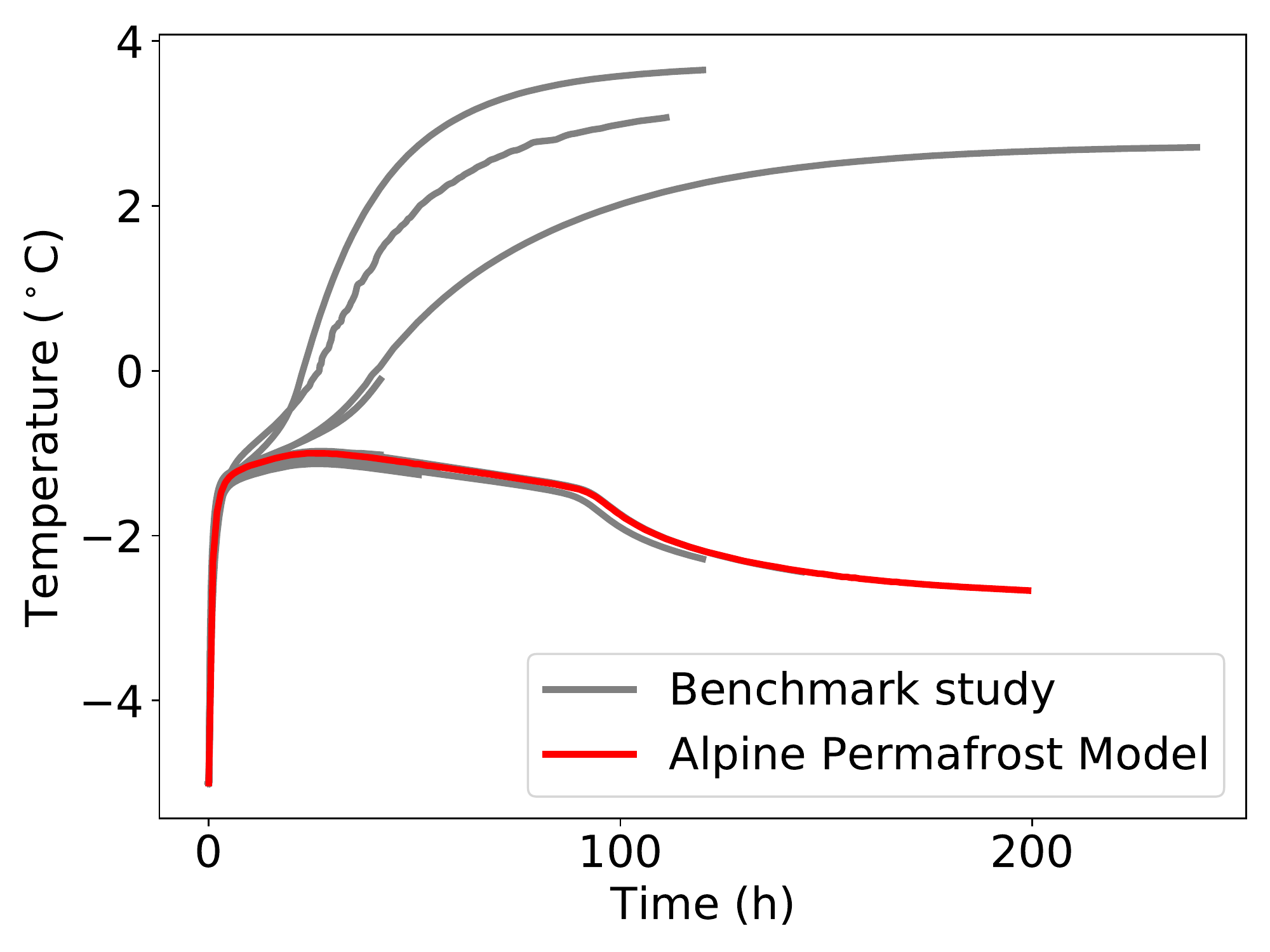}
        \caption{Hydraulic gradient: 6\%}
        \label{fig:TH3_6}
    \end{subfigure}
    
    \begin{subfigure}[b]{\textwidth}
        \centering
        \includegraphics[width=0.32\textwidth]{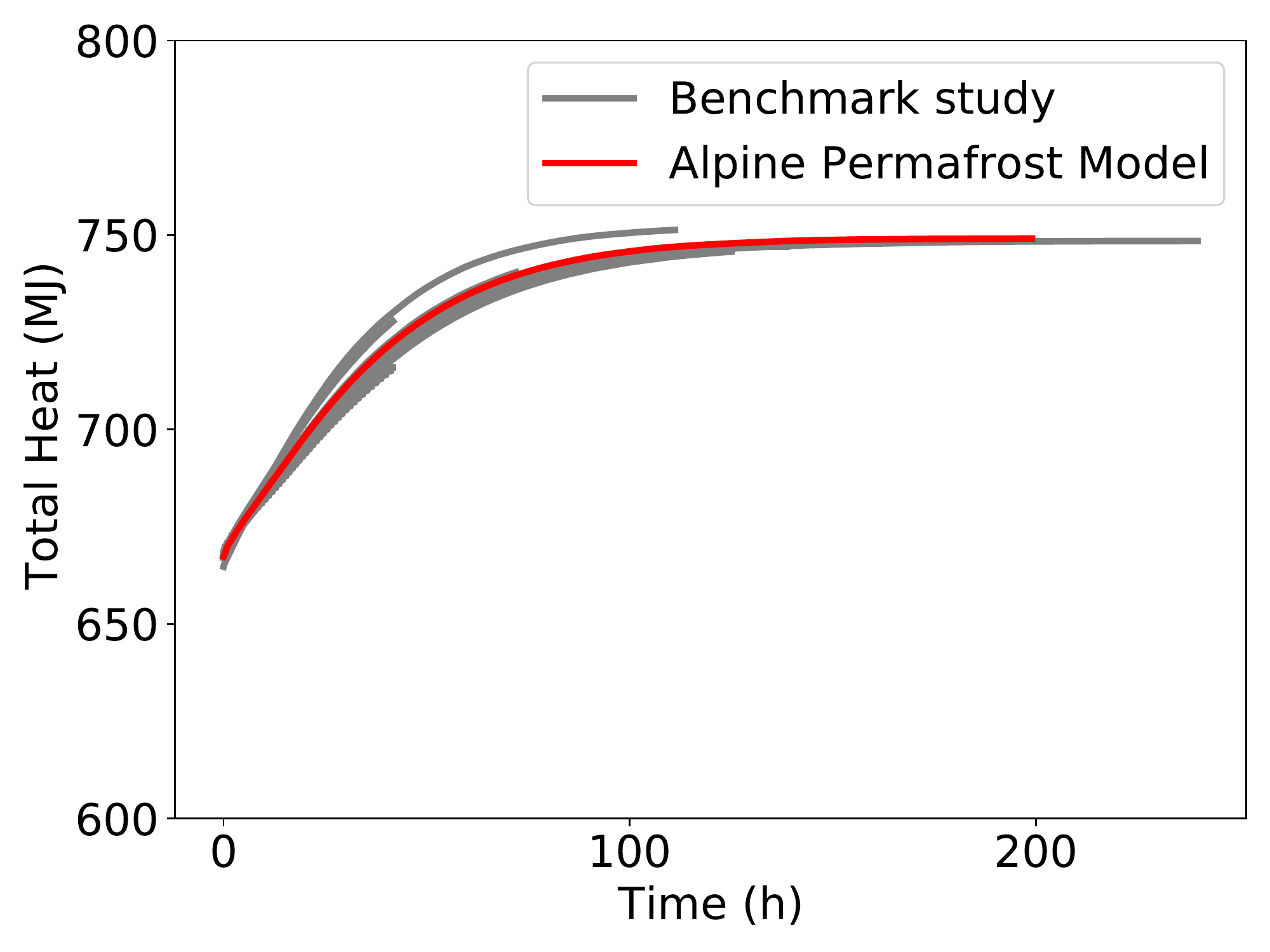}
        \hfill
        \includegraphics[width=0.32\textwidth]{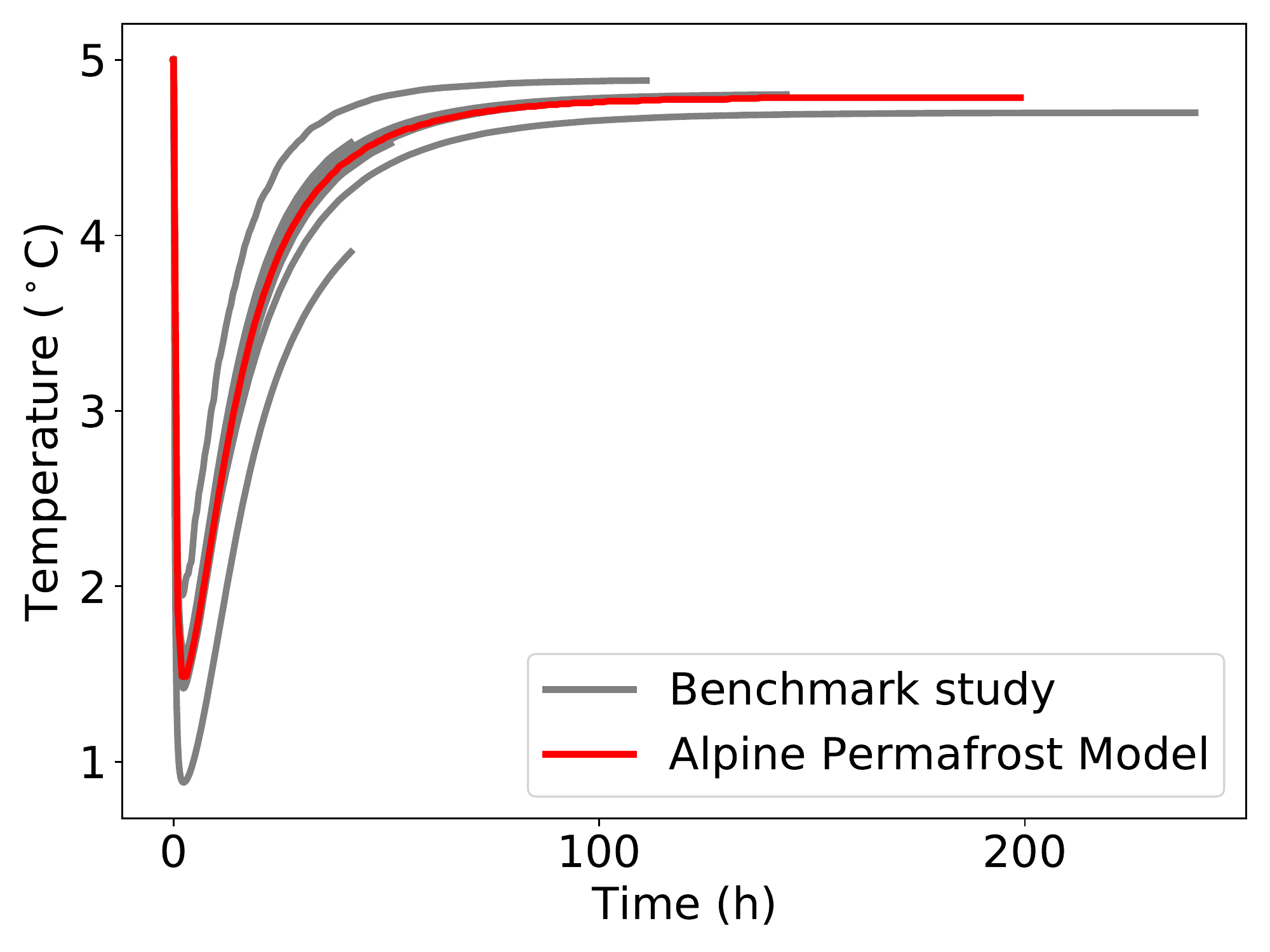}
        \hfill
        \includegraphics[width=0.32\textwidth]{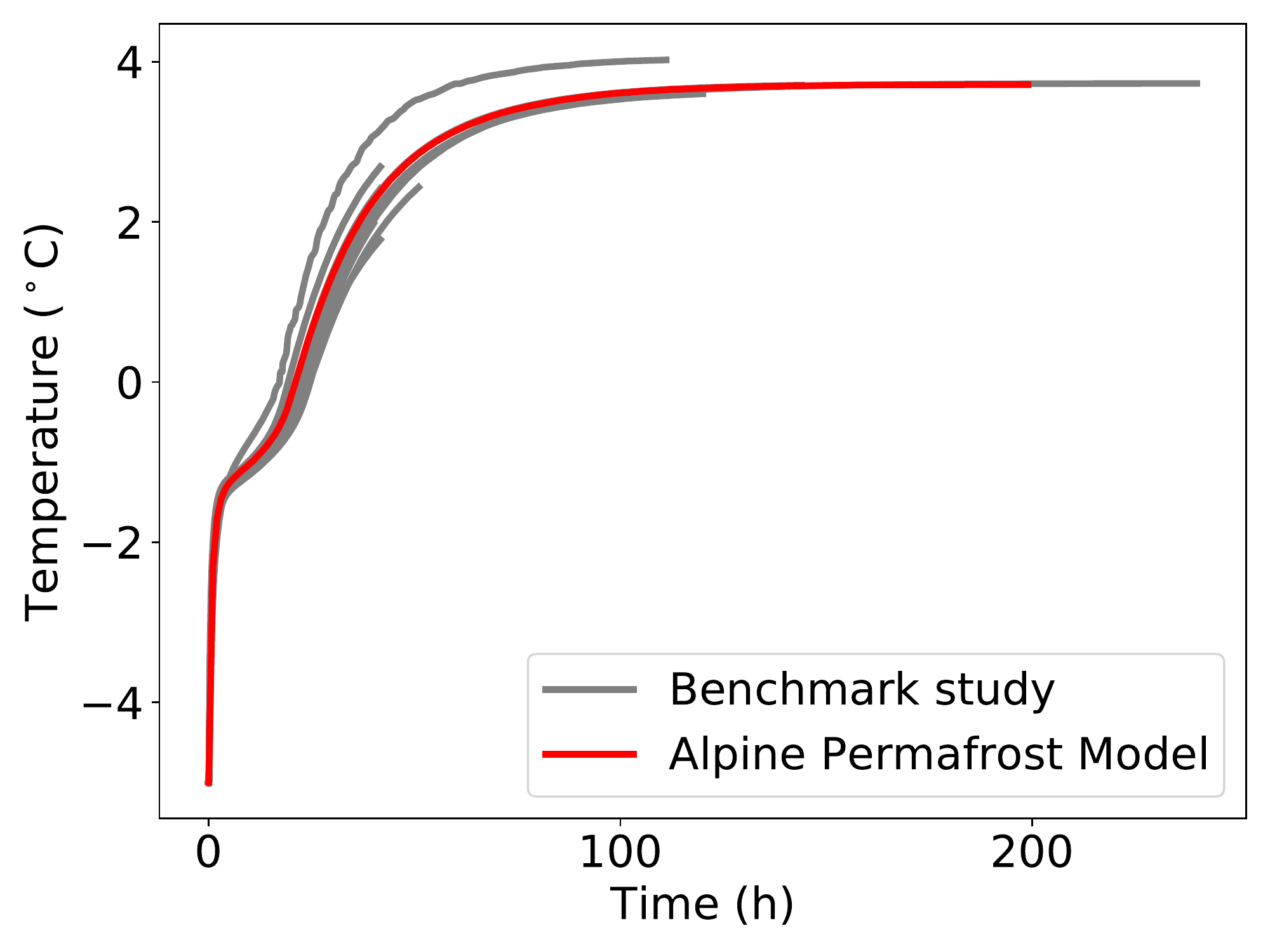}
        \caption{Hydraulic gradient: 9\%}
        \label{fig:TH3_9}
    \end{subfigure}
    \caption{Performance Measures for TH3: Talik Opening/Closure. The graphics on the left show the development of the total heat with the ones in the middle and on the right displaying the evolution of the temperature at the center of the domain and at point $x$ 18\,cm above the center. The rows (a) and (b) represent the scenarios with hydraulic gradients of 6\% and 9\%.}
    \label{fig:TH3_PM_comparison}
\end{figure}

\subsection{Discussion}
The validation studies, verify the thermo-hydraulic behavior of our model. 
In all four scenarios of the Melting of a Frozen Inclusion test case, the results of our model match those of the participating models regarding the minimal temperature and the total water volume. For the second performance measure, i.e., the total heat flux exiting the system, small differences occur. They are caused by the approximation of temperature and pressure gradients at the boundaries using a finite difference method based on the values at the cell centers. These deviations appear more significant for smaller hydraulic gradients since the error magnitude stays consistent over all four scenarios, while the order of the performance measure increases with the hydraulic gradient. 

For three out of four (3\%, 9\%, 15\%) scenarios of the Talik Opening/Closure test case, our results match those of the benchmark study. The 6\% case is really interesting, where four of the participating codes predict an opening of the talik, while the other nine forecast a closure. According to \cite{grenier2018groundwater}, the threshold value for the hydraulic gradient whether the talik opens or closes, lies between 6.3\% and 6.4\%, confirming our outcome. Our results for the 6\% scenario match those of the nine codes which also predict a closure of the talik. 

The thawing of ice and the consecutive discharge of melt water as well as the opening or closure of gaps within otherwise frozen ground are essential processes for the modeling of active layer dynamics and permafrost distributions. The validations studies prove our model's ability to depict them correctly. 
\section{Numerical case studies }
\label{sec:field_scale}

Having established the thermo-hydraulic behaviour of our approach against well-established permafrost models, we run a series of numerical case studies based on the elevation profiles of the Zugspitze (DE) and the Matterhorn (CH). In comparison to the benchmark scenarios, these alpine settings additionally include seasonal thawing and freezing, unsaturated soil, and complex topographies.  

In the European Alps, permafrost occurs starting at heights between 2400\,m and 2600\,m (e.g. \cite{haeberli1975untersuchungen, gruber2004permafrost}).  
In the case of the Zugspitze, the elevation of the simulated domain ranges from 2008\,m up to 2962\,m at the peak, and thus, should contain permafrost free as well as permanently frozen ground. 
The Zugspitzplatt, the plateau south of the summit of the Zugspitze, which is also part of the simulated domain, is characterized by uneven terrain and located well within the elevation range where the first occurrence of permafrost can be expected. In sharp contrast, the Matterhorn has a steep pyramid-like shape and its summit lies at a height of 4478\,m. Due to the cold temperature at high altitudes, the upper part of the mountain is expected to be permanently frozen. 
These two chosen profiles are markedly different in terms of elevation as well as terrain, and thus, offer a rather broad range of geophysical test conditions: The Matterhorn profile presents ideal conditions to simulate the influence of a warming scenario on the distribution of permafrost, while the Zugspitze provides a more complex topography and conditions that are ideal for studying active layer dynamics.

\subsection{Zugspitze}
\label{subsec:Zugspitze}

The Zugspitze (2962\,m) is the highest mountain of Germany and belongs to the Wetterstein mountains at the border to Austria. We reconstruct its elevation profile based on data of the Shuttle Radar Topography Mission (\url{https://earthexplorer.usgs.gov}). One data point is posted for every arc-second (roughly 30\,m), and we use linear interpolation to approximate the areas in between. The domain is discretized into 5\,m wide and 2\,m high cells, and it has a minimal depth of 100\,m, which is enhanced in steep areas to ensure a minimal cell depth of 50. The duration of the simulation is set to 250\,years, and the minimal, initial, and maximal time steps are set to $1$\,hour, 1\,day, and 10\,days, respectively. 

At the topographic surface, we impose daily average temperatures in form of Dirichlet boundary conditions. Based on the measurements of the weather station (2960\,m) near the top of the Zugspitze, which are available since August 1900 (\url{www.ncdc.noaa.gov}), we apply cyclic temperature boundary conditions in form of a sinus curve reaching its extreme values of $-11^\circ$C and $3^\circ$C in the middle of February and August, respectively. To take changes in elevation into account, we extrapolate the temperatures for the entire domain by application of a lapse rate with elevation differences calculated relative to the weather station. In fact, the lapse rate itself is subject to changes due to season and elevation, but for this study, we rely on the traditional value of $6.5^\circ$C\,km$^{-1}$ \citep{minder2010surface}. 
For simplicity, we assume a fixed atmospheric pressure ($\approx$ 1\,atm) and water saturation (80\%) at the surface throughout the simulations.
On the sides, there is no mass-flux of either fluid phase and no conductive flux. The former is also true for the bottom boundary where we simulate heat generated from within the earth in form of a geothermal gradient of $24^\circ$C\,km$^{-1}$ \citep{peters2012overview,Spooner2020}.

\subsubsection{Unsaturated setting}
\label{subsubsec:Zugspitze_homo}
\begin{figure}
    \centering
    \begin{subfigure}[b]{\textwidth}
        \centering
        \includegraphics[width=0.49\textwidth]{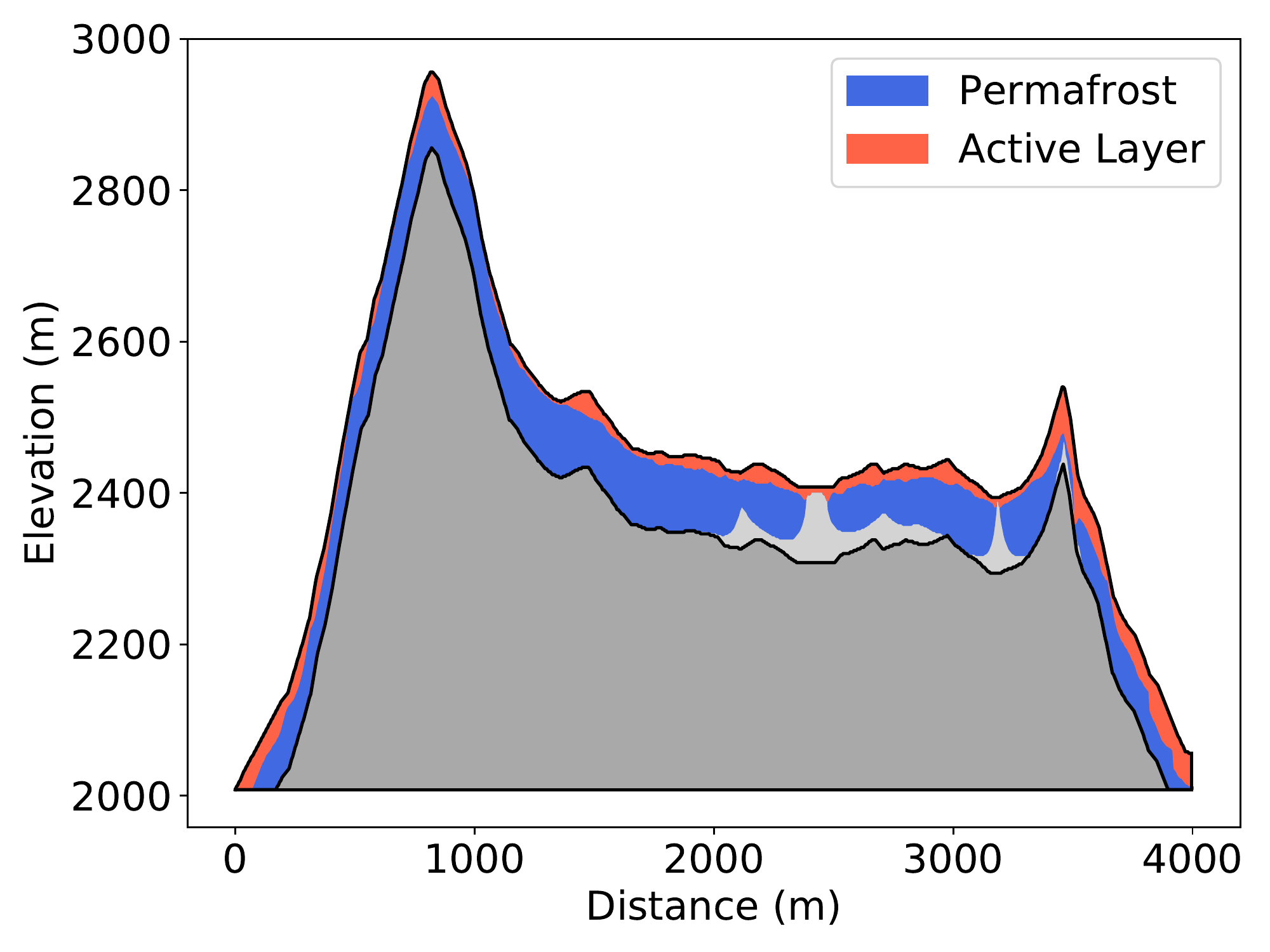}
        \hfill
        \includegraphics[width=0.49\textwidth]{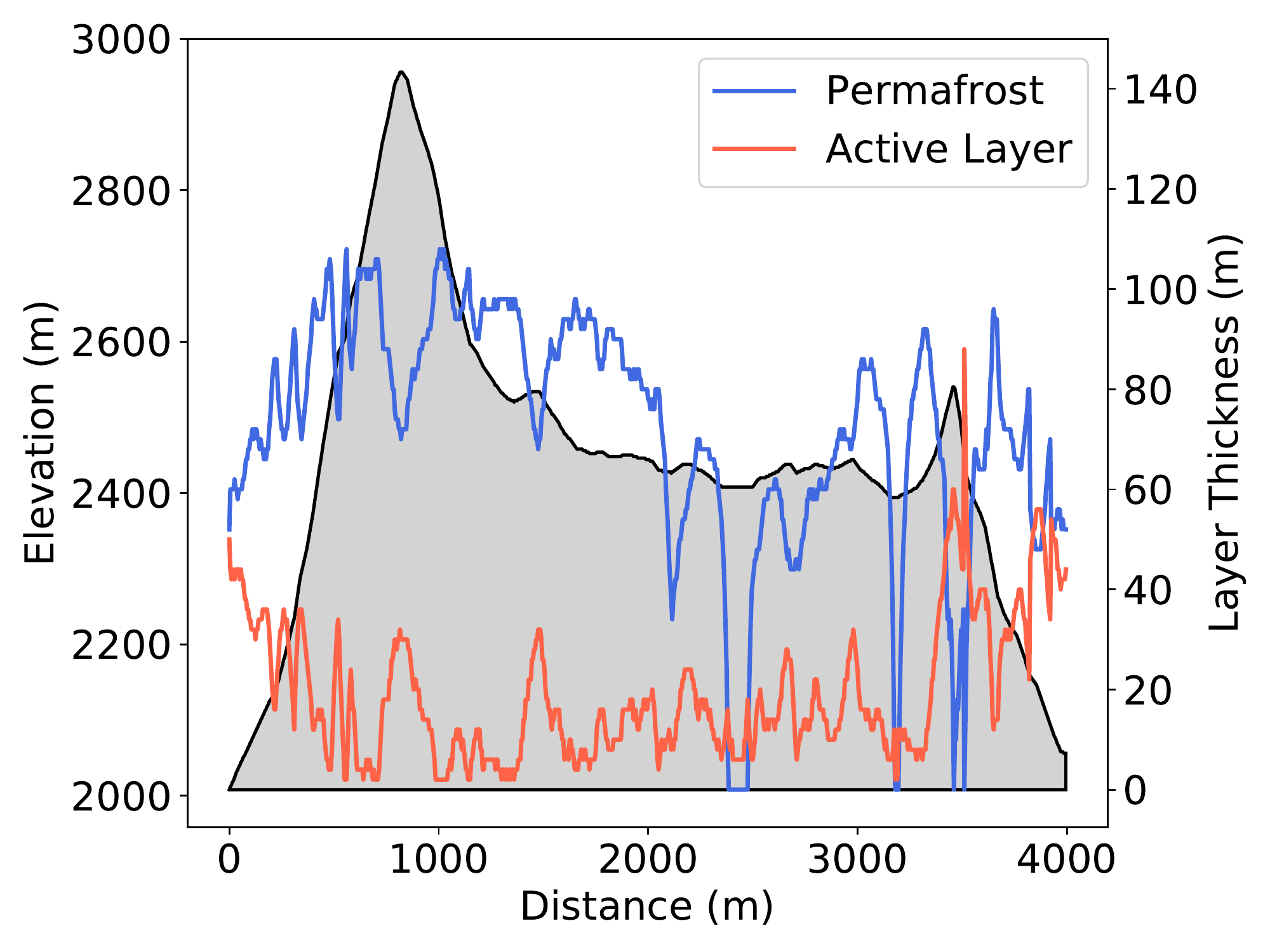}
        \caption{High permeability scenario: $K_0 = 10^{-12}$}
        \label{fig:Zugspitze_homo_12}
    \end{subfigure}

    \begin{subfigure}[b]{\textwidth}
        \centering
        \includegraphics[width=0.49\textwidth]{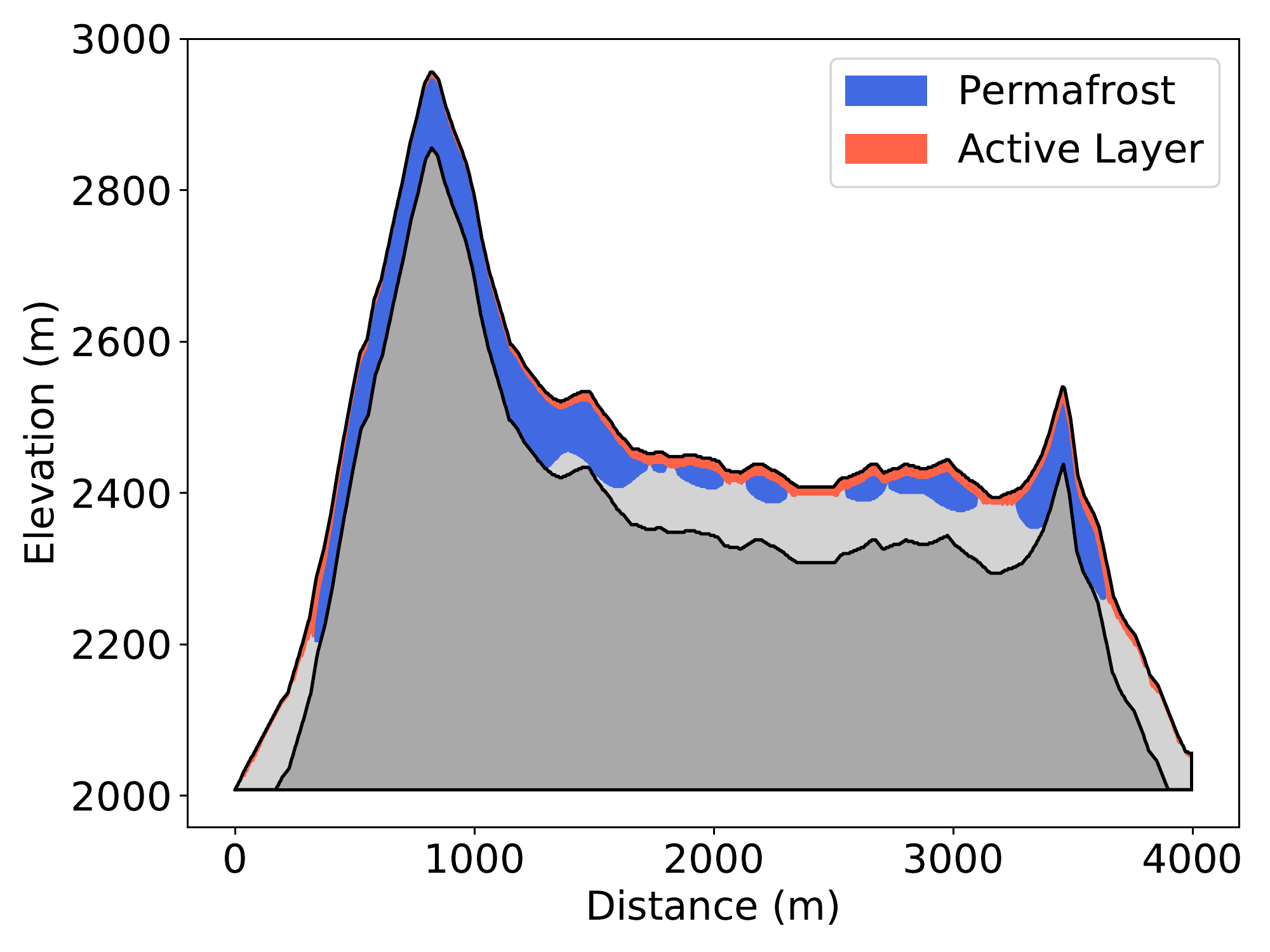}
        \hfill
        \includegraphics[width=0.49\textwidth]{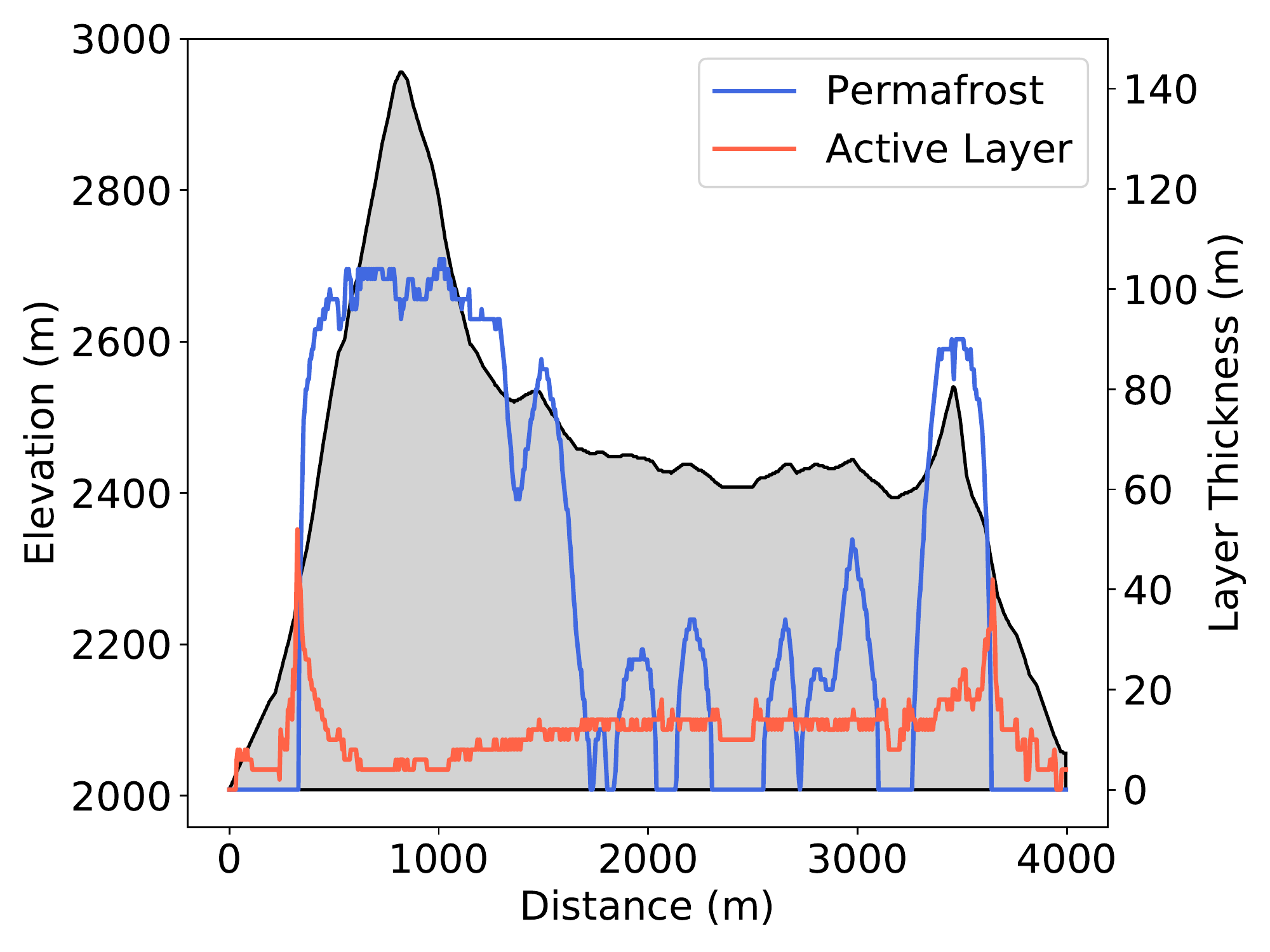}
        \caption{Low permeability scenario: $K_0 = 10^{-13}$}
        \label{fig:Zugspitze_homo_13}
    \end{subfigure}
    \caption{Distribution (left) and thickness (right) of the permafrost layer and the active layer for the unsaturated scenarios with high (top row) and low (bottom row) permeability. The mountain profile of the Zugspitze is added for orientation. }
    \label{fig:TH3_PM_comparison}
\end{figure}

The upper layer of the Wetterstein mountains is comprised of Wetterstein limestone \citep{lauber2014use}. According to \cite{krautblatter2010temperature}, such rocks have base porosity of $4.42 \pm 0.41 \%$ and permeability of $(6.16 \pm 1.11) *  10^{-12}$\,m$^2$. 
For the simulations, we use a reference porosity of $5\%$ and run two simulations with reference permeabilities of $10^{-12}$ and $10^{-13}$\,m$^2$, respectively. The composition of the simulated domain is assumed to be homogeneous for simplicity. 
The initial temperature is set to $-2$\,$^\circ$C at the surface and extended to the entire domain by application of the prescribed regional thermal gradient.
Based on the soil freezing curve, we determine the porosity, the ice concentration and the water saturation. 

For the case of a permeability of $10^{-12}$\,m$^2$, the permafrost distribution and the active layer after 250\,years are visualized in Figure \ref{fig:Zugspitze_homo_12}. 
The north-south cross section runs through the summit of the Zugspitze, on the left, and also includes the smaller peak of Plattspitzen.
The right part of the graphic shows the local distribution of permafrost and active layer, while the left image displays the thicknesses of both layers, together with the elevation profile as an orientation. 
Nearly throughout the domain, the permafrost layer reaches the bottom boundary, and thus, would also extend outside of the simulated domain. 
At the plateau in between the two peaks, the permafrost distribution is highly irregular with two taliks forming within the otherwise predominantly frozen ground. They are located below small valleys with hills on both sides. 
Also, the thickness of the active layer varies strongly. 
It is especially increased at topographically exposed positions such as local peaks and hills. 

The results for the $10^{-13}$\,m$^2$ case are shown in Figure \ref{fig:Zugspitze_homo_13}. 
At both peaks, permafrost reaches the bottom boundary of the domain, while at the plateau in between, the permafrost layer is highly discontinuous, occuring as scattered patches centered around local peaks.
The active layer shows a uniform distribution following the modelled topography, except of steep increases at the outside edges of the permafrost layer. 

\subsubsection{Saturated setting}

In order to emphasize the importance of present air and to show its influence on the distribution of permafrost and active layer dynamics, we assume the full saturation of the modelled soil for the following scenario. Therefore, the entire pore space is initially filled with water, frozen or liquid, and the imposed value for the water saturation at the boundary is set to 100\%. Otherwise, the setting of this simulation is equivalent to the unsaturated case study with a permeability of $10^{-12}$\,m$^2$. 

\begin{figure}
    \centering
    \includegraphics[width=0.49\textwidth]{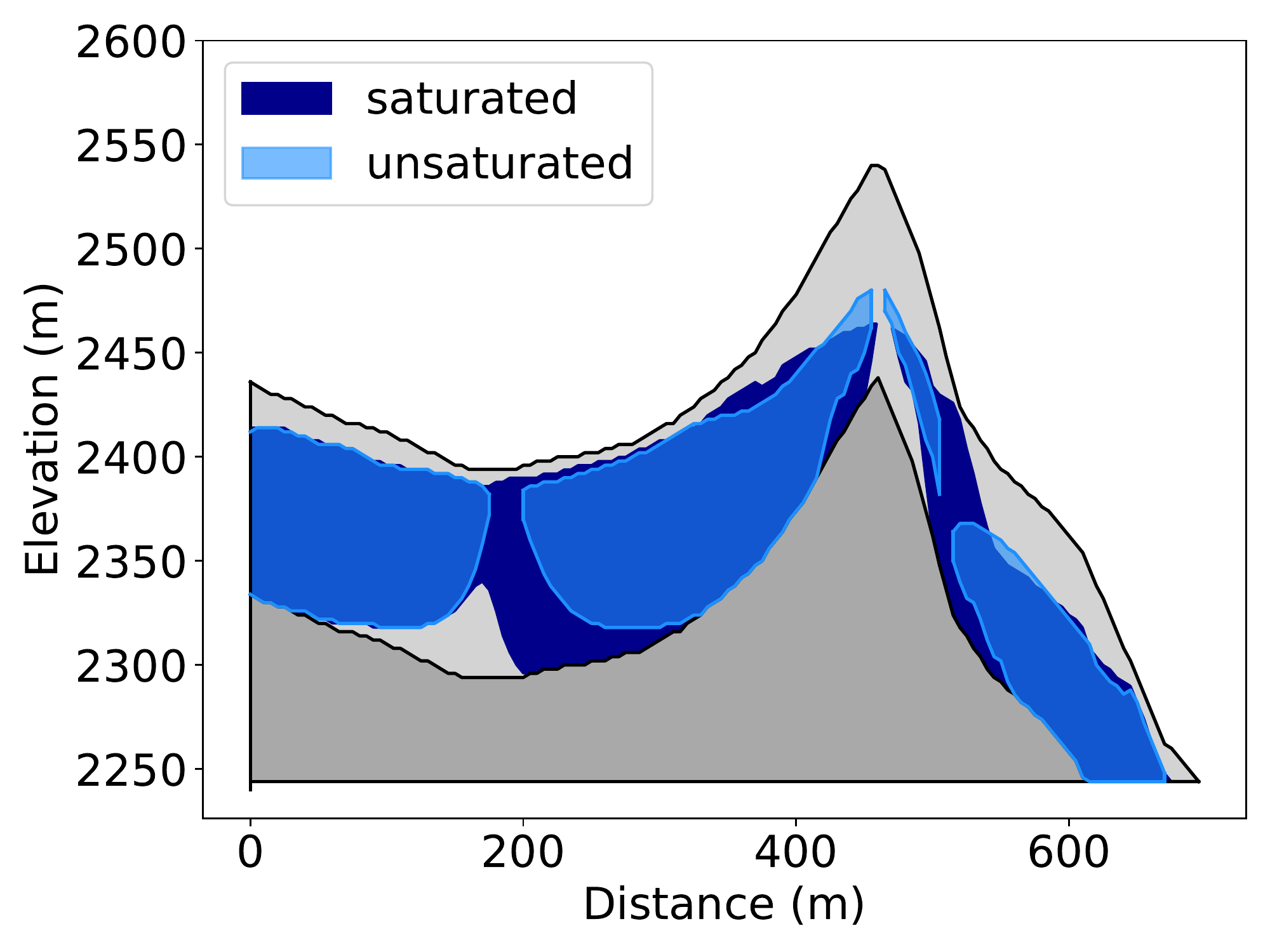}
    \hfill
    \includegraphics[width=0.49\textwidth]{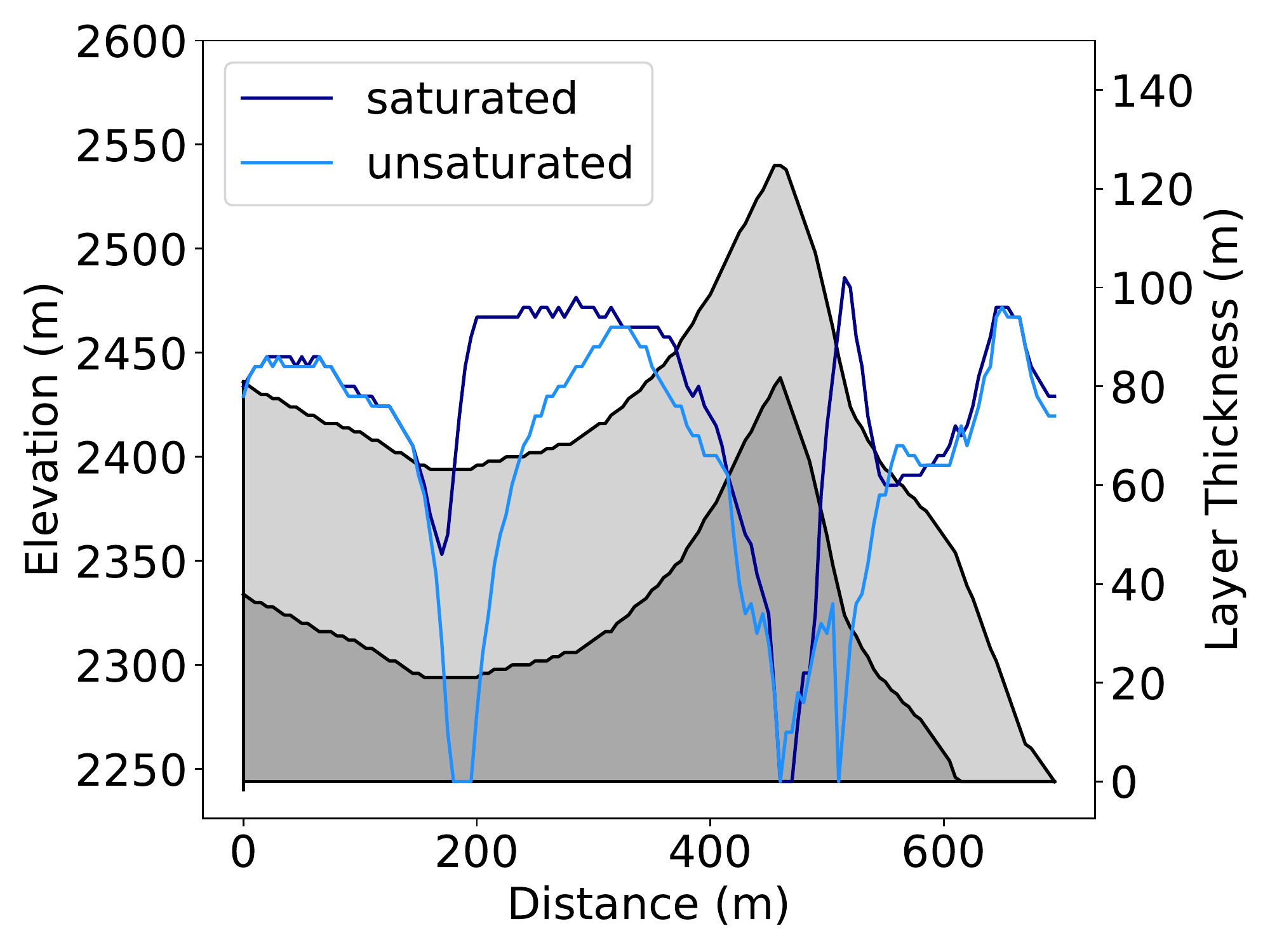}
    \caption{Comparison of the distribution (left) and thickness (right) of the permafrost layer for the saturated and the unsaturated scenario. The corresponding sector of the mountain profile of the Zugspitze is outlined in black.}
    \label{fig:Zugspitze_homo_sat}
\end{figure}

The distribution of permafrost and the thickness of the active layer develop similar to the unsaturated scenarios. The permafrost layer covers most of the simulated depth throughout the domain and shows irregularity at the uneven plateau between the peak of the Zugspitze and the ridge of the Plattspitzen, where again a talik opens. Also, the active layer thickness varies across the full length of the mountain profile with increased active layer dynamics around local peaks and other exposed position. 
While the general outcomes of the two scenarios coincide, the influence of the increased water content becomes apparent in a closer comparison. 
Figure \ref{fig:Zugspitze_homo_sat} displays the differences of the permafrost layers of both simulations at the Plattspitzen. In the case of the fully saturated soil, more ground stays permanently frozen and the permafrost layer is more interconnected. 
Locally, the thickness of the permafrost layers of the saturated and the unsaturated scenario differs by up to 92\,m. 

\subsection{Numerical case study: Matterhorn}
\label{subsec:Matterhorn}

Next, we apply our model to one of the most recognizable mountains worldwide, the Matterhorn (4478\,m) in Switzerland. Its pyramid-shaped summit mainly consists of gneiss of the Dent Blance nappe \citep{pleuger2007structural}. According to \cite{rempel2016modeling}, the permeability of this material ranges from $10^{-18}$ to $10^{-12}$\,m$^2$, and the authors use a nominal value of $10^{-14}$\,m$^2$ for their study, which is what we also use as reference permeability in our study. 
For the reference porosity, we use a value of 1\% \citep{draebing2012p}. 
We use similar boundary conditions as for the Zugspitze case studies. 
The temperature at the surface is based on the measurements of the Pian Rosa mountain top airport (3488\,m) in Italy (\url{www.ncdc.noaa.gov}), which is located 6\,km south-south-east from the peak of the Matterhorn. 
The duration of the simulations, the initial, the minimal, and the maximal time steps are the same as for Zugspitze. 
In the case of the Matterhorn, the domain has a depth of 300\,m and it is discretized into cells measuring 5\,m by 1\,m. 
\subsubsection{Baseline scenario}
\label{subsubsec:Matterhorn_homogeneous}

\begin{figure}
    \centering
    \includegraphics[width=0.49\textwidth]{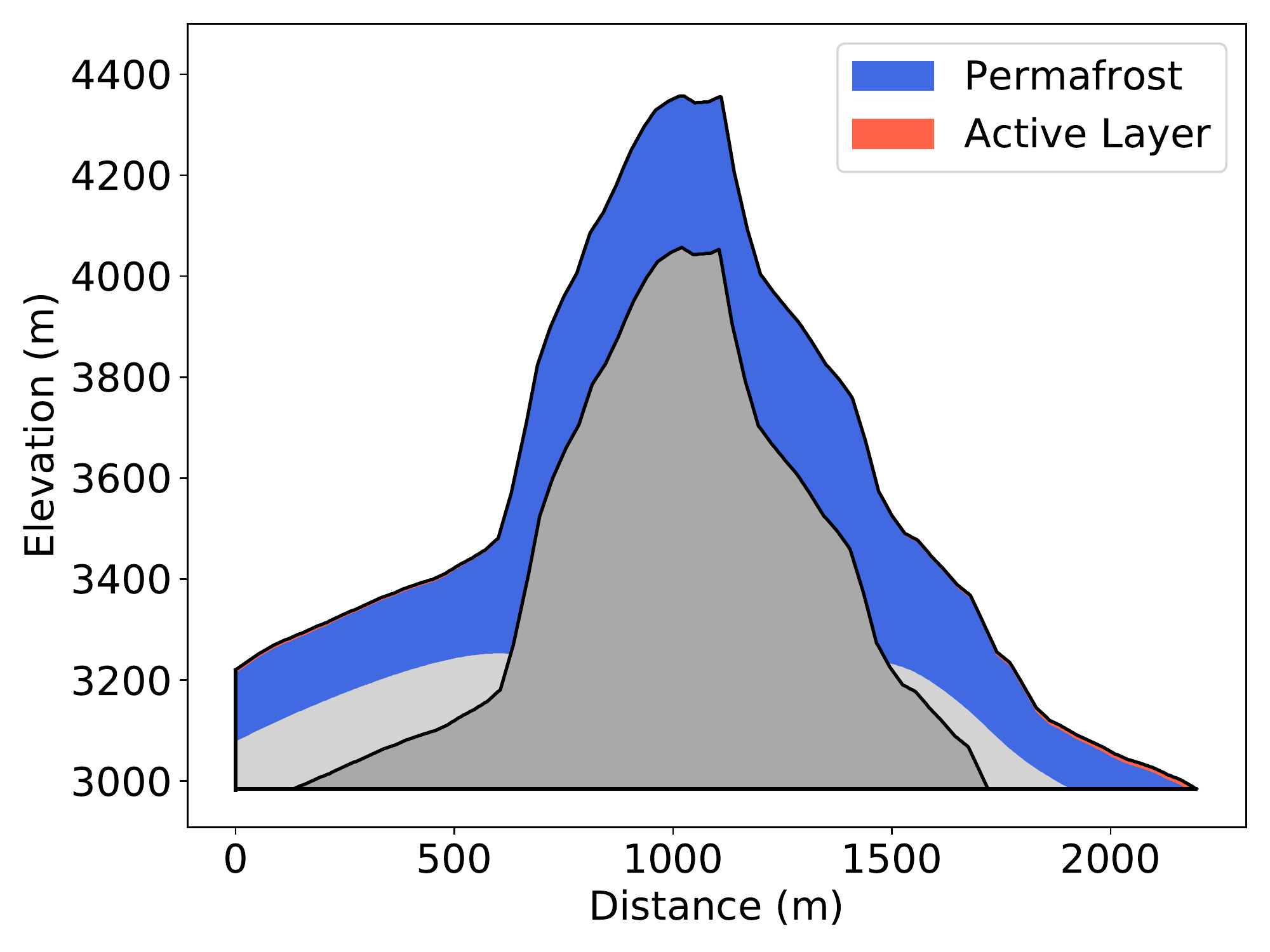}
    \hfill
    \includegraphics[width=0.49\textwidth]{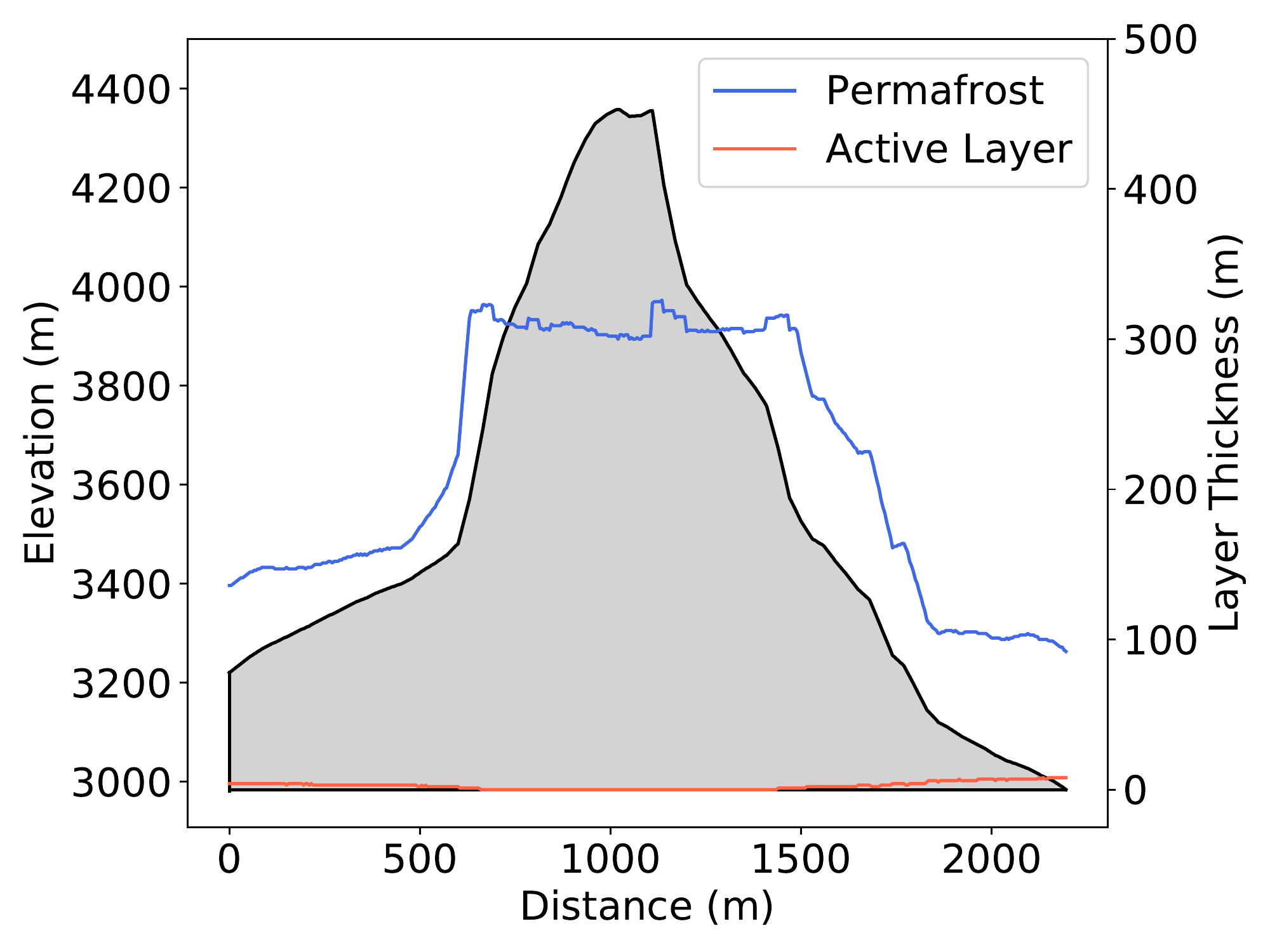}
    \caption{Permafrost distribution (left) and thickness (right) at the Matterhorn after 250 \,years.}
    \label{fig:Matterhorn_homo}
\end{figure}

Periodic temperature boundary conditions are applied at the top of the domain. Based on the temperature data described above, we consider a sinus curve with extreme values of $-13.8^\circ$C and $3.2^\circ$C at an elevation of 3488\,m. Again, we use a lapse rate of $-6.5^\circ$C\,km$^{-1}$. 

Figure \ref{fig:Matterhorn_homo} displays a north-south cross section through the peak of the Matterhorn with the distribution of active layer and permafrost at the end of the simulation. Permafrost occupies most of the simulated domain with the thickness of the permafrost layer varying between 100\,m and 300\,m. Above 3600\,m the active layer is not existent since the temperature stays consistently below 0$^\circ$C, and thus, no seasonal thawing and freezing occurs. At lower altitudes the active layer reaches depths of up to 20\,m. 

\subsubsection{Warming scenario}
\label{subsubsec:Matterhorn_warm}

\begin{figure}
    \centering
    \includegraphics[width=0.49\textwidth]{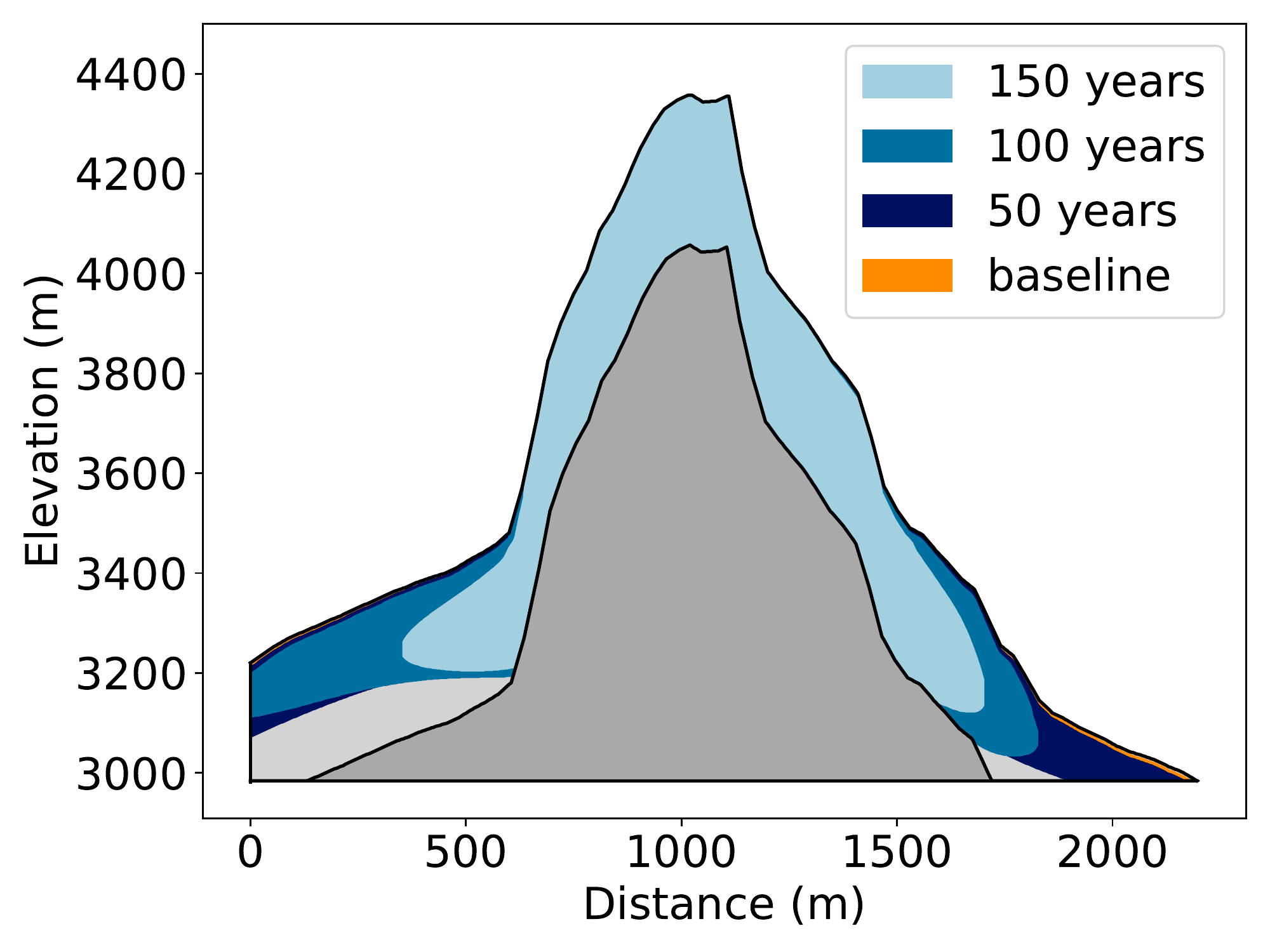}
    \hfill
    \includegraphics[width=0.49\textwidth]{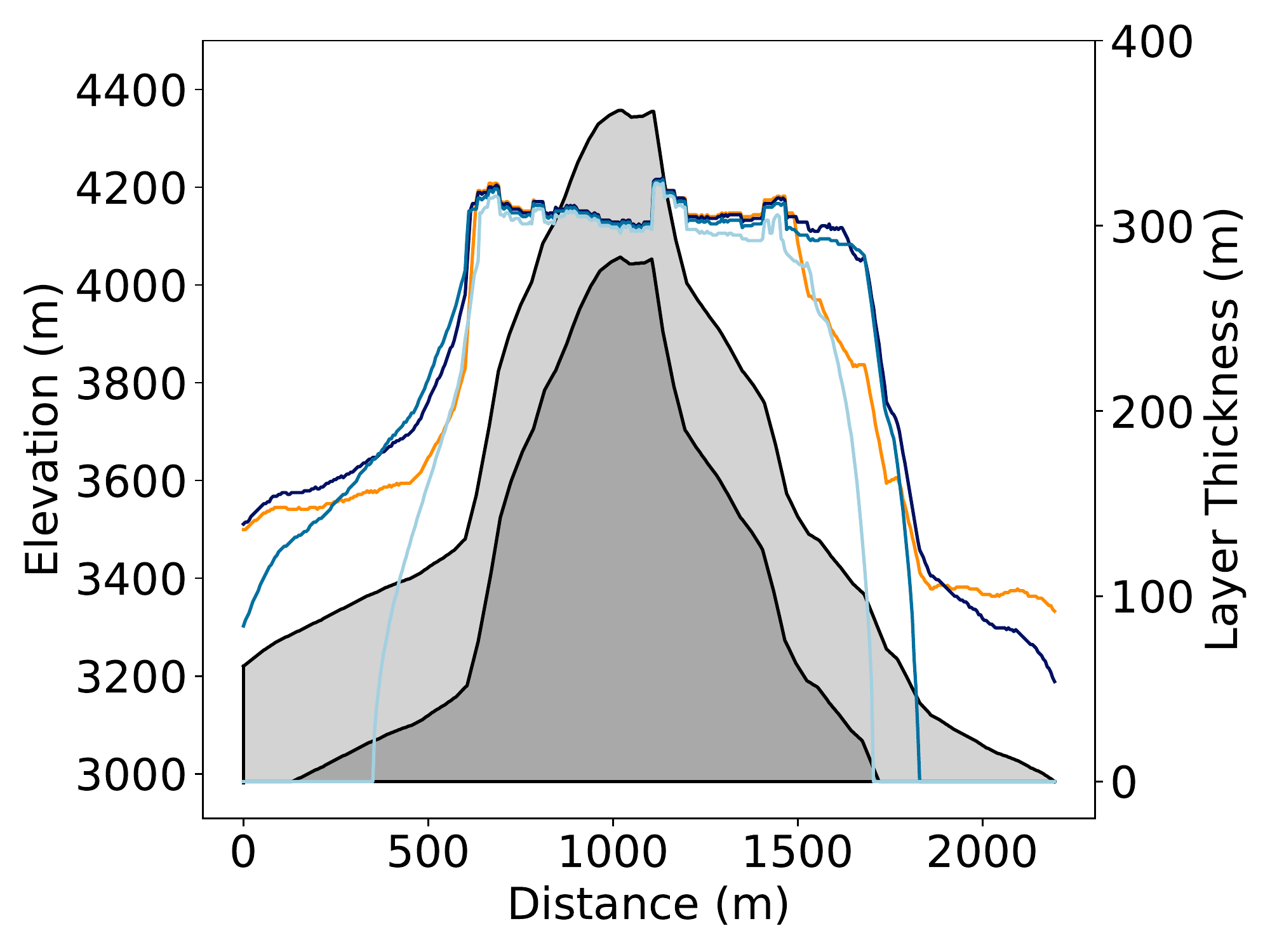}
    \caption{Evolution of permafrost distribution (left) and thickness (right) at the Matterhorn over the course of a  150\,year long warming period with a warming rate of $0.045^\circ$C dec$^{-1}$.}
    \label{fig:Matterhorn_homo_warm}
\end{figure}

Next, we simulate an idealized warming scenario. After an initialization period of 100\,years, where permafrost and active layers are allowed to form under repeated seasonal freezing and thawing, the surface temperature rises at a rate of $0.045$\,$^\circ$C dec$^{-1}$ which corresponds to the average warming of the near-surface air temperature for the European land since 1981 \citep{etzelmuller2020twenty}. In every other way, the setting of this scenario coincides with the description in \ref{subsubsec:Matterhorn_homogeneous}. 

The permafrost distribution and the thickness of the permafrost layer for the warming scenario are visualized in Figure \ref{fig:Matterhorn_homo_warm}. 
After 50\,years of warming, in low-lying regions small signs of degradation are visible in form of a gap between the surface and the underlying permafrost layer, while at higher altitudes the thickness of the permafrost layer is still increasing. 
100\,years into the warming period, permafrost has started to degrade significantly, and it no longer occurs below 3000\,m. 
At the boundaries of the permafrost layer, where the degradation is ongoing, the active layer between the surface and the consistently frozen ground is widened. 
The degradation process continues until the end of the simulation. 
After 150\,years of warming, permafrost can only be found above 3150\,m, and at its boundaries, the gap to surface is up to 100\,m wide. 
Throughout the warming period, the permafrost above 3600\,m was not affected and remained mostly unchanged. 

\subsection{Discussion}

The focal point of the numerical case studies is the sensitivity of permafrost to lateral thermal and fluid fluxes along topographic gradients. 
In one-dimensional models, which are widely used for deriving regional and basin scale estimates of permafrost depths, soil columns are only influenced by the boundary conditions at their surface and bottom. 
Thus, the temperature fields, and therefore, the thickness and distribution of the permafrost layer and the active layer essentially follow the topography of the modelled mountain.
The inherent implication is that the permafrost distribution is largely controlled by the thermal diffusivity of the sediments and hydrological drainage networks. 

Our multi-dimensional modeling results clearly show the influence of lateral fluid and thermal fluxes on permafrost depths and distribution, and their small scale manifestations, e.g. around local peaks and valleys, cannot be ignored.
Contrary to the 1D modeling approaches, convection-driven heat transfer plays an indispensable role in our two-dimensional simulations. 
The lateral fluxes lead to an increased exposure of hills and peaks to the applied weather conditions as well as the drain-off and  accumulation of liquid water in local sinks and valleys, and therefore, to the formation of taliks, patchy and discontinuous permafrost layers, and enhanced active layer dynamics around topographically exposed positions and along permafrost boundaries. 
Our findings coincide with the claims by \cite{gruber2004interpretation} and \cite{etzelmuller2013recent} regarding the importance of lateral influences for the modeling of alpine permafrost. 

The extent of fluid and heat fluxes is highly dependent on the characteristics of the modelled soil. Higher permeabilities lead to increased pore-fluid movement along topographic gradients, and thus, increase the lateral convective heat transport. 
Enhanced pore-fluid infiltration also increases the overall depth of propagation of the cold-temperature into the sediment, leading to thicker and more dynamic active layers as well as thicker permafrost layers.
On the other hand, as permeabilities decrease, the heat transport becomes more and more limited by the thermal diffusivity of the sediments, which are typically very low. 
In low permeability sediments, the active layers are thinner and more evenly distributed closely following the surface topography, as visible in Figure \ref{fig:Zugspitze_homo_13}, and the permafrost layers are also less deep and possibly even discontinuous due to limited propagation of the cold temperature into the sediment.

Naturally, the amount of water inside the porous medium influences the ice-water phase transition, and thus, the degree of saturation has to be considered. The relative ratio of the pore-filling fluids influences the density, the thermal conductivity and the heat capacities of the mixture. 
Therefore, parameters which are relevant for freezing and thawing processes change throughout the domain and over time. 
Locally, this leads to significant differences in the thickness and distribution of the permafrost layer and the formation of taliks in otherwise frozen ground. 

The sensitivity of permafrost to warming trends has been established in previous research (e.g. \cite{etzelmuller2020twenty}). 
\cite{connon2018influence} describe the process of permafrost degradation as follows: Warmer temperatures lead to the widening of the active layer and subsequently to the opening of taliks between the active layer at the surface and the permafrost underneath. 
In the following, the unfrozen gaps expand and the permafrost layer shrinks and consequently vanishes. 
This explanation coincides with the results of our warming scenario, where we observe the formation and expansion of a gap between the surface and the permafrost layer during its degradation.

\subsection{Limitations of the study}

The mountain case studies are intended to show the influence of topography driven lateral thermal and fluid fluxes on the distribution of permafrost. 
Thereby, we utilize idealized mountain settings and purposefully restrict the number of modeled surface processes to rule out alternative causes.

One limitation of the study is that the surface covers are ignored. 
The presence of glaciers influences both the temperature and the water saturation close to the surface. 
According to \cite{haeberli1983permafrost}, the temperatures at the base of glaciers are higher than at rocks which are in direct contact to the atmosphere. 
Furthermore, our model only resolves subsurface processes and does not consider surface covering snow,
and therefore, ignores its insulating effect, its influence on the ground surface temperature, the runoff of melt water, and the evolution snow covers undergo based on time and weather conditions. 
Moreover, the values imposed at the surface boundary are simplified. 
The air pressure and the water saturation are imposed as constant values, and the surface temperature is solely based on measurements of nearby weather stations, extrapolated by application of a lapse rate. The influences of other weather variables (e.g., \cite{noetzli2007three, luetschg2008sensitivity, kokelj2015increased}) such as wind, precipitation, orientation of mountains, and solar radiation are ignored. 

Finally, we use idealized reconstructions of the morphological as well as geophysical settings of the Zugspitze and the Matterhorn. The sediments are assumed to be homogeneous and the presence of different kinds of soil and rocks, cracks, and rock glaciers, and thus, the resulting porosity, permeability, and thermal conductivity variations are ignored. 

The main interest within the scope of this study was to highlight the role of lateral fluxes on the distribution of permafrost and active layer dynamics. 
Despite the applicability of our model and its realization in DUNE to three-dimensional problem settings, we limited our numerical test cases to only 2D, which was sufficient to qualitatively show the influences of lateral fluid and convection-driven heat fluxes along topographic gradients. 
\section{Conclusion and Outlook}
\label{sec:conclusion}

We present an extended multi-physics model for subsurface permafrost in alpine regions. The validation studies demonstrate the ability of our model to portray phase-change in combination with flow processes correctly and to predict the opening and closure of gaps of unfrozen ground within permafrost regions. The thermal-hydraulic behavior of the model coincides with that of well-established soil freezing models.

From the results of the numerical case studies, we draw the following conclusions: 
The complex topography of mountainous regions leads to lateral fluxes along its topographic gradients, which influence the distribution and thickness of permafrost as well as the active layer dynamics. 
At local peaks and other topographically exposed positions, the lateral pore-fluid movement and the resulting convection-driven heat transport result in increased active layer dynamics, while the accumulation of liquid water in local sinks and valleys leads to a highly irregular permafrost distribution and even to the formation of taliks in otherwise frozen ground. 

In low permeability sediments, fluid velocities are low and the depth and rate of penetration of surface temperature into the sediment are limited. This leads to shallow and largely uniform active layers, which closely follow the surface topography, and to thinner, more irregular, and less connected permafrost layers. 
On the contrary, in the case of higher permeabilities, the active layer is thicker and its shape shows noticeable deviations from the mountain course. The effects of convection-driven heat fluxes begin to dominate resulting in a thicker and highly irregular permafrost distribution. 

Finally, changes in the degree of saturation alter the composition and properties of the pore-fluid mixture, and thus, influence thermal and fluid fluxes as well as freezing and thawing processes. While an increased pore-water content leads to a thicker and more connected permafrost layer, drier conditions result in a less regular and more patchy distribution of permafrost. Even small changes in the degree of saturation can cause large local differences in the thickness of the permafrost layer. 

The results of this study and the presented model are not only limited to alpine permafrost environments, but are also applicable to the hydrogeology of Arctic regions. 
Recently, one-dimensional, thermo-hydraulic models for unsaturated flow were used to study the formation and expansion of taliks in the Northwest Territories, Canada \citep{devoie2019taliks}. 
Besides soil freezing in unsaturated conditions, the model is generally well suited to describe freezing and thawing processes in combination with mobile gas phases. 
Therefore, sub-permafrost methane concentrations such as in Svalbard, Norway \citep{hodson2020sub}) and northwest Alaska \citep{sullivan2021influence} provide another possible future area of application.
\section{Acknowledgements}

We would like to thank Dr. Christophe Grenier for his input on the recreation of the validation studies. We gratefully acknowledge the support of the German Science Foundation (DFG) for funding part of this work through grant WO 671/11-1. GC acknowledges funding from the DFG Research Group (FOR2793/1) Sensitivity of High Alpine Geosystems to Climate Change since 1850 (SEHAG) through grant CH 981/3-1.

\newpage
\appendix
\section{Benchmark Results}
\label{apx:performance measures}

\subsection{Melting of a Frozen Inclusion}

\begin{figure}[H]
    \centering
    \begin{subfigure}[b]{0.32\textwidth}
        \includegraphics[width=\textwidth]{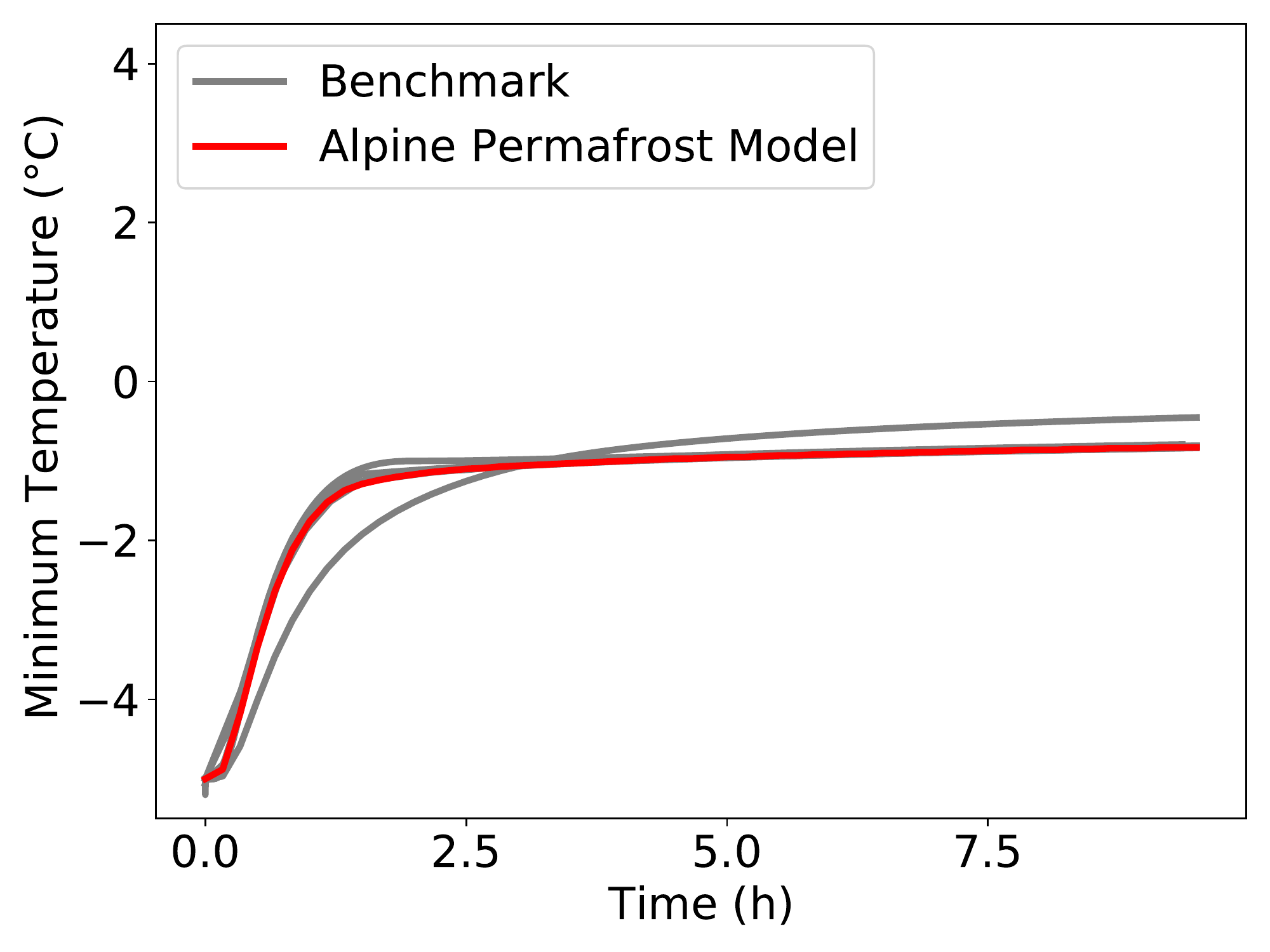}
    \end{subfigure}
    \hfill
    \begin{subfigure}[b]{0.32\textwidth}
        \includegraphics[width=\textwidth]{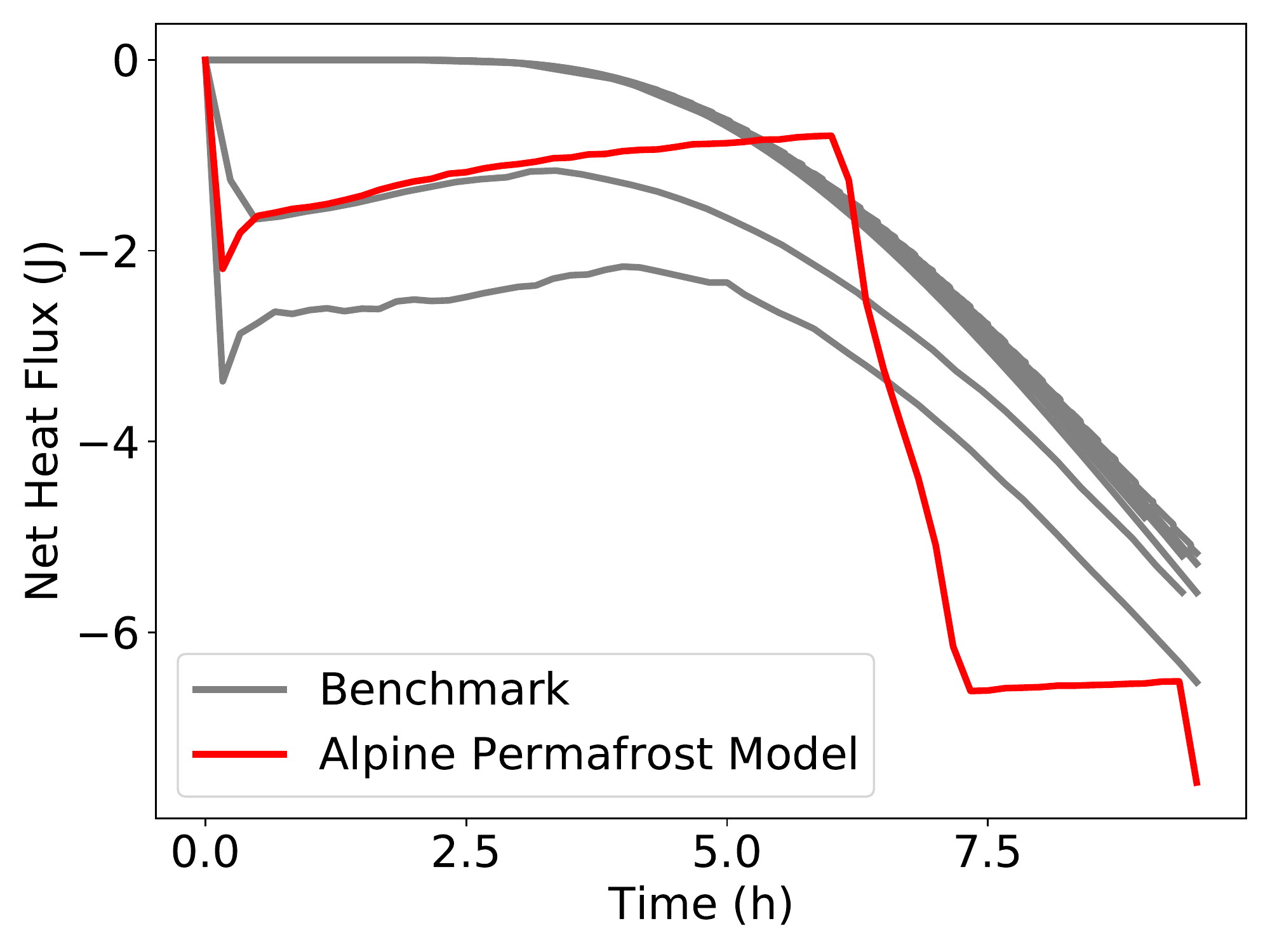}
    \end{subfigure}
    \hfill
    \begin{subfigure}[b]{0.32\textwidth}
        \includegraphics[width=\textwidth]{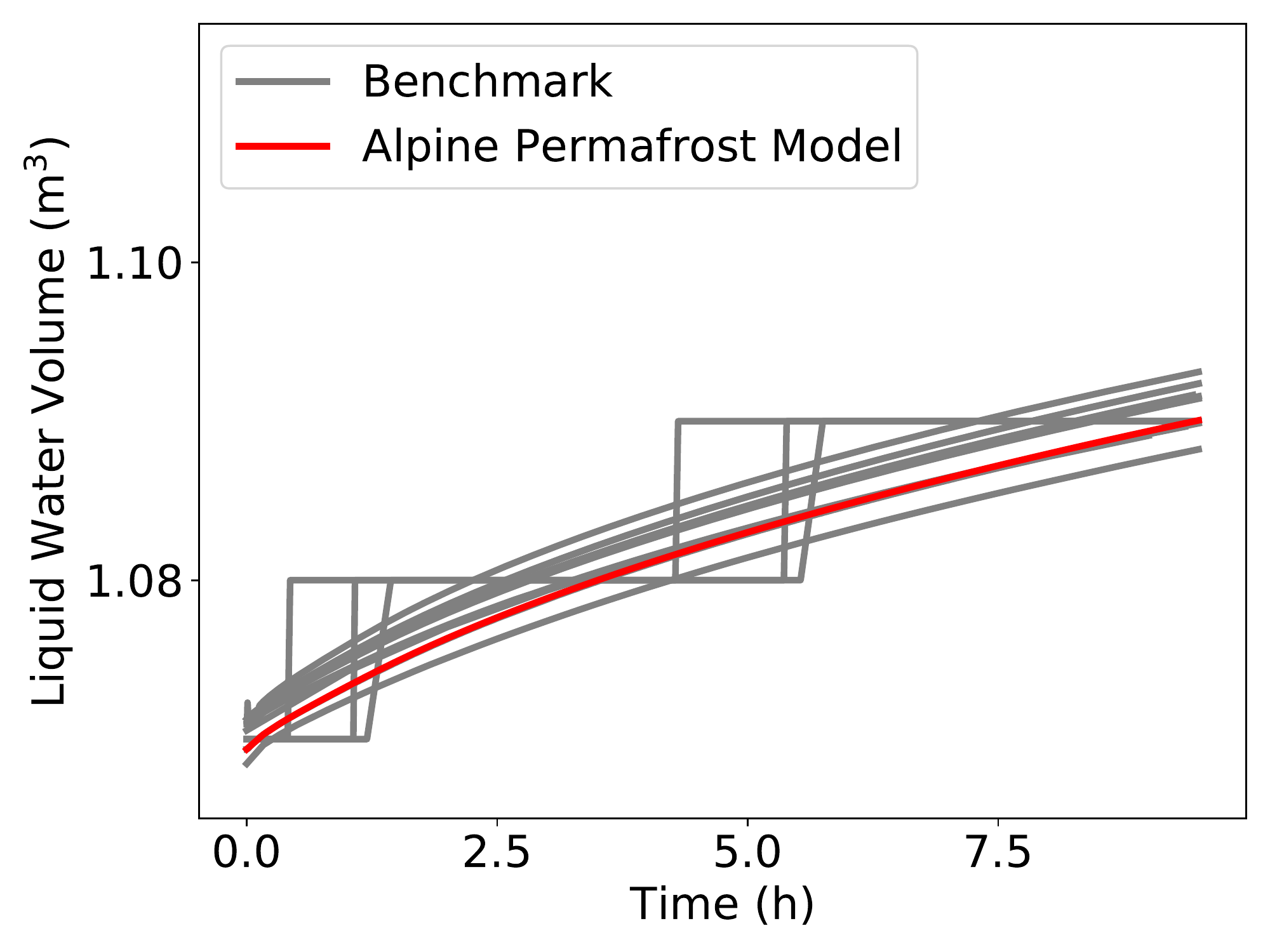}
    \end{subfigure}
    \caption{Hydraulic gradient: 0\% }
\end{figure}

\begin{figure}[H]
    \centering
    \begin{subfigure}[b]{0.32\textwidth}
        \includegraphics[width=\textwidth]{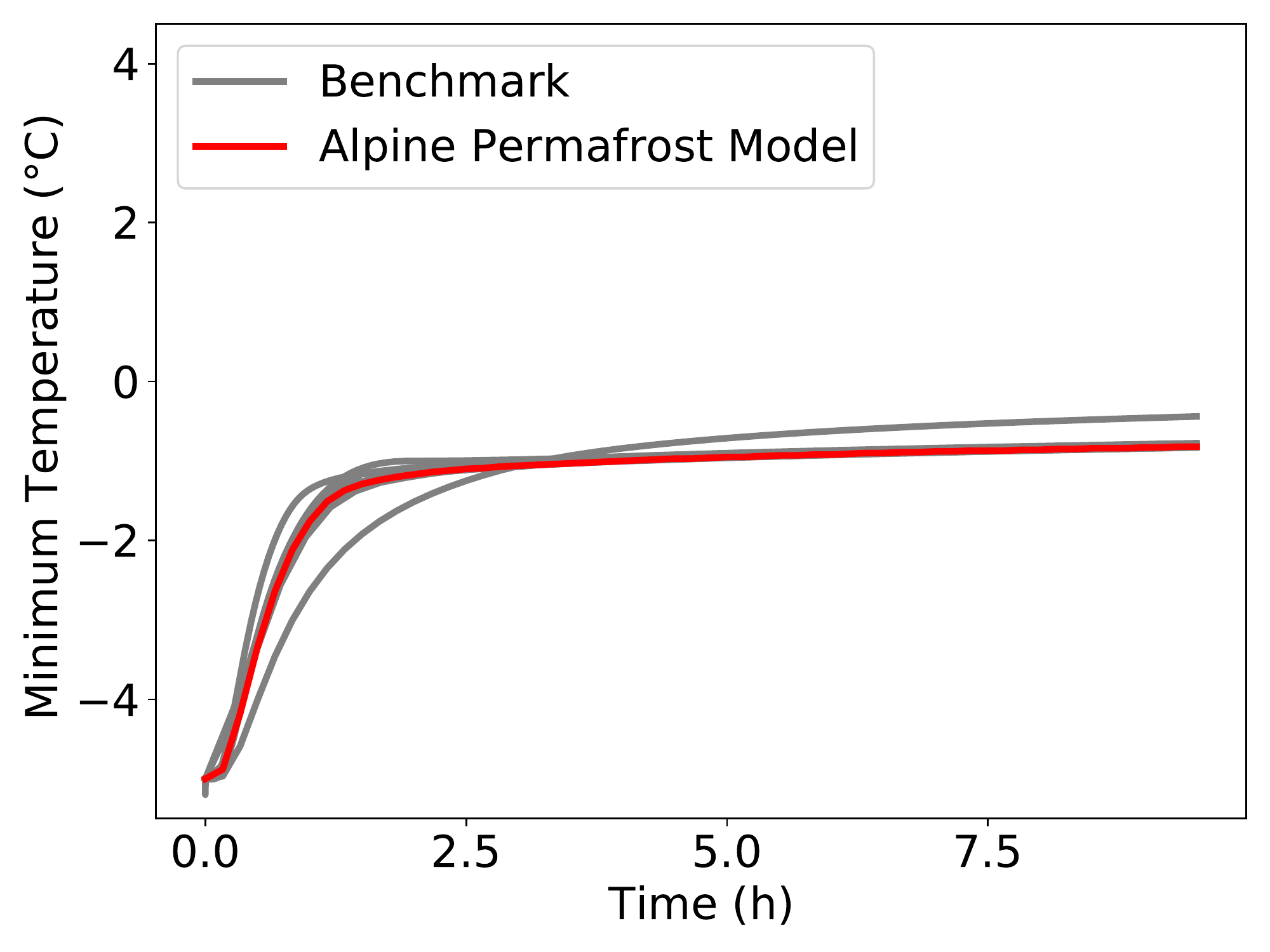}
    \end{subfigure}
    \hfill
    \begin{subfigure}[b]{0.32\textwidth}
        \includegraphics[width=\textwidth]{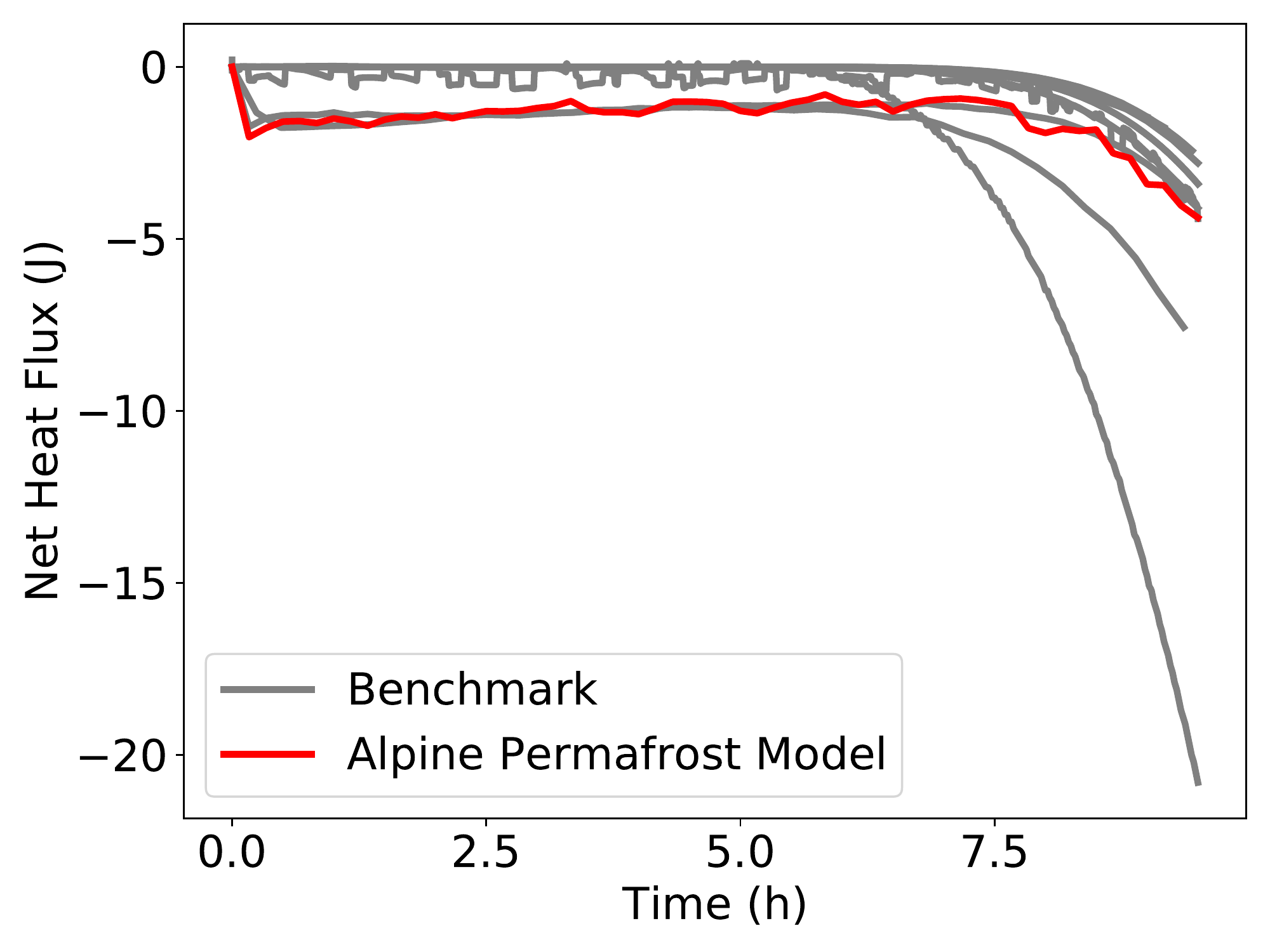}
    \end{subfigure}
    \hfill
    \begin{subfigure}[b]{0.32\textwidth}
        \includegraphics[width=\textwidth]{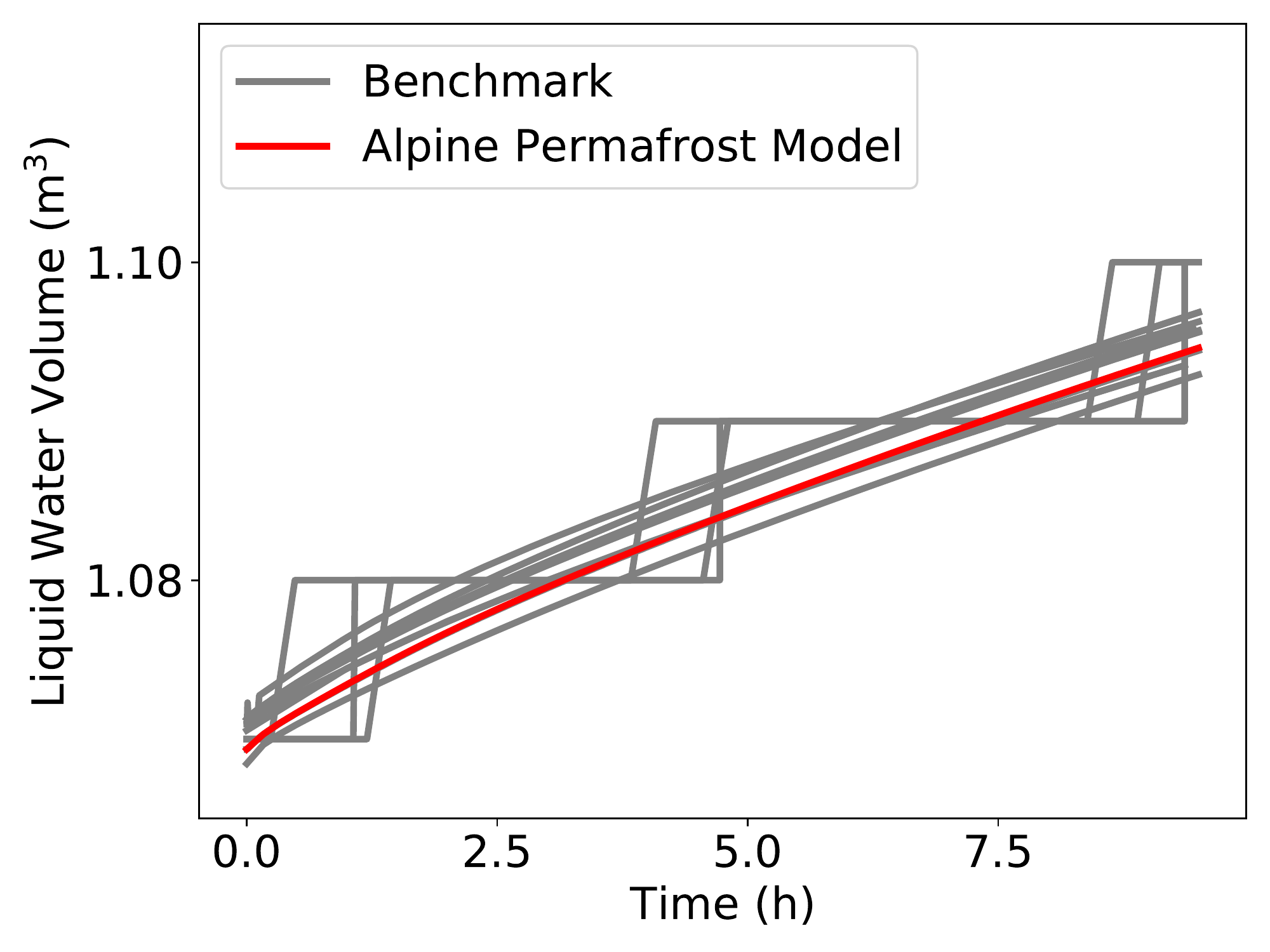}
    \end{subfigure}
    \caption{Hydraulic gradient: 3\% }
\end{figure}

\begin{figure}[H]
    \centering
    \begin{subfigure}[b]{0.32\textwidth}
        \includegraphics[width=\textwidth]{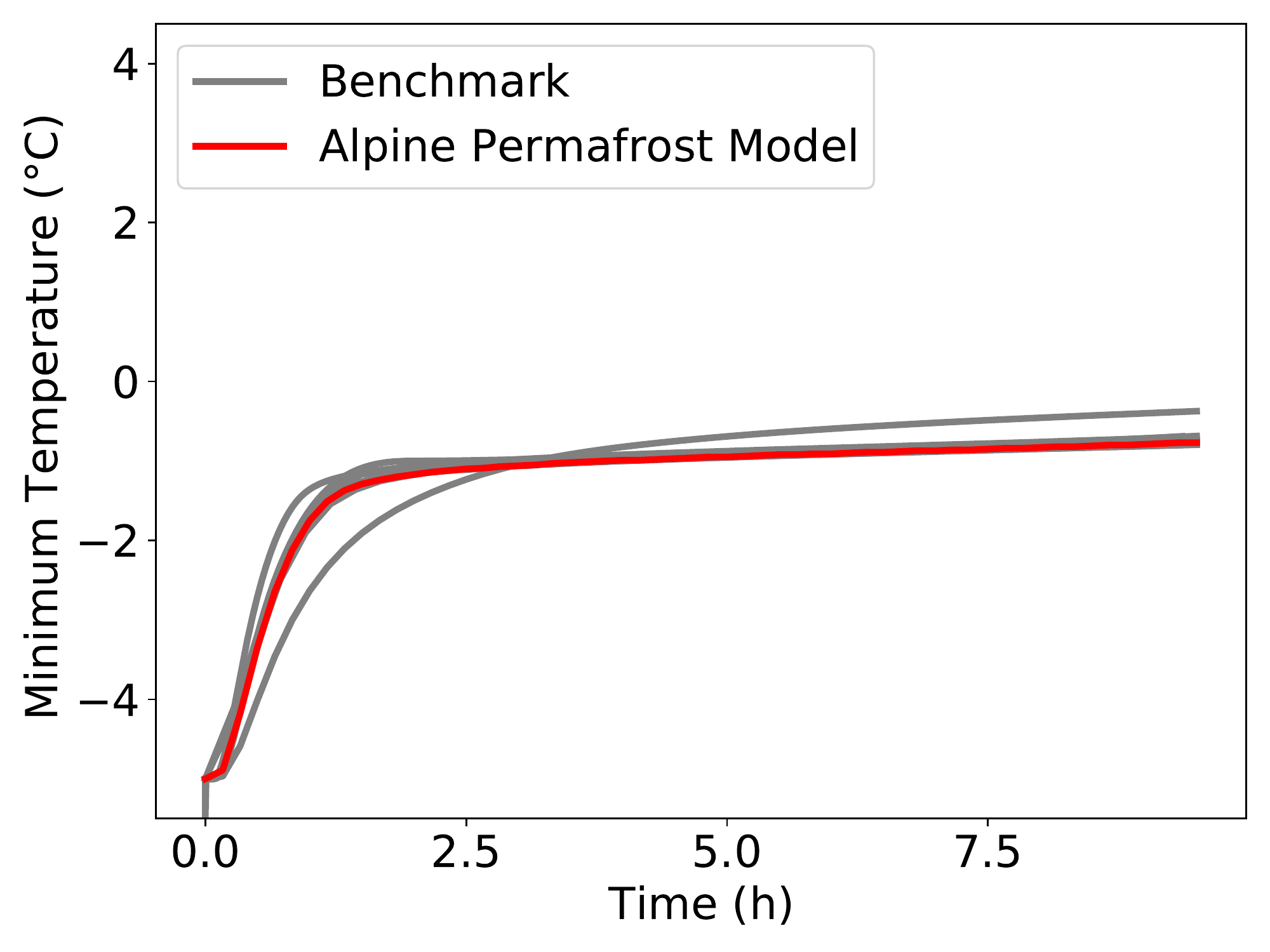}
    \end{subfigure}
    \hfill
    \begin{subfigure}[b]{0.32\textwidth}
        \includegraphics[width=\textwidth]{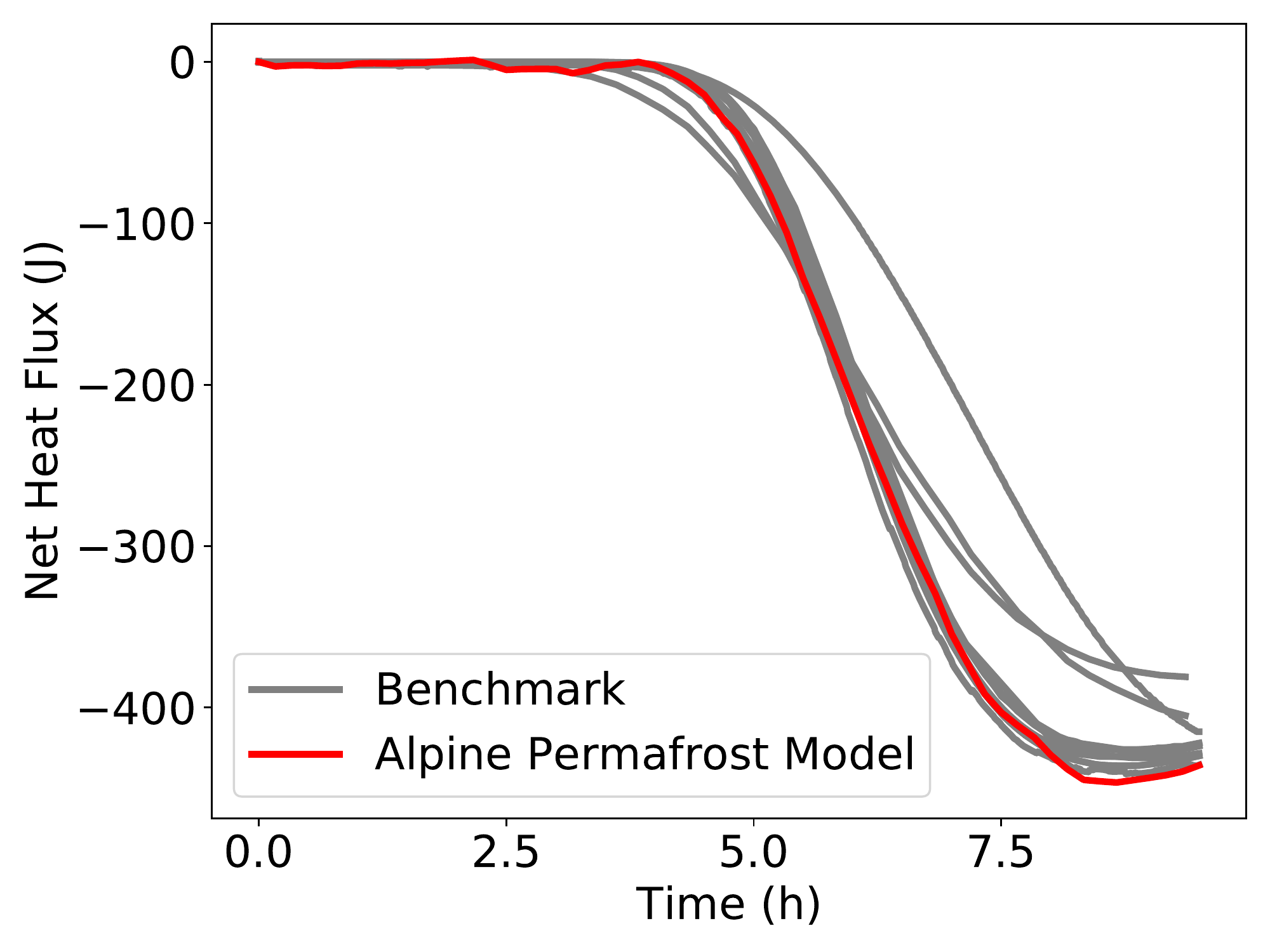}
    \end{subfigure}
    \hfill
    \begin{subfigure}[b]{0.32\textwidth}
        \includegraphics[width=\textwidth]{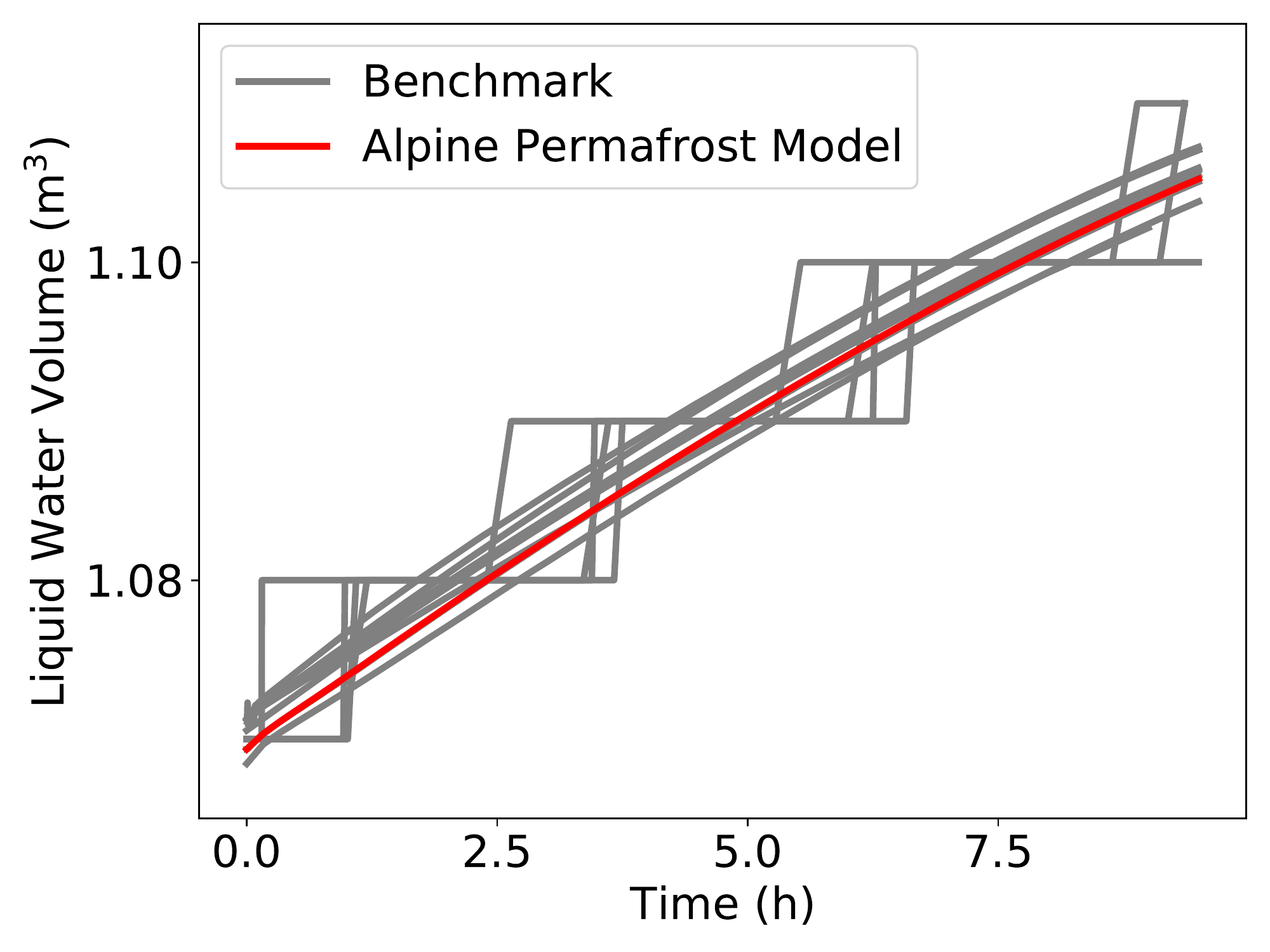}
    \end{subfigure}
    \caption{Hydraulic gradient: 9\% }
\end{figure}

\begin{figure}[H]
    \centering
    \begin{subfigure}[b]{0.32\textwidth}
        \includegraphics[width=\textwidth]{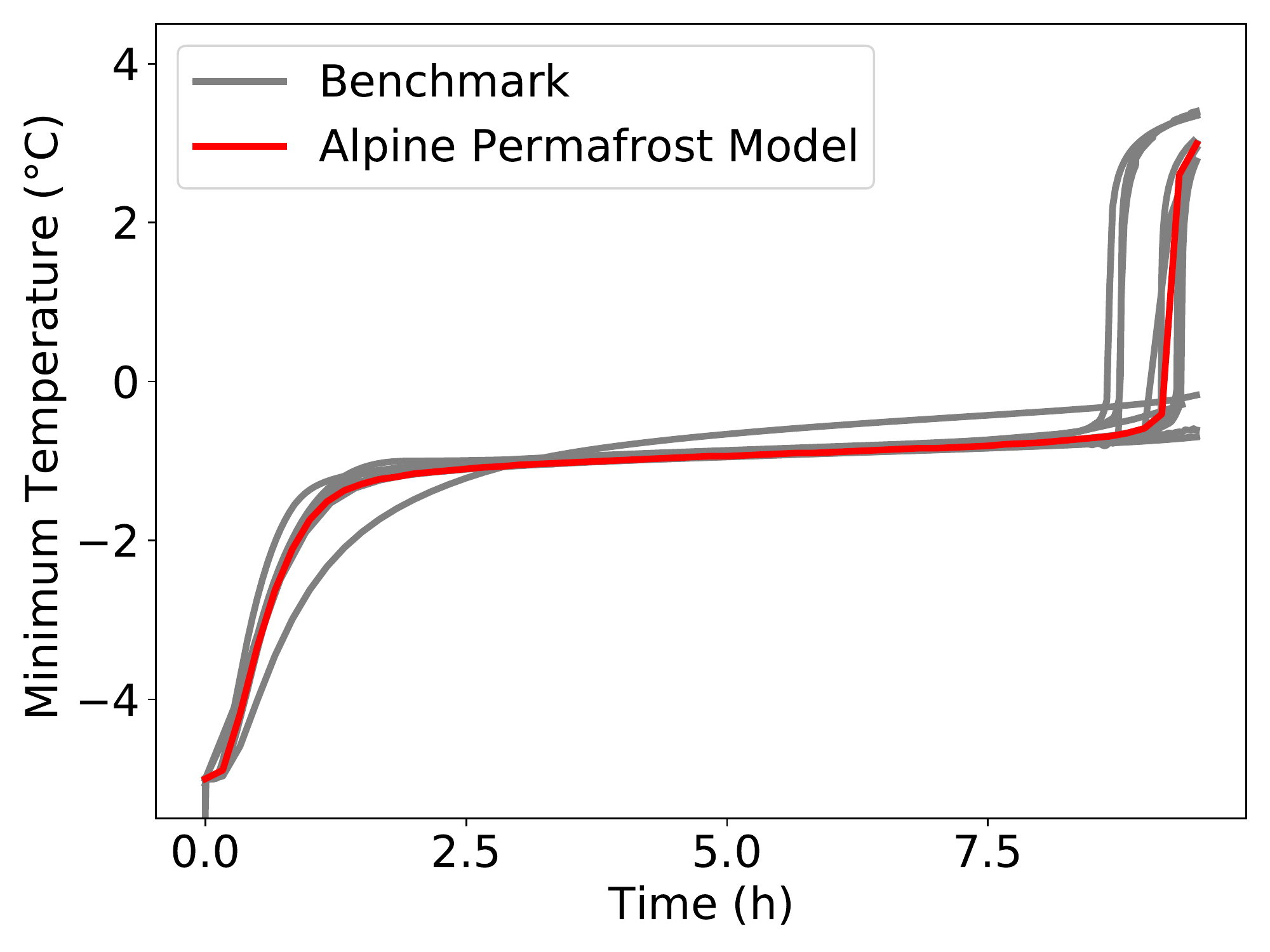}
    \end{subfigure}
    \hfill
    \begin{subfigure}[b]{0.32\textwidth}
        \includegraphics[width=\textwidth]{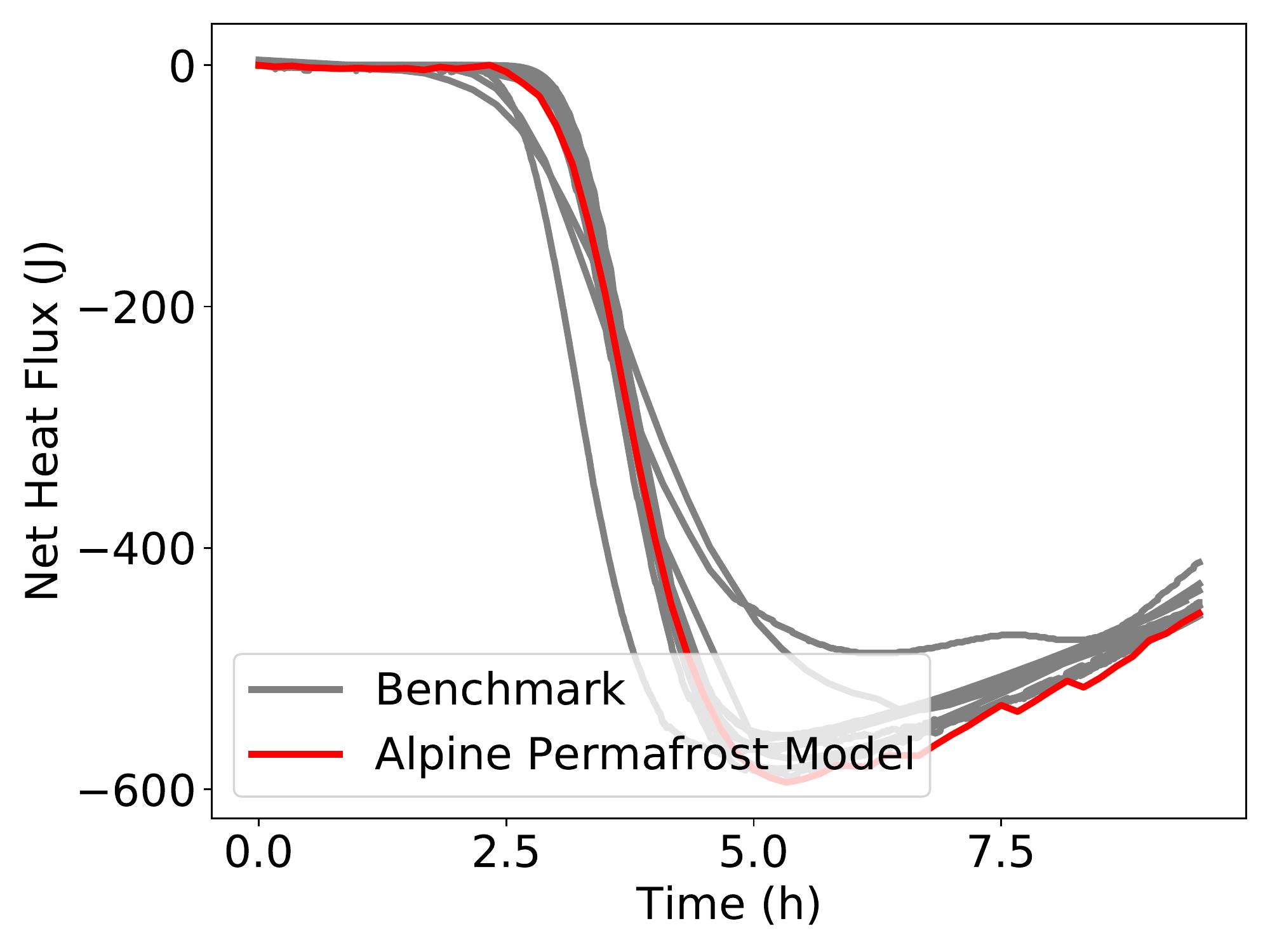}
    \end{subfigure}
    \hfill
    \begin{subfigure}[b]{0.32\textwidth}
        \includegraphics[width=\textwidth]{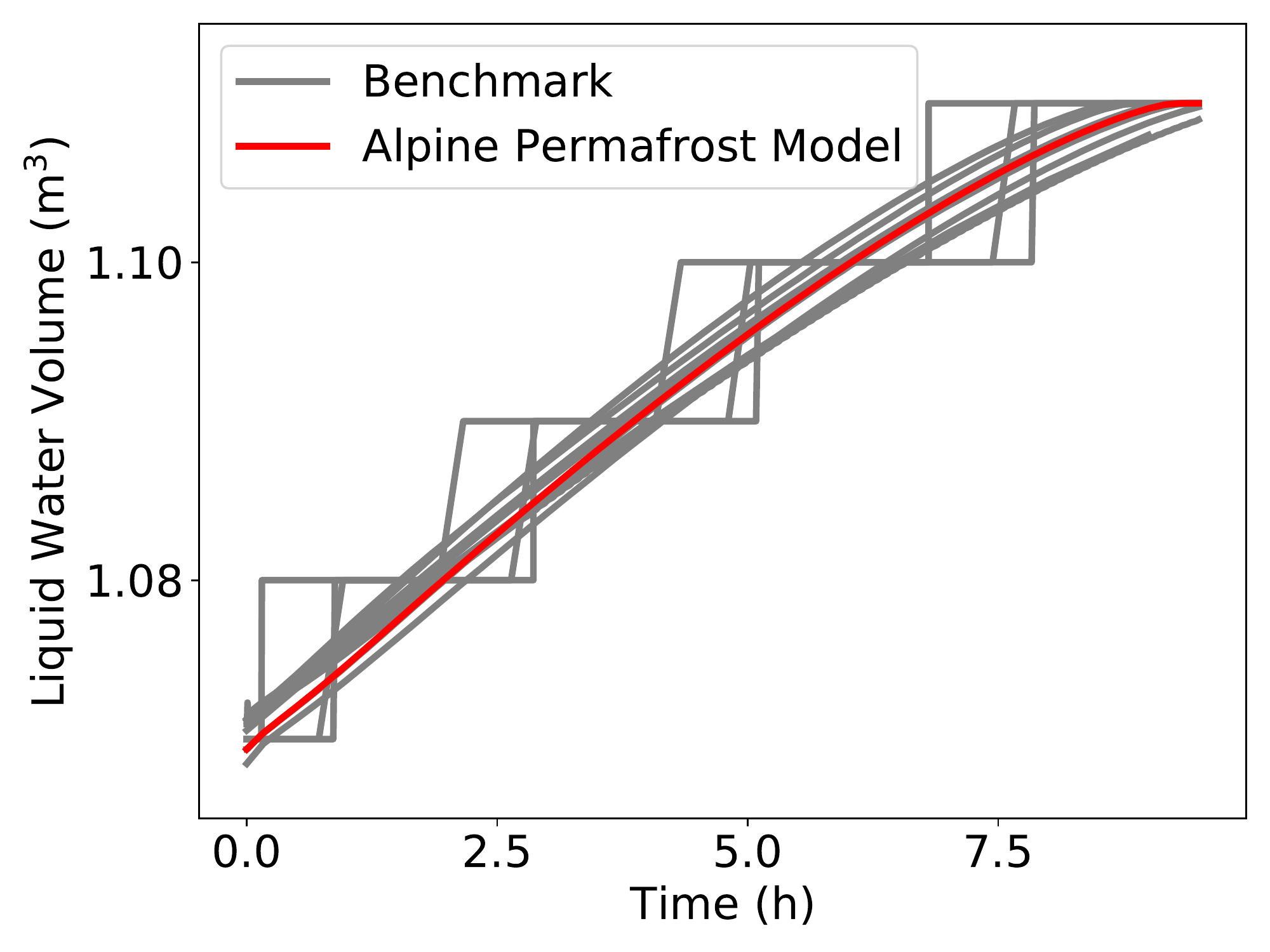}
    \end{subfigure}
    \caption{Hydraulic gradient: 15\% }
\end{figure}

\begin{figure}[H]
    \centering
    \begin{subfigure}[b]{0.49\textwidth}
        \centering
        \includegraphics[width=\textwidth]{figures/TH2/TH2_Temperature_field_00_8h.pdf}
        \caption{Hydraulic gradient: 0\% }
    \end{subfigure}
    \begin{subfigure}[b]{0.49\textwidth}
        \centering
        \includegraphics[width=\textwidth]{figures/TH2/TH2_Temperature_field_03_8h.pdf}
        \caption{Hydraulic gradient: 3\% }
    \end{subfigure}
    \begin{subfigure}[b]{0.49\textwidth}
        \centering
        \includegraphics[width=\textwidth]{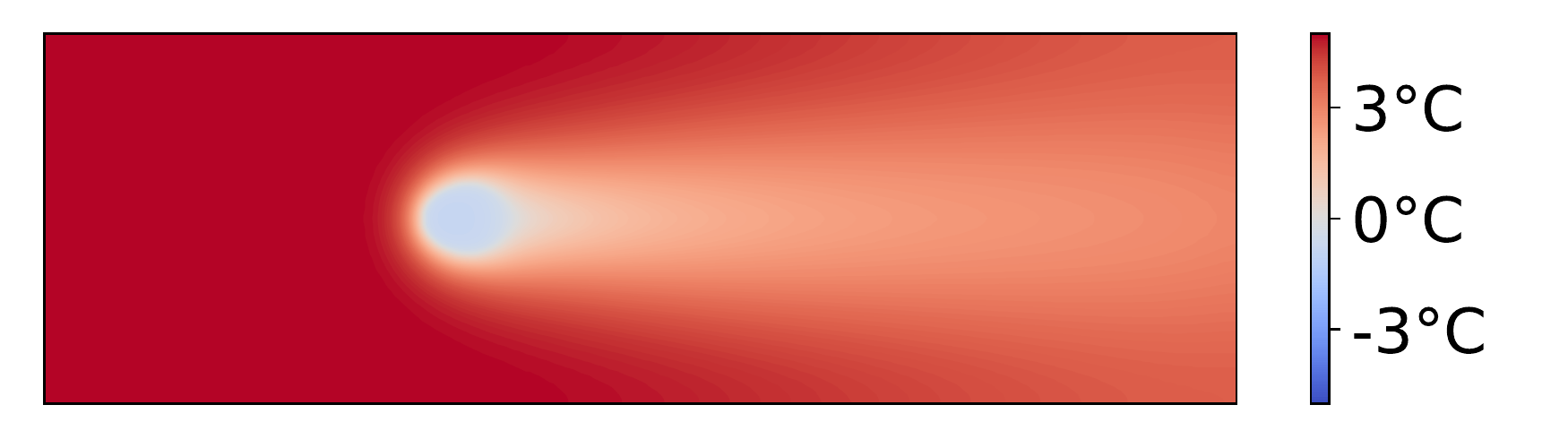}
        \caption{Hydraulic gradient: 9\% }
    \end{subfigure}
    \begin{subfigure}[b]{0.49\textwidth}
        \centering
        \includegraphics[width=\textwidth]{figures/TH2/TH2_Temperature_field_15_8h.pdf}
        \caption{Hydraulic gradient: 15\% }
    \end{subfigure}
    \caption{Temperature fields of the four simulation scenarios after 8\,hours. }
\end{figure}

\subsection{Talik Opening/Closure}

\begin{figure}[H]
    \centering
    \begin{subfigure}[b]{0.32\textwidth}
        \includegraphics[width=\textwidth]{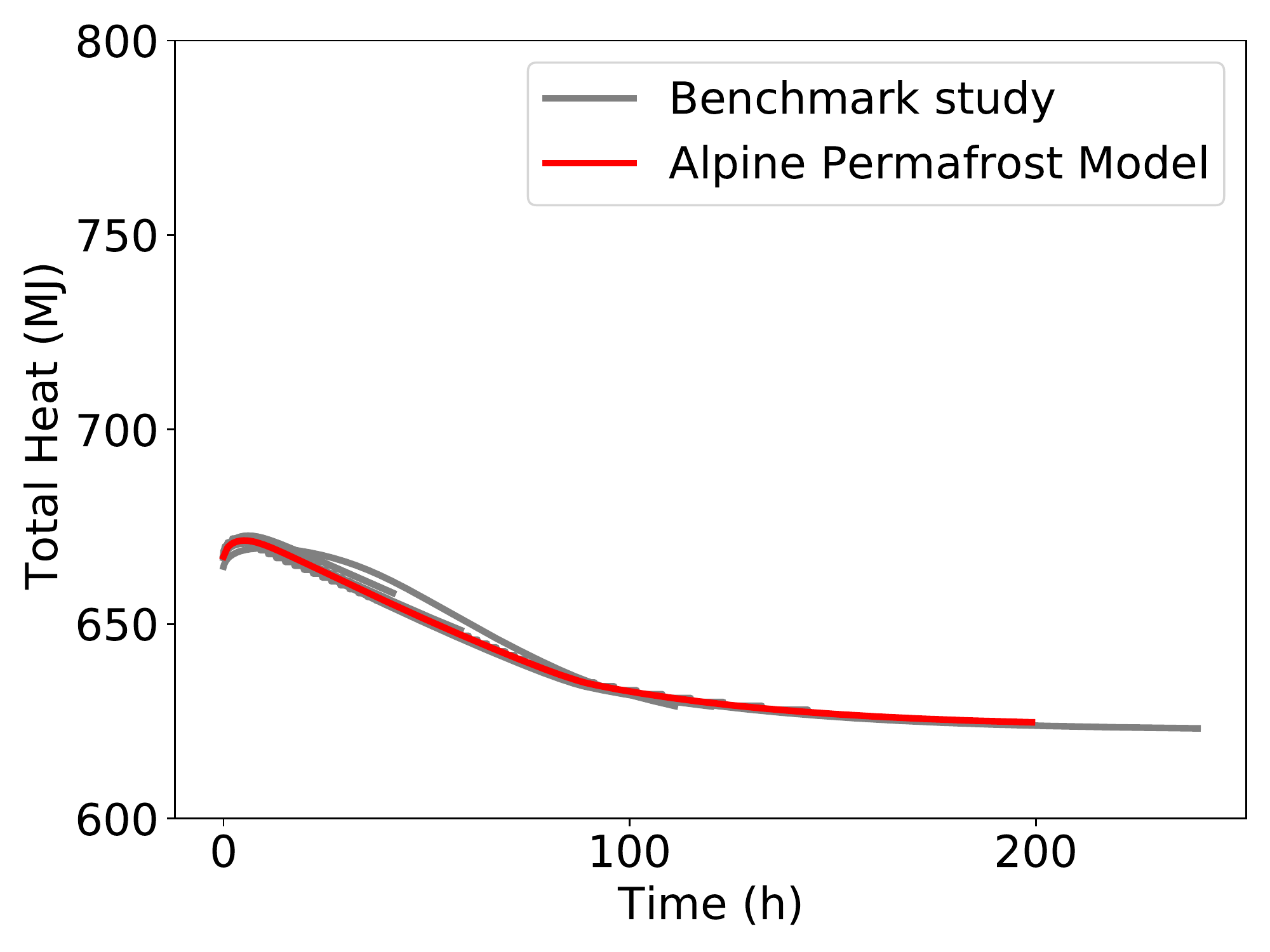}
    \end{subfigure}
    \hfill
    \begin{subfigure}[b]{0.32\textwidth}
        \includegraphics[width=\textwidth]{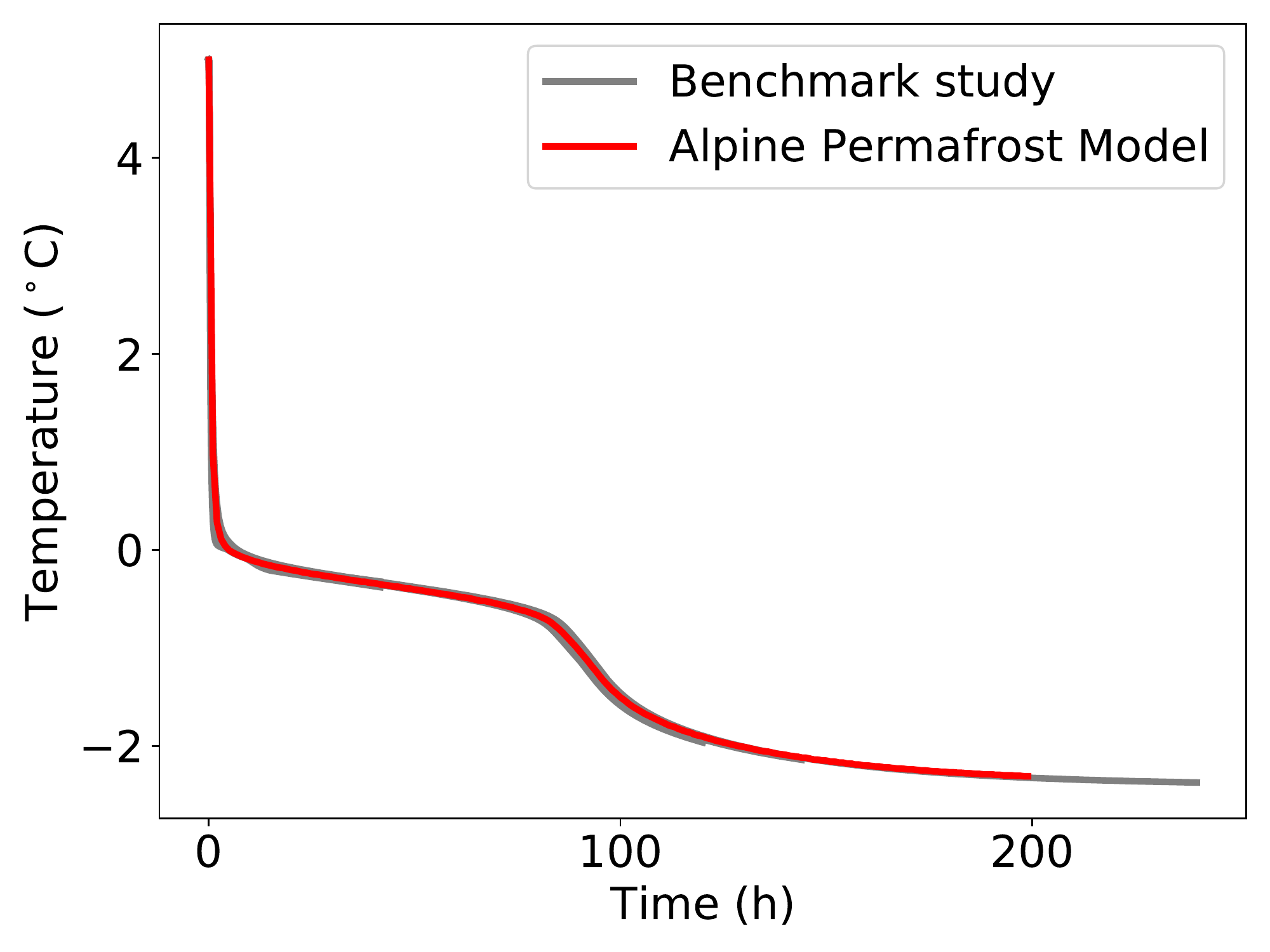}
    \end{subfigure}
    \hfill
    \begin{subfigure}[b]{0.32\textwidth}
        \includegraphics[width=\textwidth]{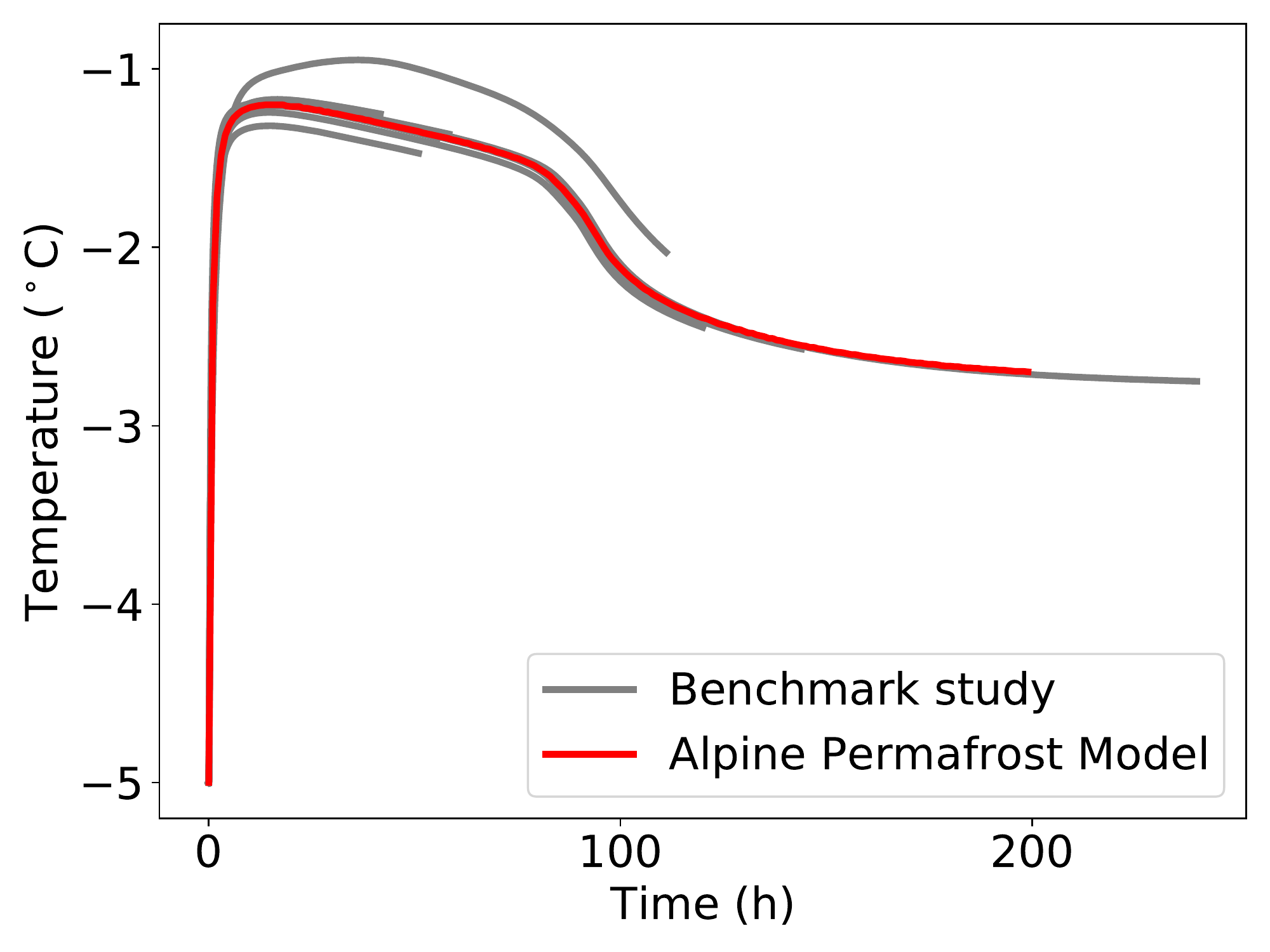}
    \end{subfigure}
    \caption{Hydraulic gradient: 3\% }
\end{figure}

\begin{figure}[H]
    \centering
    \begin{subfigure}[b]{0.32\textwidth}
        \includegraphics[width=\textwidth]{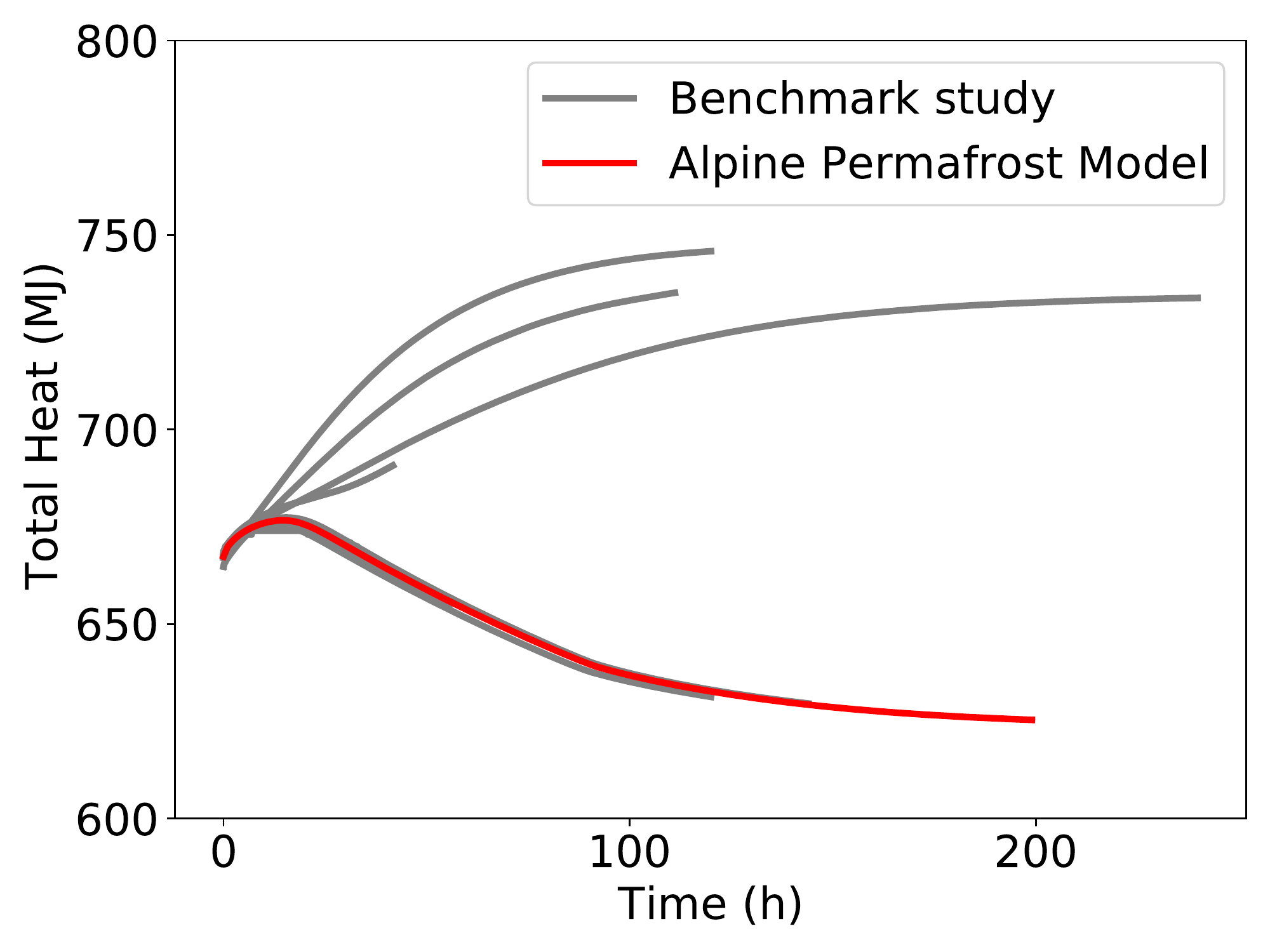}
    \end{subfigure}
    \hfill
    \begin{subfigure}[b]{0.32\textwidth}
        \includegraphics[width=\textwidth]{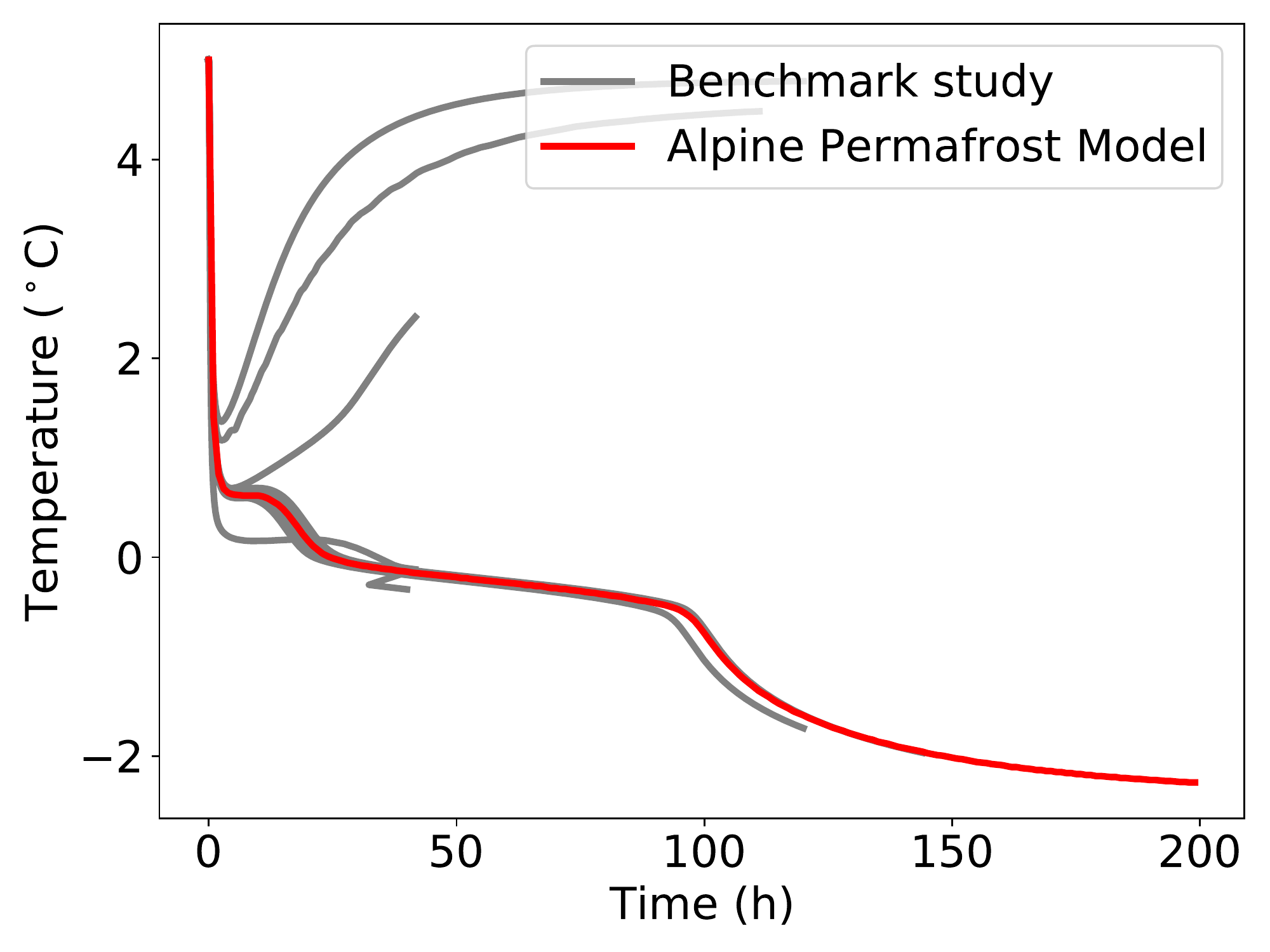}
    \end{subfigure}
    \hfill
    \begin{subfigure}[b]{0.32\textwidth}
        \includegraphics[width=\textwidth]{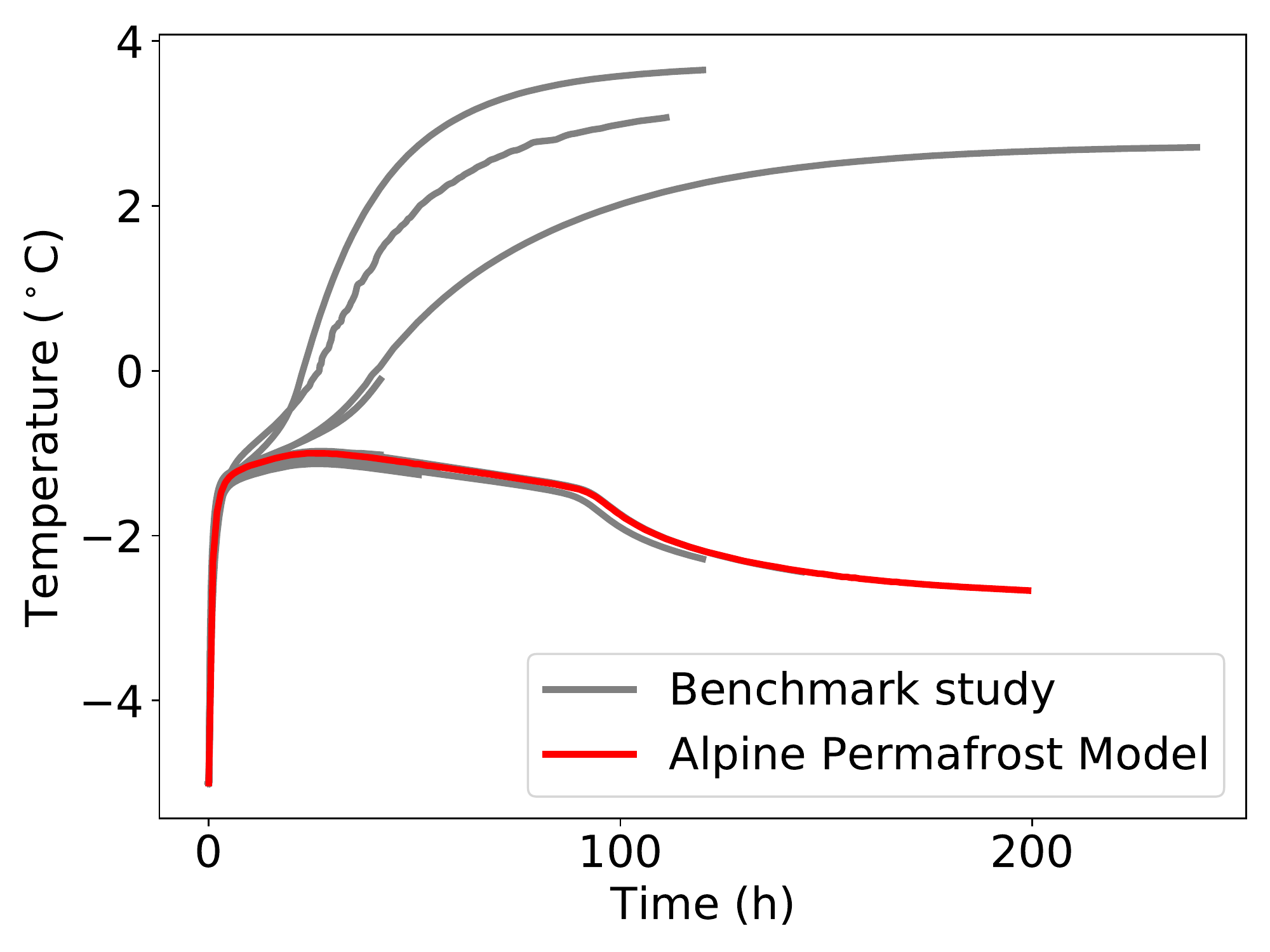}
    \end{subfigure}
    \caption{Hydraulic gradient: 6\% }
\end{figure}

\begin{figure}[H]
    \centering
    \begin{subfigure}[b]{0.32\textwidth}
        \includegraphics[width=\textwidth]{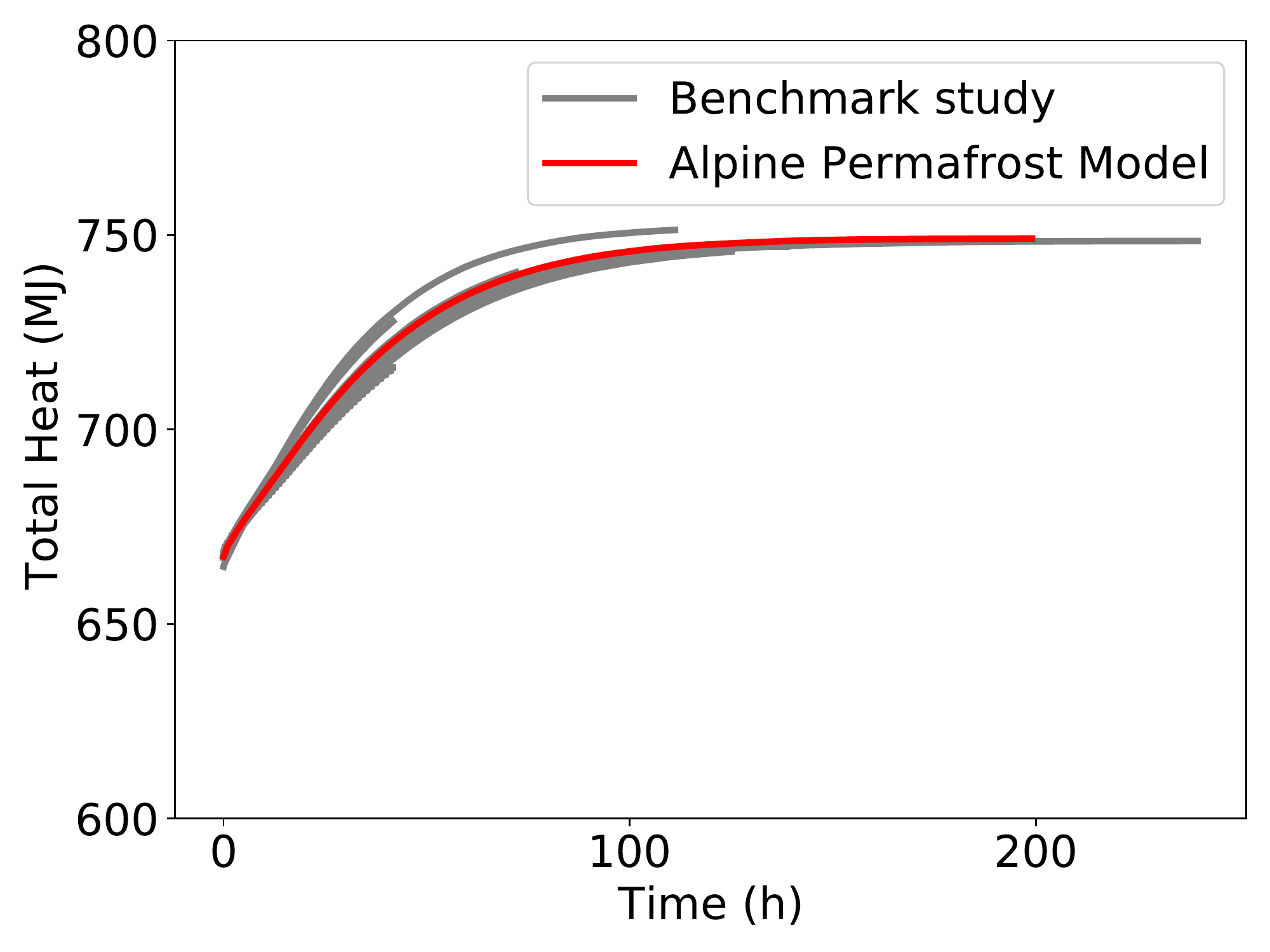}
    \end{subfigure}
    \hfill
    \begin{subfigure}[b]{0.32\textwidth}
        \includegraphics[width=\textwidth]{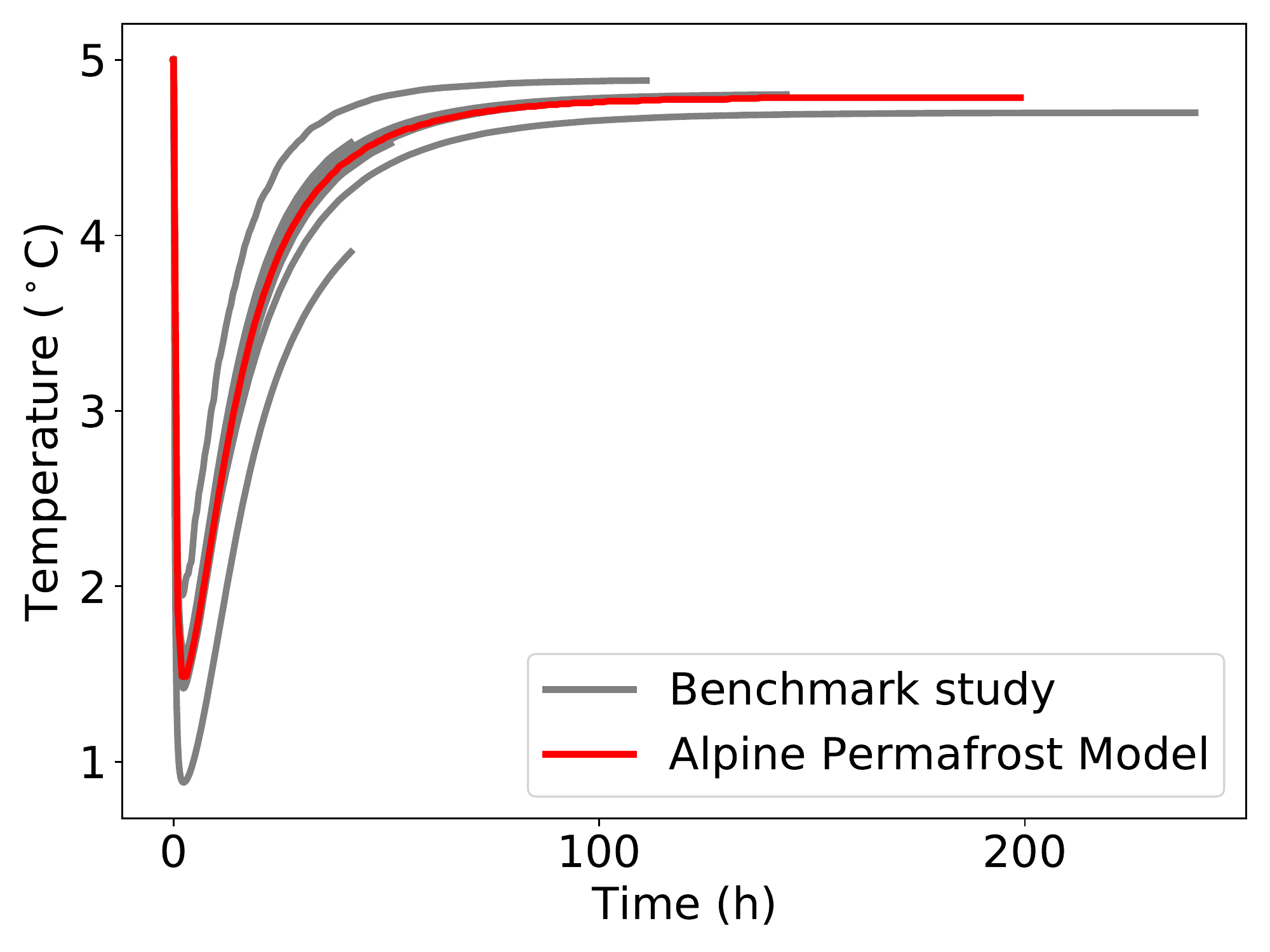}
    \end{subfigure}
    \hfill
    \begin{subfigure}[b]{0.32\textwidth}
        \includegraphics[width=\textwidth]{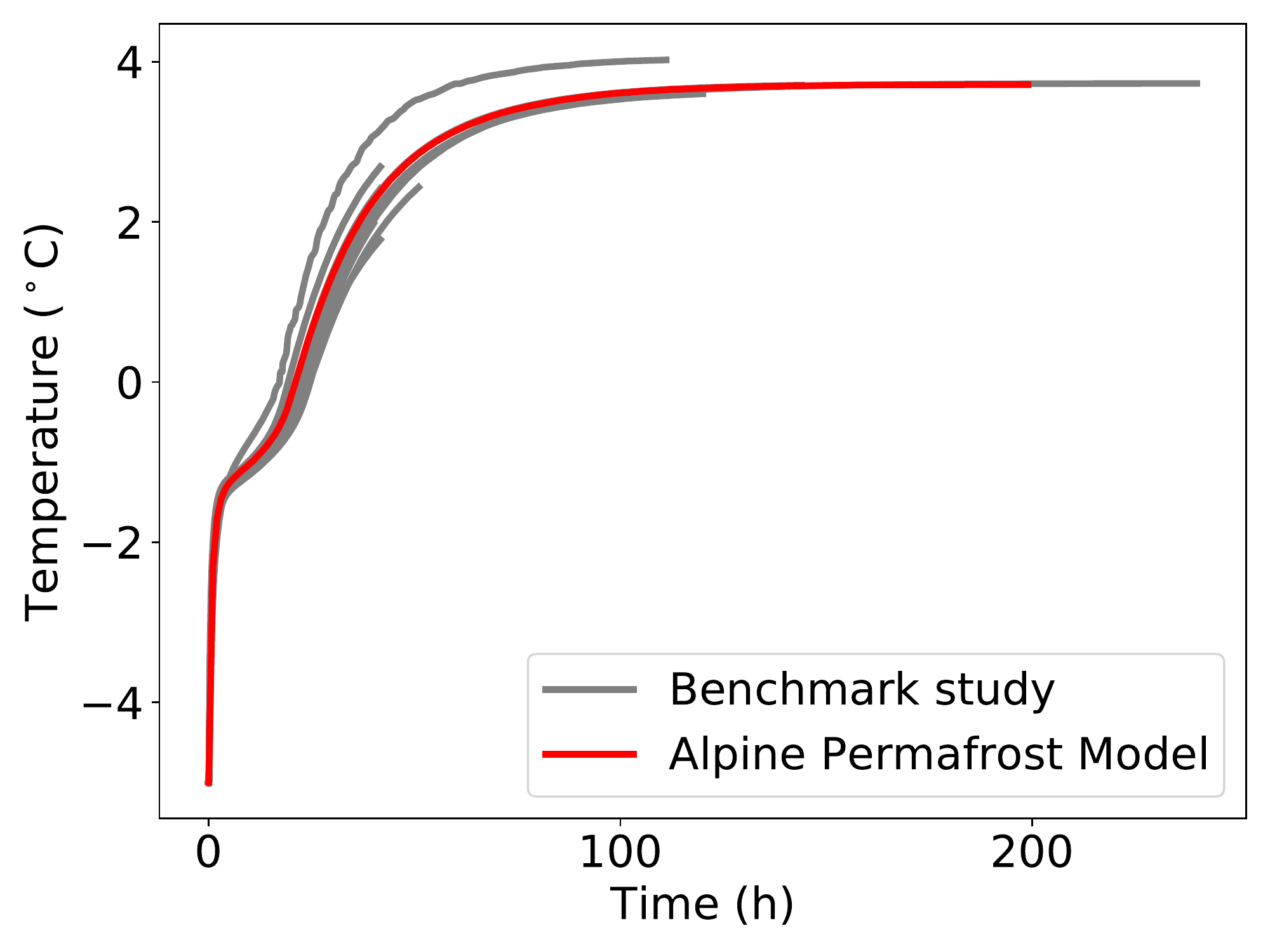}
    \end{subfigure}
    \caption{Hydraulic gradient: 9\% }
\end{figure}

\begin{figure}[H]
    \centering
    \begin{subfigure}[b]{0.32\textwidth}
        \includegraphics[width=\textwidth]{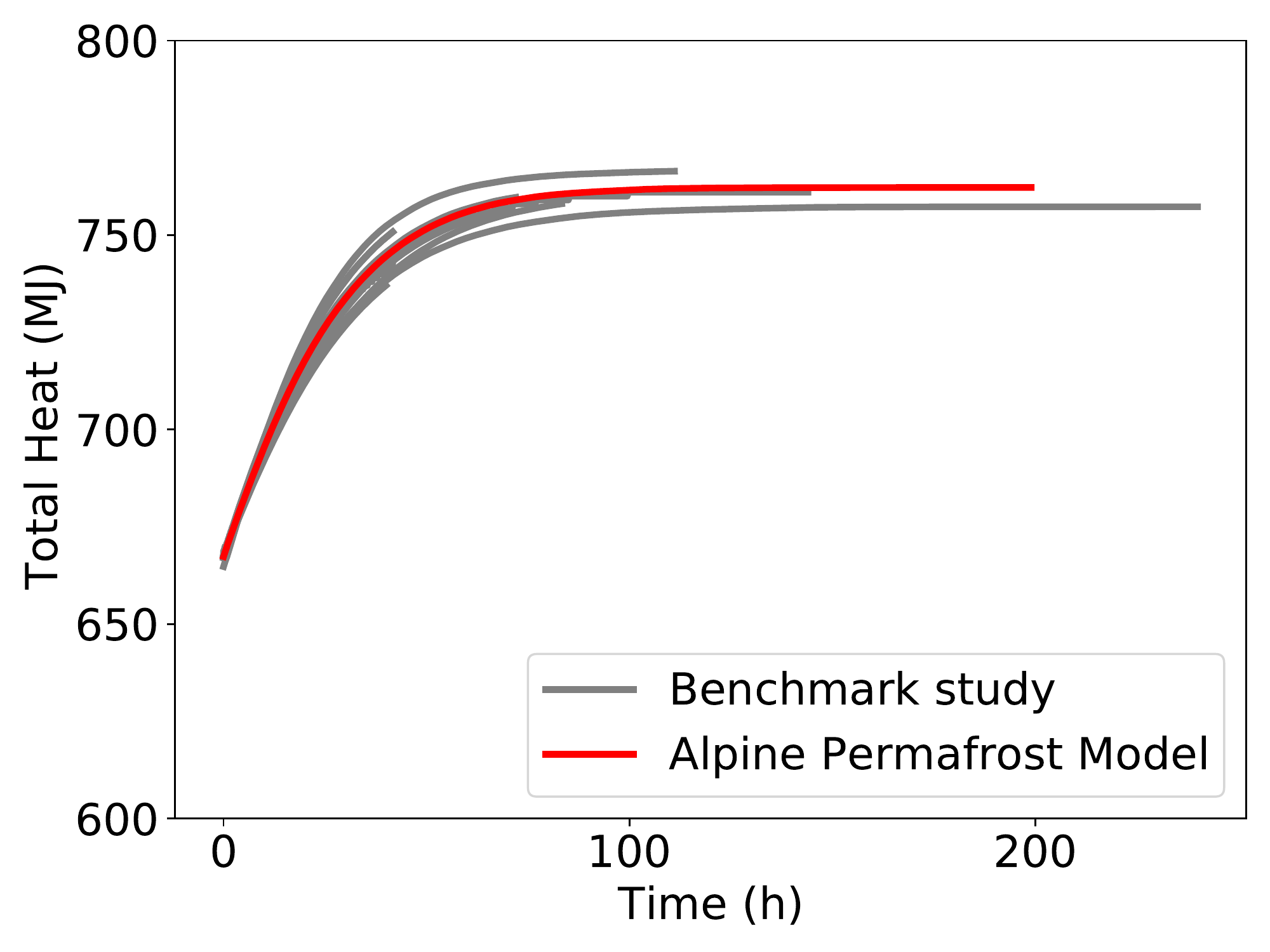}
    \end{subfigure}
    \hfill
    \begin{subfigure}[b]{0.32\textwidth}
        \includegraphics[width=\textwidth]{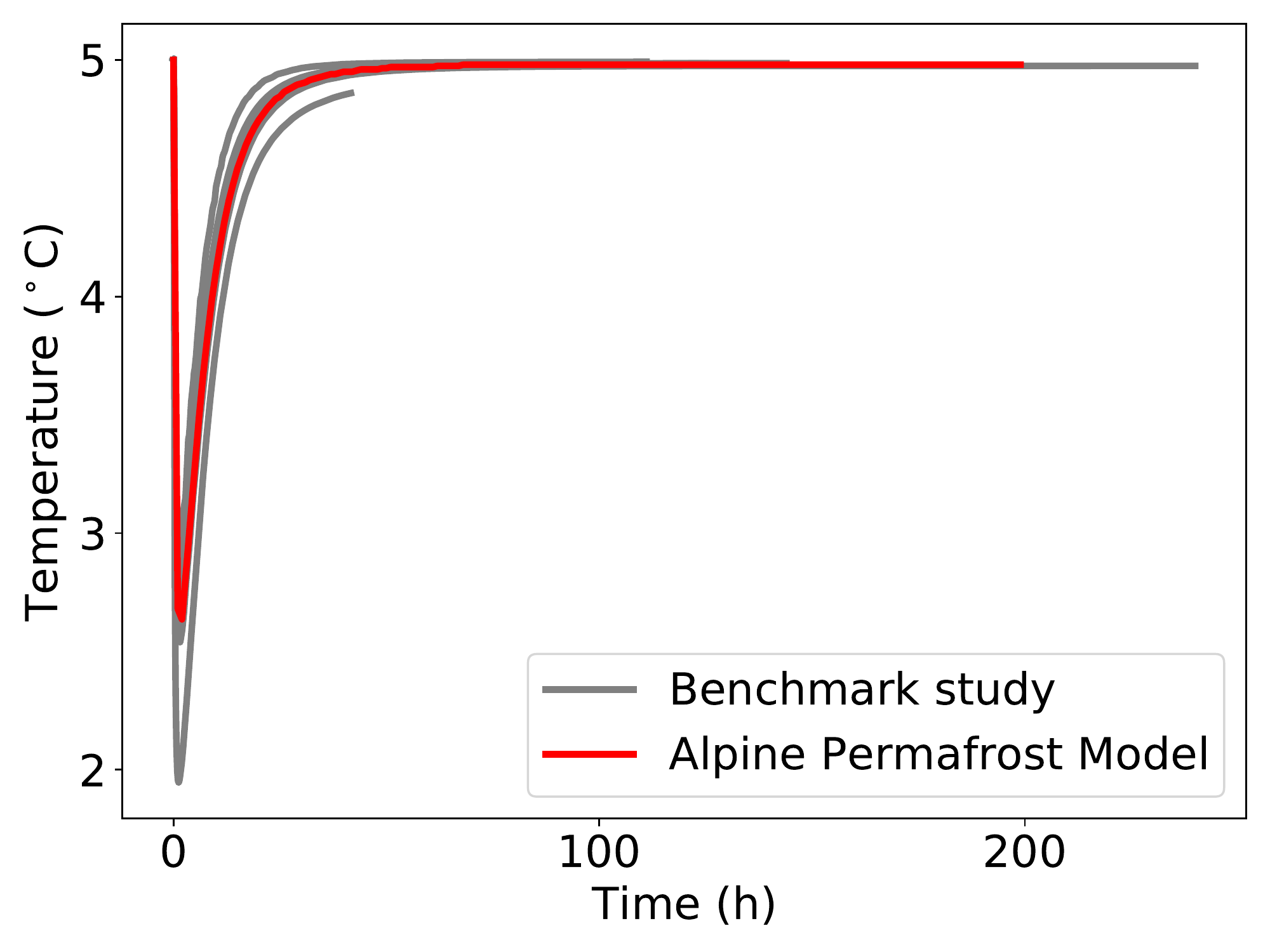}
    \end{subfigure}
    \hfill
    \begin{subfigure}[b]{0.32\textwidth}
        \includegraphics[width=\textwidth]{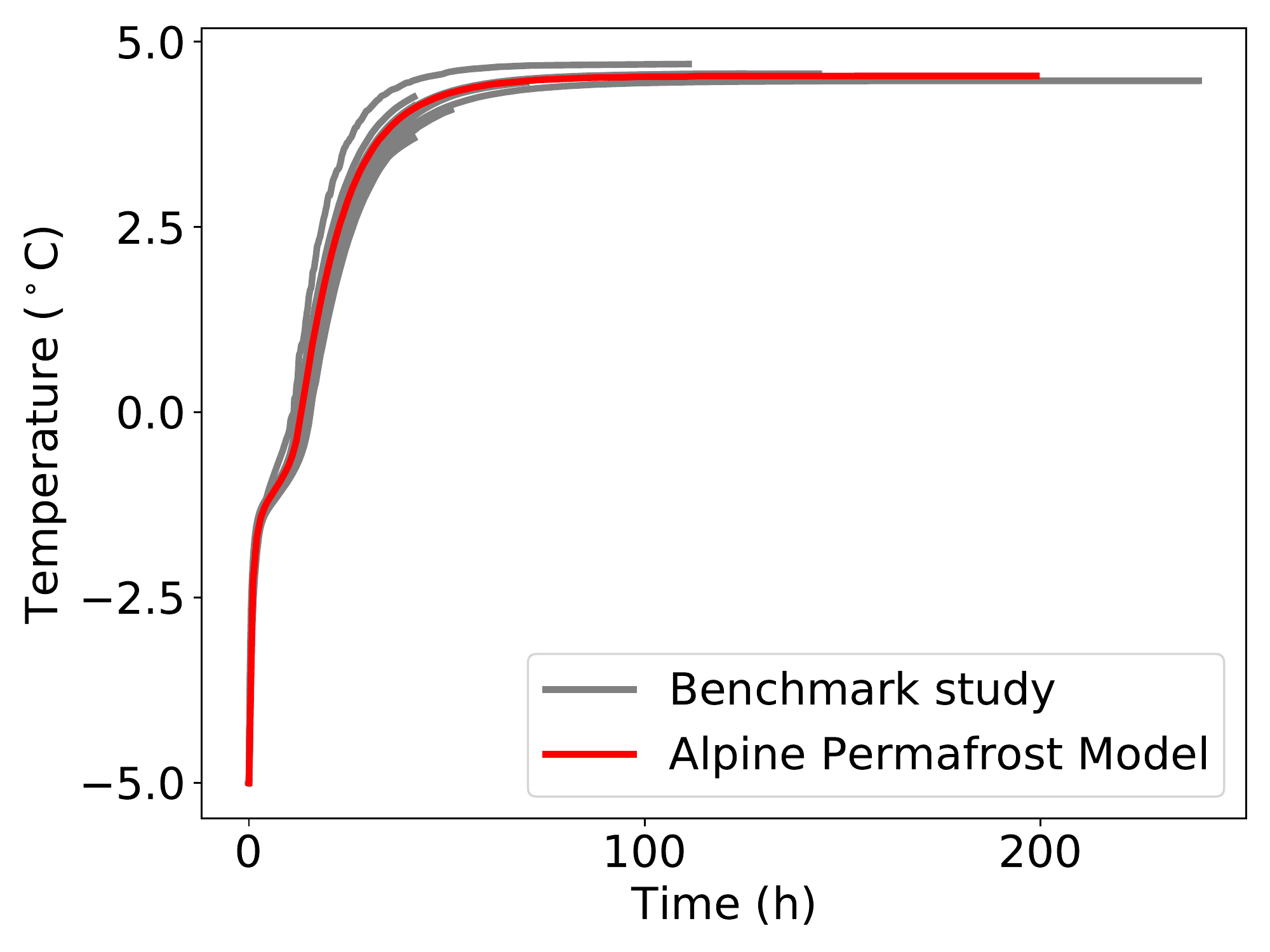}
    \end{subfigure}
    \caption{Hydraulic gradient: 15\% }
\end{figure}

\begin{figure}[H]
    \centering
    \begin{subfigure}[b]{0.24\textwidth}
        \centering
        \includegraphics[width=\textwidth]{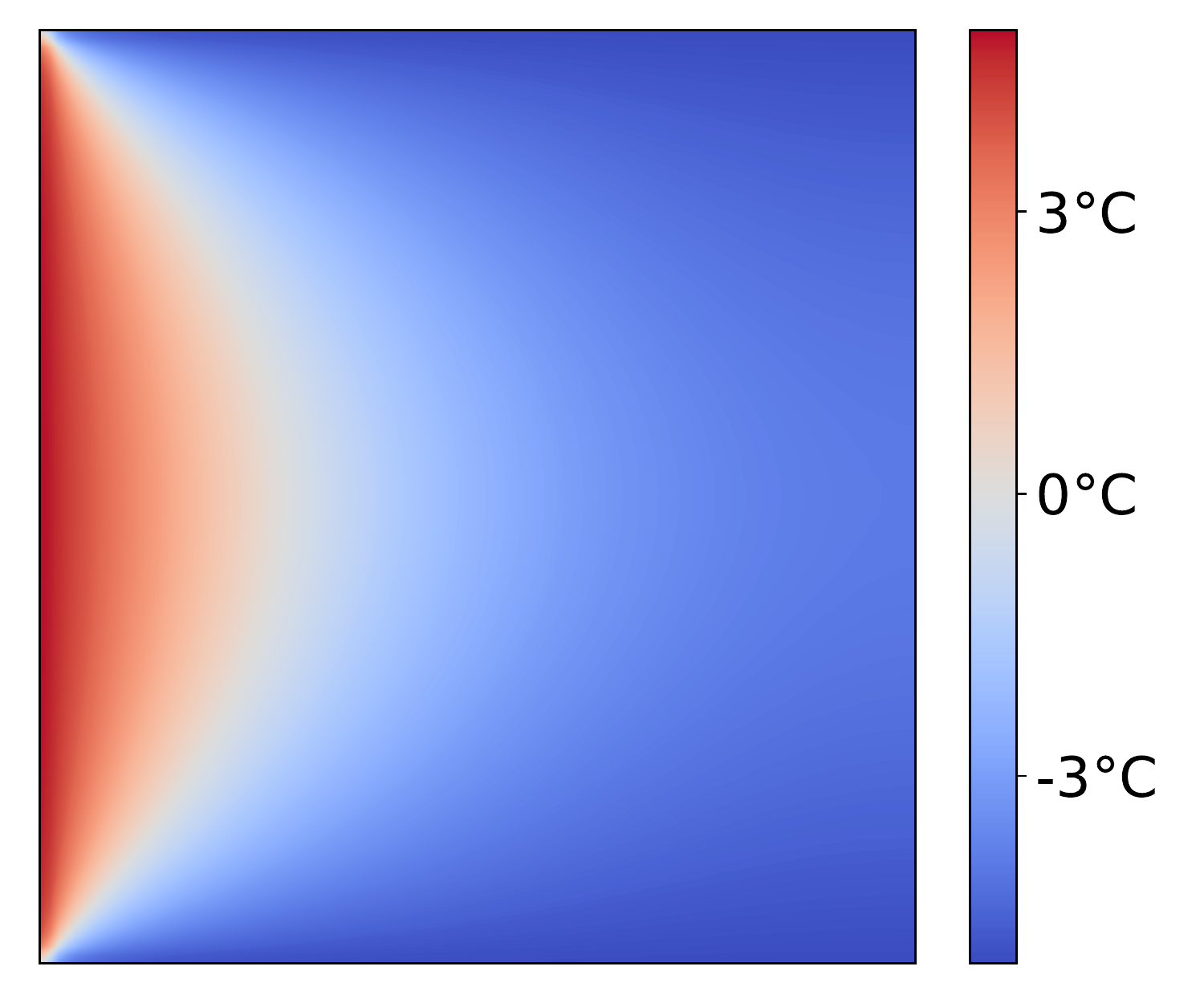}
        \caption{Hydraulic gradient: 3\% }
    \end{subfigure}
    \hfill
    \begin{subfigure}[b]{0.24\textwidth}
        \centering
        \includegraphics[width=\textwidth]{figures/TH3/TH3_Temperature_field_06_200h.pdf}
       \caption{Hydraulic gradient: 6\% }
    \end{subfigure}
    \hfill
    \begin{subfigure}[b]{0.24\textwidth}
        \centering
        \includegraphics[width=\textwidth]{figures/TH3/TH3_Temperature_field_09_200h.pdf}
       \caption{Hydraulic gradient: 9\% }
    \end{subfigure}
    \hfill
    \begin{subfigure}[b]{0.24\textwidth}
        \centering
        \includegraphics[width=\textwidth]{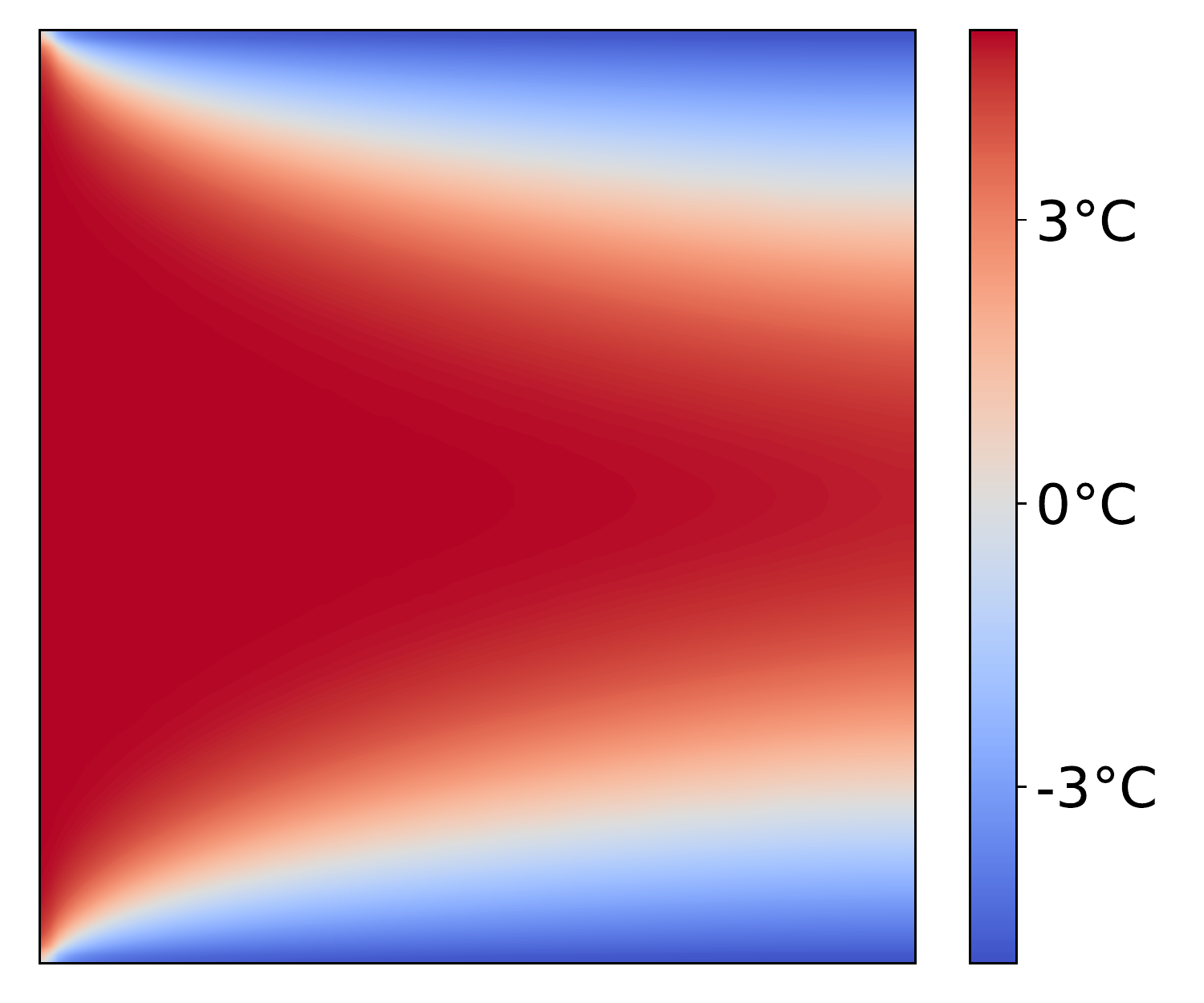}
        \caption{Hydraulic gradient: 15\% }
    \end{subfigure}
    \caption{Temperature fields of the four simulation scenarios after 200\,hours. }
\end{figure}
\section{Parameter Tables}
\label{apx:parameter table}

\subsection{Shared Parameters}

\begin{center}
    \begin{tabular}{|c|p{5cm}|c|c|}
        \hline 
        \textbf{Symbol} & \textbf{Description} & \textbf{Value} & \textbf{Unit} \\ \hline
        
        $\rho_a$ & density of air & 1.284 & $kg.m^{-3}$\\
        $\rho_w$ & density of water & 1000 & $kg.m^{-3}$\\ 
        $\rho_i$ & density of ice & 920 & $kg.m^{-3}$\\ 
        $\rho_s$ & density of soil & 2650 & $kg.m^{-3}$\\ 
        \hline 
        
        $K^T_a$ & thermal conductivity of air & 0.02428 & $W.m^{-1}.K^{-1}$\\ 
        $K^T_w$ & thermal conductivity of water & 0.6 & $W.m^{-1}.K^{-1}$\\  
        $K^T_i$ & thermal conductivity of ice & 2.14 & $W.m^{-1}.K^{-1}$\\ 
        $K^T_s$ & thermal conductivity of soil & 9.0 & $W.m^{-1}.K^{-1}$\\ 
        \hline 
        
        $C^P_a$ & isobaric heat capacity of air & $1.0038 * 10^3$ & $J.kg^{-1}.K^{-1}$\\ 
        $C^P_w$ & isobaric heat capacity of water& 4182 & $J.kg^{-1}.K^{-1}$\\ 
        $C^P_i$ & isobaric heat capacity of ice& 2060 & $J.kg^{-1}.K^{-1}$\\ 
        $C^P_s$ & isobaric heat capacity of soil& 835 & $J.kg^{-1}.K^{-1}$\\  
        \hline 
        
        $C^V_a$ & isochoric heat capacity of air & $0.7167 * 10^3$ & $J.kg^{-1}.K^{-1}$\\ 
        $C^V_w$ & isochoric heat capacity of water & 4182 & $J.kg^{-1}.K^{-1}$\\ 
        $C^V_i$ & isochoric heat capacity of ice & 2060 & $J.kg^{-1}.K^{-1}$\\ 
        $C^V_s$ & isochoric heat capacity of soil & 835 & $J.kg^{-1}.K^{-1}$\\  
        \hline 
        
        $\mu_a$ & viscosity of air & $1.725 * 10^{-5}$ & $kg.m^{-1}.s^{-1}$\\ 
        $\mu_w$ & viscosity of water & $0.001793 * a(T)$ & $kg.m^{-1}.s^{-1}$ \\
        $a(T)$ & & $\left( 1 + b T + c T^2 \right)^{-1}$ & (--) \\
        $b$ & & 0.337 & (--) \\
        $c$ & & 0.000221 & (--) \\ 
        \hline
        
        $SFC(T)$ & soil freezing curve & $0.05 + 0.95 * \exp ( (T_f-T) / W)$ & (--) \\ 
        $W$ & & $50$ & $K$\\ 
        \hline 
        
        $L_f$ & latent heat of freezing & 334000 &$J.kg^{-1}$ \\ 
        $T_f$ & freezing point of water & 273.15 & $K$\\  
        $P_b$ & entry pressure & 5000 & $Pa$ \\ 
        $g$ & gravitation & $9.81$ & $m.s^{-2}$ \\ 
        $S_{ar}$ & residual saturation of air & 0 & (--) \\
        $S_{wr}$ & residual saturation of water & 0 & (--) \\
        $\lambda$ & distribution of the pore size & 1.2 & (--)\\ 
        $\beta$ & parameter in Civan's power law & 1 & (--)\\ 
        \hline 
        
    \end{tabular}
\end{center}

\subsection{Initial Conditions}
\begin{center}
    \begin{tabular}{|c|p{5cm}|c|c|}
        \hline 
        \textbf{Primary Variable} & \textbf{Description} & \textbf{Value} & \textbf{Unit} \\ \hline
        $P_a$ & air pressure & 101325 & Pa \\ 
        $T$ & temperature & $-2 + 0.024 * depth$ & $K$ \\
        $\phi$ & porosity & $\phi_0 - (1 - SFC(T)) * \left(V_w + V_i\right) / V_{REV}$ & (--) \\
        $S_w$ & water saturation & $SFC(T) * \left(\left(V_w + V_i\right) / V_{REV}\right) / \phi $ & (--) \\ 
        $c_i$ & ice concentration & $(\left(V_w + V_i\right) / V_{REV} - \phi * S_w) / (1 - \phi)$ & (--)\\ 
        \hline 
        
    \end{tabular}
\end{center}

\subsection{Boundary Conditions}
\begin{center}
    \begin{tabular}{|c|l|c|c|c|}
        \hline 
        \textbf{Boundary} & \textbf{Type} & \textbf{Variable} & \textbf{Value} & \textbf{Unit}\\ \hline
        \multirow{3}{*}{Surface} & Dirichlet & $P_a$ & 101325 & Pa\\ 
        & Dirichlet & T &  $T_{mean} - T_{amp} * \cos \left( 2 \pi \left( t / \left(60^2 * 24 * 365\right) + 45/365\right)\right)$ & K \\ 
        & Dirichlet & $S_w$ & $S_{w, surface}$& (--) \\ \hline 
        
        \multirow{3}{*}{Left and Right} & Neumann & $\nabla P_a \cdot \mathbf{n} $ & 0 & Pa\\ 
        & Neumann & $\nabla T \cdot \mathbf{n} $ & 0 & K \\ 
        & Neumann & $\nabla P_w \cdot \mathbf{n} $ & 0 & (--)\\ \hline 
        
        \multirow{3}{*}{Bottom} & Neumann & $\nabla P_a \cdot \mathbf{n} $ & 0 & Pa\\ 
        & Neumann & $\nabla T \cdot \mathbf{n} $ & 0.024 & K \\ 
        & Neumann & $\nabla P_w \cdot \mathbf{n} $ & 0 & (--) \\ \hline 
    \end{tabular}
\end{center}

\subsection{Zugspitze}
\begin{center}
     \begin{tabular}{|l|c|c|c|}
        \hline
        \textbf{Scenario} & \textbf{Parameter} & \textbf{Value} & \textbf{Unit} \\ \hline
        
        \multirow{7}{*}{\shortstack[l]{Zugspitze \\ unsaturated \\ high permeability}} & $T_{amp}$ & 7 & K \\ 
        & $T_{mean}$ & $269.15 - 6.5 * (z - z_{WS}) $ & K \\
        & $z_{WS}$ & 2960 & m \\  
        & $S_{w,surface}$ & 0.8 & (--) \\ 
        & $\left(V_w + V_i\right) / V_{REV}$ & 0.8 & (--) \\  
        & $K_0$ & $10^{-12}$ & $m^2$ \\ 
        & $\phi_0$ & 0.05 & (--) \\ \hline 
        
        \multirow{7}{*}{\shortstack[l]{Zugspitze \\ unsaturated \\ low permeability}} & $T_{amp}$ & 7 & K \\ 
        & $T_{mean}$ & $269.15 - 6.5 * (z - z_{WS}) $ & K \\
        & $z_{WS}$ & 2960 & m \\  
        & $S_{w,surface}$ & 0.8 & (--) \\ 
        & $\left(V_w + V_i\right) / V_{REV}$ & 0.8 & (--) \\  
        & $K_0$ & $10^{-13}$ & $m^2$ \\ 
        & $\phi_0$ & 0.05 & (--) \\ \hline 
        
        \multirow{7}{*}{\shortstack[l]{Zugspitze \\ saturated}} & $T_{amp}$ & 7 & K \\ 
        & $T_{mean}$ & $269.15 - 6.5 * (z - z_{WS}) $ & K \\
        & $z_{WS}$ & 2960 & m \\  
        & $S_{w,surface}$ & 1 & (--) \\ 
        & $\left(V_w + V_i\right) / V_{REV}$ & 1 & (--) \\  
        & $K_0$ & $10^{-12}$ & $m^2$ \\ 
        & $\phi_0$ & 0.05 & (--) \\ \hline 
     \end{tabular}
\end{center}

\subsection{Matterhorn}
\begin{center}
     \begin{tabular}{|l|c|c|c|}
        \hline
        \textbf{Scenario} & \textbf{Parameter} & \textbf{Value} & \textbf{Unit} \\ \hline
        
        \multirow{7}{*}{Matterhorn} & $T_{amp}$ & 8 & K \\ 
        & $T_{mean}$ & $267.35 - 6.5 * (z - z_{WS}) $ & K \\ 
        & $z_{WS}$ & 3488 & m \\  
        & $S_{w,surface}$ & 0.8 & (--) \\ 
        & $\left(V_w + V_i\right) / V_{REV}$ & 0.8 & (--) \\  
        & $K_0$ & $10^{-14}$ & $m^2$ \\ 
        & $\phi_0$ & 0.01 & (--) \\ \hline 
        
        \multirow{8}{*}{\shortstack[l]{Matterhorn \\ with warming}} & $T_{amp}$ & 8 & K \\ 
        & $T_{mean}$ & $267.35 - 6.5 * (z - z_{WS}) + 0.045 * q $ & K \\ 
        & $z_{WS}$ & 3488 & m \\ 
        & $q$ & $\max \{ 0, t / \left(60^2 * 24 * 365\right) - 100\} $ & years \\  
        & $S_{w,surface}$ & 0.8 & (--) \\ 
        & $\left(V_w + V_i\right) / V_{REV}$ & 0.8 & (--) \\  
        & $K_0$ & $10^{-14}$ & $m^2$ \\ 
        & $\phi_0$ & 0.01 & (--) \\ \hline 
     \end{tabular}
\end{center}

\newpage 
\bibliography{refs}

\end{document}